\documentclass[11pt]{article}
\usepackage[linktocpage=true]{hyperref}
\usepackage[all,knot,matrix,arrow]{xy}
\usepackage{color}
\usepackage[textwidth = 430 pt, textheight = 630 pt]{geometry}
\definecolor{MyDarkBlue}{rgb}{0.15,0.25,0.45}
\usepackage{epsfig,rotating}
\usepackage{amsmath,amssymb}
\usepackage{amsfonts}
\usepackage{mathrsfs}
\usepackage{bbm}
\usepackage{bm}
\usepackage[T1]{fontenc}
\usepackage{hyperref}
\hypersetup{
colorlinks=true,
citecolor=MyDarkBlue,
linkcolor=MyDarkBlue,
urlcolor=MyDarkBlue,
pdfauthor={Christian S\"amann and Richard J. Szabo},
pdftitle={Groupoids, Loop Spaces and Quantization of 2-Plectic Manifolds},
pdfsubject={hep-th}
breaklinks=true
}

\let\fn\footnote
\renewcommand{\footnote}[1]{\linespread{1.1}\fn{#1}\linespread{1.29}}

\usepackage{fancyhdr}
\usepackage[left]{lineno}

\makeatletter\renewcommand{\section}{\@startsection
{section}{1}{\z@}{-3.5ex plus -1ex minus
    -.2ex}{2.3ex plus .2ex}{\bf }}
\makeatletter\renewcommand{\subsection}{\@startsection{subsection}{2}{\z@}{-3.25ex
plus -1ex minus
   -.2ex}{1.5ex plus .2ex}{\it }}
\makeatletter\renewcommand{\subsubsection}{\@startsection{subsubsection}{3}{-2.45ex}{-3.25ex
plus -1ex minus -.2ex}{1.5ex plus .2ex}{\it }}
\renewcommand{\thesection}{\arabic{section}}
\renewcommand{\thesubsection}{\arabic{section}.\arabic{subsection}}
\renewcommand{\@seccntformat}[1]{\@nameuse{the#1}.~~}

\renewcommand{\theequation}{\thesection.\arabic{equation}}
\makeatletter \@addtoreset{equation}{section}

\renewenvironment{thebibliography}[1]
     {\baselineskip=16pt plus 2pt minus 1pt
      \section*{\large\refname
        \@mkboth{\MakeUppercase\refname}{\MakeUppercase\refname}}%
     \list{\@biblabel{\@arabic\c@enumiv}}%
           {\settowidth\labelwidth{\@biblabel{#1}}%
            \leftmargin\labelwidth
            \advance\leftmargin\labelsep
            \@openbib@code
            \usecounter{enumiv}%
            \let\p@enumiv\@empty
            \renewcommand\theenumiv{\@arabic\c@enumiv}}%
      \sloppy
      \clubpenalty4000
      \@clubpenalty \clubpenalty
      \widowpenalty4000%
      \sfcode`\.\@m}

\newcommand{\appendices}{
\section*{Appendix}\label{appendices}\setcounter{subsection}{0}
\addcontentsline{toc}{section}{Appendix}
\setcounter{equation}{0}
\makeatletter
\renewcommand{\theequation}{\Alph{subsection}.\arabic{equation}}
\renewcommand{\thesubsection}{\Alph{subsection}}
\@addtoreset{equation}{subsection}
\makeatother
}

\usepackage{epsfig,rotating}
\usepackage{amsmath,amssymb}
\usepackage{amsfonts}
\usepackage{mathrsfs}
\usepackage{bbm}
\usepackage{bm}

\usepackage{graphicx}
\usepackage{xypic}

\def\slasha#1{\setbox0=\hbox{$#1$}#1\hskip-\wd0\hbox to\wd0{\hss\sl/\/\hss}}

\def\periodb#1{\setbox0=\hbox{$#1$}#1\hskip-\wd0\hbox to\wd0{-}}

\newcommand{\delder}[1]{\frac{\delta}{\delta #1}}   		

\newcommand{\unit}{\mathbbm{1}}   			
\newcommand{\id}{\mathrm{id}}   			

\newcommand{\CA}{\mathcal{A}}    			

\newcommand{\pd}{\dot{p}}
\newcommand{\xd}{\dot{x}}
\newcommand{\xdd}{\ddot{x}}

\newcommand{\CC}{\mathcal{C}}

\newcommand{\CCL}{\mathscr{L}}
\newcommand{\CD}{\mathcal{D}}
\newcommand{\CCD}{\mathscr{D}}

\newcommand{\CG}{\mathcal{G}}

\newcommand{\CH}{\mathcal{H}}
\newcommand{\CCH}{\mathscr{H}}

\newcommand{\CJ}{\mathcal{J}}
\newcommand{\CK}{\mathcal{K}}
\newcommand{\CCK}{\mathscr{K}}
\newcommand{\CL}{\mathcal{L}}

\newcommand{\CO}{\mathcal{O}}

\newcommand{\CP}{\mathcal{P}}

\newcommand{\CR}{\mathcal{R}}
\newcommand{\CS}{\mathcal{S}}
\newcommand{\CT}{\mathcal{T}}

\newcommand{\CU}{\mathcal{U}}
\newcommand{\CV}{\mathcal{V}}

\newcommand{\frg}{\mathfrak{g}}				

\newcommand{\frH}{\mathfrak{H}}

\newcommand{\sfp}{{\sf p}}
\newcommand{\sfs}{{\sf s}}
\newcommand{\sft}{{\sf t}}
\newcommand{\sfm}{{\sf m}}

\newcommand{\mbf}[1]{{\boldsymbol {#1} }}

\newcommand{\mbfT}{{\mbf T}}

\newcommand{\nablatr}{{{\Large\blacktriangledown}}}

\newcommand{\FT}{\mathbbm{T}}     			
\newcommand{\FR}{\mathbbm{R}}     			
\newcommand{\FC}{\mathbbm{C}}     			
\newcommand{\RZ}{\mathbbm{Z}}     			
\newcommand{\CPP}{{\mathbbm{C}P}}    			

\newcommand{\lambdah}{\hat{\lambda}}

\newcommand{\dd}{\mathrm{d}}     			
\newcommand{\dpar}{\partial}     			
\newcommand{\delb}{{\overline{\delta}}}	     		
\newcommand{\embd}{{\hookrightarrow}}     		
\newcommand{\de}{\mathrm{e}}     			
\newcommand{\di}{\mathrm{i}}     			
\newcommand{\eps}{{\varepsilon}}			

\newcommand{\ald}{{\dot{\alpha}}}     			
\newcommand{\bed}{{\dot{\beta}}}
\newcommand{\gad}{{\dot{\gamma}}}
\newcommand{\ded}{{\dot{\delta}}}

\newcommand{\thetad}{{\dot{\theta}}}

\newcommand{\eand}{{\qquad\mbox{and}\qquad}}     		
\newcommand{\ewith}{{\qquad\mbox{with}\qquad}}
\newcommand{\efor}{{\qquad\mbox{for}\qquad}}
\newcommand{\eon}{{\qquad\mbox{on}\qquad}}

\newcommand{\der}[1]{\frac{\dpar}{\dpar #1}}   		
\newcommand{\dder}[1]{\frac{\dd}{\dd #1}}   		
\newcommand{\derr}[2]{\frac{\dpar #1}{\dpar #2}}   	
\newcommand{\tr}{\,\mathrm{tr}\,}     			
\newcommand{\pr}{\mathsf{pr}}     			

\newcommand{\agl}{\mathfrak{gl}}     			

\newcommand{\sU}{\mathsf{U}}     			

\newcommand{\sSU}{\mathsf{SU}}

\newcommand{\sMat}{\mathsf{Mat}}

\newcommand{\bfF}{\boldsymbol{F}}
\newcommand{\sDiff}{\mathsf{Diff}}

\newcommand{\sPU}{\mathsf{PU}}

\newcommand{\sEnd}{\mathsf{End}\,}

\newcommand{\remark}[1]{}     				
\def\tyng(#1){\hbox{\tiny$\yng(#1)$}}			
\def\tyoung(#1){\hbox{\tiny$\young(#1)$}}			

\newcommand{\beq}{\begin{eqnarray}}
\newcommand{\eeq}{\end{eqnarray}}

\newcommand{\clidf}{\Omega_{{\rm cl},\RZ}}
\newcommand{\Pair}{{\sf Pair}}
\newcommand{\hol}{{\sf hol}}

\newenvironment{conditions}{
\vspace{-2mm}\begin{itemize}
\setlength{\itemsep}{-1mm}
}{\vspace{-2mm}\end{itemize}}

\renewcommand*\l@section[2]{%
  \ifnum \c@tocdepth >\z@
    \addpenalty\@secpenalty
    \addvspace{0.8em \@plus\p@}%
    \setlength\@tempdima{1.5em}%
    \begingroup
      \parindent \z@ \rightskip \@pnumwidth
      \parfillskip -\@pnumwidth
      \leavevmode \bfseries
      \advance\leftskip\@tempdima
      \hskip -\leftskip
      #1\nobreak\hfil \nobreak\hb@xt@\@pnumwidth{\hss #2}\par
    \endgroup
  \fi}

\begin{document}

\begin{titlepage}
\begin{flushright}
EMPG--12--24\\ NI--12034--BSM
\end{flushright}
\vskip 2.0cm
\begin{center}
{\LARGE \bf Groupoids, Loop Spaces \\[0.3cm] and Quantization of 2-Plectic Manifolds}
\vskip 1.5cm
{\Large Christian S\"amann and Richard J.\ Szabo}
\setcounter{footnote}{0}
\renewcommand{\thefootnote}{\arabic{thefootnote}}
\vskip 1cm
{\em Department of Mathematics\\
Heriot-Watt University\\
Colin Maclaurin Building, Riccarton, Edinburgh EH14 4AS, U.K.}\\[0.1cm] and\\[0.1cm]
{\em Maxwell Institute for Mathematical Sciences, Edinburgh, U.K.}
\\[0.5cm]
{Email: {\ttfamily C.Saemann@hw.ac.uk , R.J.Szabo@hw.ac.uk}}
\end{center}
\vskip 2.0cm
\begin{center}
{\bf Abstract}
\end{center}
\begin{quote}
We describe the quantization of 2-plectic manifolds as they arise in
the context of the quantum geometry of M-branes and non-geometric flux
compactifications of closed string theory. We review the groupoid
approach to quantizing Poisson manifolds in detail, and then extend it to
the loop spaces of 2-plectic manifolds, which are naturally symplectic
manifolds. In particular, we discuss the groupoid quantization of the
loop spaces of $\FR^3$, $\FT^3$ and $S^3$, and derive some interesting
implications which match physical expectations from string theory and M-theory.
\end{quote}
\end{titlepage}

\tableofcontents


\bigskip

\section{Introduction and summary}

\paragraph{Higher quantization.}

The quantization of general physical systems is a long-standing problem. It is the inverse operation of taking the classical limit of a quantum mechanical system. As the classical limit cannot be expected to yield an injective map between quantum and classical systems, it is not surprising that quantization is not unique. Correspondingly, there is a wealth of different approaches to quantization. In this paper we will be interested in the groupoid approach, which also employs the techniques of geometric quantization. We will limit ourselves to the kinematical problem of quantization, i.e.,\ the construction of an algebra of quantum operators from a classical phase space.

Aside from their quantum physics origin, there are many reasons for
studying quantized phase spaces. In particular, the description of
certain string theory backgrounds often requires generalized notions
of geometry that can be regarded as quantized symplectic
manifolds. As we will review below, both string theory as well as recent developments in
M-theory also suggest to consider analogous extensions for
multisymplectic manifolds. These are manifolds endowed with closed,
non-degenerate differential forms of degrees larger than two. In the
context of Nambu mechanics~\cite{Nambu:1973qe}, multisymplectic manifolds serve as
multiphase spaces leading to dynamical systems with time evolution governed by more than one Hamiltonian.

The development of quantization methods for multisymplectic manifolds
is still only in its infancy; only the issue of prequantization
has been explored in detail. Recall that to prequantize a symplectic manifold
one constructs a principal circle bundle (or an associated line
bundle) over the manifold which is endowed with a connection whose
curvature is proportional to the symplectic form. These structures are used in the definition of the Atiyah algebroid, whose sections yield a faithful representation of the Poisson algebra of the symplectic manifold. In the simplest multisymplectic setting, the case of 2-plectic manifolds, the prequantum circle bundle is replaced by a prequantum gerbe. Furthermore, these manifolds come with a Lie 2-algebra that essentially takes over the role of the Poisson algebra \cite{Baez:2008bu}. A faithful representation of this Lie 2-algebra can be constructed on sections of a Courant algebroid that is naturally derived from the multisymplectic structure \cite{Rogers:2010ac,Rogers:2010sc}. While the prequantization of 2-plectic manifolds seems clear, the incorporation of a polarization turning the prequantization into an actual quantization is, however, far from obvious.

Here we will exploit the proposal of Hawkins~\cite{Hawkins:0612363} of
how to include a polarization into the groupoid approach to
quantization. In the general groupoid approach, see e.g.\
\cite{MR1103911,Crainic:0403269,Crainic:0403268}, one starts from the
observation that any Poisson manifold naturally gives rise to a Lie
algebroid. There is a standard technique for quantizing duals of Lie
algebras as convolution algebras of the Lie groups obtained from
integrating the Lie algebras. This technique can be extended to Lie
algebroids, which yields convolution algebras of their integrating Lie
groupoids. Quantization in this sense therefore corresponds to integration. The Lie algebroid of a Poisson manifold, however, has twice the dimension of the original manifold. To correct for this, one needs to introduce a polarization and in \cite{Hawkins:0612363} it was shown how to do this. By using the resulting twisted polarized convolution algebras, one can use the groupoid approach to obtain Moyal 
planes, noncommutative tori, as well as any geometrically quantized K\"ahler manifold.

The groupoid approach is particularly interesting for the quantization of multisymplectic manifolds for two reasons. Firstly, multisymplectic manifolds are in a certain sense ``categorified'' symplectic manifolds, and the groupoid approach seems very suitable for a categorification. Secondly, Hawkins' version of the groupoid approach directly constructs an algebra of operators, avoiding the introduction of a Hilbert space. This is useful as the Lie 2-algebra of a 2-plectic manifold should yield some nonassociative algebra of quantum operators, because its Jacobiator (or associator) is in general non-vanishing. Therefore a model of quantum operators as endomorphisms on some linear Hilbert space seems to be excluded.

Instead of categorifying Hawkins' approach, we will follow a different
route in this paper which goes back to ideas due to Gaw\k{e}dzki
\cite{Gawedzki:1987ak} and Brylinski \cite{0817647309}. The loop space
of any 2-plectic manifold carries a natural symplectic structure. In
the case of three-dimensional 2-plectic manifolds, this symplectic
structure can moreover be extended to a K\"ahler structure. So instead
of quantizing the 2-plectic manifold directly, we can apply the
prescriptions of groupoid quantization to the loop space of the
manifold. However, it
should be stressed that the difficulties of dealing with
multisymplectic manifolds and prequantum gerbes is traded for the
difficulties of working with the corresponding infinite-dimensional loop spaces.

There is yet a fourth way of considering the problem of
quantizing multisymplectic manifolds. As stated above, multisymplectic
manifolds may give rise to multiphase spaces in Nambu mechanics. Just
as a symplectic form yields a Poisson structure, certain
multisymplectic forms encode a Nambu-Poisson structure. It has been
suggested to use Nambu-Poisson structures as starting points for
quantization and to quantize them in terms of Filippov's $n$-Lie
algebras \cite{Filippov:1985aa}. This approach, however, has not
delivered complete results so far; see~\cite{DeBellis:2010pf} and
references therein.

\paragraph{Physical context.}

The treatment of this paper is mostly technical, so let us briefly
summarize some of the physical settings in which the implications of
our loop space quantization are applicable. For a more detailed
discussion of our motivation from the perspective of string theory,
see~\cite{Saemann:2012ex}.

Our approach to the quantization of 2-plectic manifolds from the
perspective of their corresponding loop spaces is natural from several
different points of view. In the context of M-theory, it was realized
some time ago by~\cite{Bergshoeff:2000jn,Kawamoto:2000zt} that
canonical quantization of an open membrane boundary on an M5-brane in
a constant $C$-field background leads to a
noncommutative loop space. This is a functional analog of the
noncommutative geometry that arises in canonical quantization of open
strings ending on D-branes in a constant $B$-field background; as the M2-brane boundaries are strings,
this describes a noncommutative closed string geometry.
Further motivation for this approach comes from the fact
that loop spaces have been recently used very successfully in lifting
the Nahm construction of monopoles to self-dual strings, in a manner that is closely related to certain quantized
2-plectic
manifolds~\cite{Gustavsson:2008dy,Saemann:2010cp,Palmer:2011vx}. 

A separate context in which our approach is applicable is to the
description of non-geometric three-form flux
compactifications of closed string theory. Loop space
quantization is very natural from the point of view of closed string sigma-models,
and it yields a tractable framework in which to analyze
nonassociative structures which have been found recently. 
In certain duality frames,
our results reproduce the noncommutative geometry probed by closed
strings on a three-torus $\FT^3$ which wind around one of the
cycles, as derived in~\cite{Lust:2010iy} through canonical
quantization and closed string mode expansions; in particular, we
nicely reproduce the picture of~\cite{Bouwknegt:2004ap} for the
quantized $\FT^3$ with $Q$-flux (T-fold) as a fibration of
noncommutative two-tori over the circle $S^1$. The Lie 2-algebraic nature of the
nonassociative geometry found through closed string vertex operators in flat space
in~\cite{Blumenhagen:2011ph} is contained in our quantization of the
loop space of $\FR^3$, while our quantization of the three-sphere
$S^3$ should naturally contain the nonassociative structures which
have been found in the $\sSU(2)$ WZW
model~\cite{Blumenhagen:2010hj}.

Some of the categorified ingredients of higher quantization also have
natural physical relevance. Given a gerbe with integral curvature
$H$ on a manifold $M$, the generalized tangent bundle $E=
TM\oplus T^*M$ has the structure of an exact Courant algebroid with
the $H$-twisted Courant-Dorfman bracket. This relationship between abelian gerbes and Courant
algebroids is useful for understanding prequantization
conditions. Thus Courant algebroids offer a unifying picture of approaches to quantization of 2-plectic manifolds; recall that they are the natural analogs of the Atiyah Lie algebroid for a $\sU(1)$-gerbe
over~$M$.

Courant algebroids also naturally appear in string phase spaces. This
exemplifies the relation between exact Courant algebroids and
{gerbes}, since a gerbe on $M$ with connective structure naturally
determines a $\sU(1)$-valued 2-holonomy over surfaces $S\subset M$. Dimensional reduction over torus bundles of exact
Courant algebroids yields non-exact Courant algebroids with additional
symmetries that can be understood in terms of T-duality. In
particular, T-duality (including the B{\"u}scher rules) can be understood
as an isomorphism of Courant algebroids over T-dual spaces. In this
way Courant algebroids offer a geometric description of non-geometric
fluxes and T-folds; see~\cite{Baraglia:2011dg} for further
details. The relevance of Courant algebroids for the geometrization
and quantization of
non-geometric $R$-flux backgrounds in terms of nonassociative geometry is also stressed in~\cite{Mylonas:2012pg}.

Lie 2-algebras have further recently proven useful as gauge
 structures in certain M-brane
 models~\cite{Palmer:2012ya,Saemann:2012uq}. As explained
 in~\cite{Saemann:2012ex}, these M2-brane models are closely related
 to the quantization of 2-plectic manifolds, in particular of
 $S^3$. Lie 2-algebras and their integrating Lie 2-groups also play an
 important role in the nonassociative quantum geometry of closed
 string $R$-flux
 compactifications~\cite{Mylonas:2012pg} in the same manner as the
 quantizations spelled out in the present paper.

\paragraph{Summary.}

We now come to a summary of the results of this
paper. Preliminary accounts of the material in this article can be
found in the
papers~\cite{Samann:2011zq,Saemann:2011yi,Saemann:2012ex}, to which
the reader can turn for a more concise and relaxed presentation of
our ideas and results, as well as further physical background and
applications.

In the following we review all ingredients in the combination of the
loop space and the groupoid approaches to quantization in detail. As
an explicit example for the groupoid approach, we work out the
quantization which yields the standard
$\kappa$-Minkowski space. When it comes to loop spaces, the difficulty
in using groupoids is that the convolution algebra requires the
introduction of a measure. Unfortunately, there is no natural
reparametrization invariant measure on loop space and we therefore
cannot expect to arrive at a complete picture. We can, however, glean
some interesting consequences from following the groupoid approach as
far as possible. We will focus on three particular examples:
three-dimensional Euclidean space $\FR^3$, the three-torus $\FT^3$ and
the three-sphere $S^3$. In each case we demonstrate that the loop
space quantization reduces to the standard geometric
quantizations of $\FR^2$, $\FT^2$ and $S^2$, respectively, via an
M-theory type dimensional reduction with the loops winding around the
compactified direction.

In the case of the real affine space $\FR^3$, we find that the groupoid structure receives
corrections in powers of the Poisson tensor that are unexpected from
the corresponding structures on $\FR^2$. They combine into a twist
element that yields an operator product which agrees with the noncommutative deformations of loop space derived directly from M-theory in~\cite{Bergshoeff:2000jn,Kawamoto:2000zt,Berman:2004jv}.

In the case of the torus $\FT^3$, the loop space decomposes into
different winding sectors. It is therefore necessary to introduce a
Bohr-Sommerfeld quantization condition that essentially agrees with
results expect from a closed string sigma-model analysis. Moreover,
the Bohr-Sommerfeld variety exhibits an interesting non-trivial
structure which reproduces exactly the $Q$-space noncommutative
geometry of closed string zero modes from~\cite{Lust:2010iy} as well
as the noncommutative torus fibration of~\cite{Bouwknegt:2004ap}.

Finally, the case of the sphere $S^3$ follows more closely that of the
groupoid quantization of $S^2$, where one identifies the space of global
holomorphic sections of the prequantum line bundle over the pair
groupoid with the Hilbert space. Just like line bundles over $S^2$,
abelian (bundle) gerbes over $S^3$ are characterized up to isomorphism by an integer. When transgressed to loop space, they yield a
special class of line bundles. One can readily define the global
holomorphic sections of these line bundles. The explicit description
of the space of such sections, however, is highly non-trivial. Using
twistor techniques we manage to identify a special class of sections that reduce to the
sections of the prequantum line bundles over $S^2$ when following the
projection $\pi:S^3\rightarrow S^2$ of the Hopf fibration.

Our analysis leaves a number of open problems. First of all, it would be interesting to describe the full space of global holomorphic sections of the transgressed gerbes over $S^3$. Defining some notion of square integrability on these sections would require revisiting the issue of constructing suitable measures on loop spaces. If such measures were found and proven to be natural, one could also resolve the issue of how to define the groupoid convolution algebras that should appear in our quantization of loop spaces. Most importantly, however, one should explore a more direct approach to the quantization of 2-plectic manifolds. Just as the quantization of a symplectic manifold yields a prequantum line bundle and eventually a Hilbert space, the quantization of a 2-plectic manifold yields a prequantum gerbe and presumably a categorified or 2-Hilbert space. Correspondingly, the Lie algebra of quantum operators that forms a deformed representation of the Poisson algebra of a symplectic manifold should 
be replaced by a Lie 2-algebra of quantum operators forming a deformed representation of the Lie 2-algebra replacing the Poisson algebra in the case of a 2-plectic manifold.

\paragraph{Outline.}

This paper is structured as follows: In \S2 we review the
groupoid approach to quantization in some detail, in particular
Hawkins' proposal which incorporates polarization. In \S3 we
explain the extension of Poisson and symplectic geometry as well as
prequantization to the categorified context. We review the K\"ahler
geometry of certain loop spaces and show how they allow us to reduce
categorified prequantization to ordinary prequantization. In \S4--\S6 we apply Hawkins' version of the groupoid approach to the
quantization of the loop spaces of $\FR^3$, $\FT^3$ and $S^3$,
respectively. Three appendices at the end of the paper collect
relevant definitions and some technical details for the reader's
convenience. In Appendix~A we give all necessary details concerning
Lie groupoids, Lie algebroids and integrating symplectic groupoids. In
Appendix~B we collect various features of gerbes in the different
contexts that we need them in the main text of the paper. In
Appendix~C we present details on Courant algebroids within the context
of categorified geometric quantization.

\section{Quantization of Poisson manifolds and symplectic groupoids\label{QuantPoisson}}

The groupoid approach to quantization was introduced by Weinstein as a
means of extending the framework of geometric quantization of
symplectic manifolds to generic Poisson manifolds; in a certain sense,
explained below and in Appendix~\ref{app:A},
Poisson manifolds are the infinitesimal objects associated with
symplectic groupoids.
In this section we will review the quantization of Poisson manifolds
in the framework of Lie groupoids as
presented by Hawkins in \cite{Hawkins:0612363}, who clarified the
notion of polarization in this context. We are mainly interested in the
application of this quantization procedure to $\FR^d$, $\FT^d$, $S^2$, and
$\kappa$-deformed spaces. These are spaces for which many notions like that
of groupoid polarization and the derivation of a twist element for the
convolution algebra simplify. The review here will therefore be rather
concise on the technical side. Instead, we will present a number of
explicit examples in detail.

\subsection{Quantization of Lie algebra duals\label{sec:dualquant}}

The motivation behind the groupoid approach to quantization is the
observation that the quantization of the dual of a Lie algebra yields
a noncommutative algebra of functions which can be identified with the twisted convolution $C^*$-algebra of the integrating Lie group. Consider the two {\em Kirillov-Kostant-Souriau} (KKS) {\em Poisson
  structures} on $M=\frg^*$, the dual of a Lie algebra
$\frg$. Interpreting elements of $\frg$ as linear functions on $M$,
one defines $\{g_1,g_2\}_\pm(x):=\pm\, \langle x,[g_1,g_2]\rangle$ for
$g_1,g_2\in\frg$, $x\in M$. These
brackets extend to polynomial functions via the Leibniz
identity for the Lie bracket. To obtain all smooth functions on
$\frg^*$, we consider the completion of the polynomial algebra with respect to a suitable norm. The resulting linear Poisson structures on $\CC^\infty(\frg^*)$ are invariant under the
coadjoint action of the corresponding connected Lie group $G$.

Quantization corresponds to a map from (a subset of) the algebra of smooth
functions $\CC^\infty(\frg^*)$ on the dual
Lie algebra to the reduced convolution $C^*$-algebra $\CC_r^*(G)$ of the
integrating Lie group which is $G$-invariant. Roughly speaking, the
steps are as follows: We start from the functions in $\CC^\infty(\frg^*)$, which we  Fourier transform to $\CC^\infty(\frg)$ and subsequently exponentiate to a group $G$. There, we can convolve and finally follow the
steps back to $\CC^\infty(\frg^*)$. Below we flush out some details of
this construction, following~\cite{JSTOR:2374874,Kathotia}.

The basic idea is to identify the generators $g^i$ of $\frg$ as
coordinates for $\frg^*$ via duality. For small $t\in\FR$, this gives a
one-to-one correspondence between group elements $\exp(t\, g)\in G$,
$g\in\frg$, and functions of the form $\de^{t\,x}$ on $\frg^*$. Pulling
back the group multiplication to these functions thus yields a quantization of
$\CC^\infty(\frg^*)$. The group multiplication is computed via the usual
Baker-Campbell-Hausdorff formula
\beq
\exp(g)\, \exp(h)=\exp\big(H(g,h)\big)
\eeq
with $g,h\in\frg$. Here
\beq
H(g,h)= g+h+\mbox{$\frac12$}\, [g,h]+\mbox{$\frac1{12}$}\,
\big(\big[g\,,\,[g,h]\big]+\big[[g,h]\,,\, h\big]\big) +\dots
\eeq
is the Hausdorff series whose terms lie in $\frg$. For nilpotent Lie algebras, this construction
is equivalent to Kontsevich's quantization of the linear Poisson
structure on $\frg^*$ and to the quantization by the universal
enveloping algebra of $\frg$; in this case the exponential map is a
diffeomorphism from the whole of $\frg$ to $G$.

At the level of Schwartz functions, the Fourier transform is a linear
map $\CS(\frg^*)\to \CS(\frg)$ which sends $f\in\CS(\frg^*)$ to
$\tilde f\in\CS(\frg)$ with
\beq
\tilde f(g)= \int_{\frg^*}\, \dd x\ \de^{-2\pi\, \di\, \langle
  x,g\rangle}\, f(x)
\eeq
for $g\in\frg$, where $\dd x$ is the
translationally invariant Haar measure for the additive
group structure on $\frg^*$. For the inverse Fourier transform, we use
the exponential map to ``identify'' $\frg$ with
$G$. Clearly, $\exp:\frg\to G$ is not a
bijection in general, so this can only be done on an open set $U$ in $\frg$
where the exponential map is a diffeomorphism. To circumvent this
problem, we restrict to a dense subset of functions
$\CS_c(\frg^*)\subset \CS(\frg^*)$
which under Fourier transformation maps
to the space $\CC_c^\infty(\frg)$ of smooth functions with compact
support on $\frg$. This also ensures integrability with respect to the left-invariant Haar measure
\beq\label{eq:liHaar}
\dd\mu(s) = \lambda(g)\ \dd g \ewith \lambda(g):=
\big|\det(\dd\, \exp)_g\big| \ ,
\eeq
where $s=\exp(g)$ and $\dd g$ is the
translationally invariant Haar measure on $\frg$. For nilpotent
algebras, one can take the set $U$ to be all of $\frg$ and
replace $\CC_c^\infty(\frg)\subset L^1(G,\dd\mu)$ with the larger
Schwartz space $\CS(\frg)$;
in this case $\det(\dd\, \exp)_g=1$ for all $g\in\frg$.

In this picture, the inverse Fourier transform
of $\tilde f\in \CC_c^\infty(\frg)$ is defined by
\beq
W(\tilde f\, ):= \int_G\, \dd\mu(s)\ W(s) \, \tilde f(s)~,
\eeq
where $W(s)= \exp\big(2\pi\, \di\, \langle x,g\rangle\big)$ for
$s=\exp(g)$ obeys $W(s)\, W(t)= W(s\,t)$; it thus represents $G$ with the
various identifications made above. The $C^*$-algebra generated by
$W(\tilde f\, )$ for $\tilde f\in \CC_c^\infty(\frg)$ is the image of the
representation induced by $W$ of the reduced $C^*$-algebra
$\CC_r^*(G)$. This can be seen by explicitly computing
\beq
W(\tilde f\, )\, W(\tilde f'\,)= \int_G\, \dd \mu(s)\ W(s)\, \Big(\, \int_G\, \dd\mu(t)\ \tilde f(t)\,
\tilde f'(t^{-1}\, s)\, \Big) = W(\tilde f\circledast \tilde f'\,) \ ,
\eeq
where
\beq
(\tilde f\circledast \tilde f'\,)(s) := \int_G\,\dd\mu(t)\ \tilde
f(t)\, \tilde f'(t^{-1}\, s)
\eeq
is the usual convolution product on $\CC_r^*(G)$. It follows that
$W$ defines an algebra homomorphism.

Rewriting all of this in terms of the identifications made above, the deformed
product $f\star f':=W(\tilde f\circledast \tilde f'\,)$ of functions $f,f'\in\CS_c(\frg^*)$ thereby reads
\beq
(f\star f'\,)(x) &=& \int_{\frg}\, \dd g\ \lambda(g)\ \int_\frg\, \dd g'\
\lambda (g'\,) \ \de^{2\pi\,
  \di\, \langle x,g'\, \rangle}\, \tilde f(g)\, \tilde
f'\big(H(-g,g'\, )\big) \label{fstarfp} \\[4pt] &=& \int_\frg\, \dd g\ \lambda(g)\ \int_\frg\, \dd g'\
\lambda (g'\,) \ \de^{2\pi\,
  \di\, \langle x,g+g'\, \rangle}\ \de^{2\pi\,
  \di\, \langle x,H(-g,g'\,)+g-g'\, \rangle}\, \tilde f(g\,)\, \tilde
f'(g'\, ) \ . \nonumber
\eeq

\paragraph{Example.} The $\kappa$-Minkowski algebra is the solvable Lie algebra $\frg_\kappa$ of
dimension $d$ with generators $g^0,g^1,\dots,g^{d-1}$ obeying
\beq
[g^0,g^i]=\di\, \kappa^{-1} \, g^i \eand [g^i,g^j]=0
\label{kappaminkalg}\eeq
for $i,j=1,\dots,d-1$ and $\kappa>0$. Generally, if $[g,h]=z\, h$ for
$z\in\FC$, then the Hausdorff series can be summed explicitly to give the
braiding identity
\beq
\exp(g)\, \exp(h)=\exp\big(\de^z\, h\big)\, \exp(g) \ .
\eeq
It follows that
\beq
\exp\big(\di\, c_\mu\, g^\mu\big) =
\exp\big(\di\,c_0\, g^0\big)\, \exp\big(\di\, c_i'\,
g^i \big) \ewith c_i':=\frac\kappa{c_0}\,
\big(1-\de^{-\kappa^{-1} \, c_0}\big)\, c_i \ .
\eeq
For $p_0,p_1,\dots,p_{d-1}\in\FR$, we write
\beq
W(p_0,\vec p\,)= V_{\vec p}\, U_{p_0}~,
\eeq
where $\vec p=(p_1,\dots,p_{d-1})$, while
\beq
U_{p_0}=\exp\big(\di\, p_0\, g^0\big) \eand V_{\vec
  p}= \exp\big(-\di\, p_i\, g^i\big) \ .
\eeq
These elements generate the non-abelian $\kappa$-Minkowski
group $G_\kappa$~\cite{Agostini:2005mf} with the group law
\beq
W(p_0,\vec p\,)\, W(p'_0,\vec p\,'\,) = W(p_0+p_0',\vec
p+\de^{-\kappa^{-1} \,
  p_0}\, \vec p\,'\,) \ .
\eeq
This group law is that of the crossed product
$G_\kappa \cong \FR^{d-1}\rtimes_{\alpha_\kappa} \FR$ with the twisting group isomorphism
$\alpha_\kappa(p_0)\vec p:= \de^{-\kappa^{-1} \,
  p_0}\, \vec p$ for $\vec p\in\FR^{d-1}$. It is connected and
simply-connected, hence $G_\kappa$ is uniquely determined by its Lie algebra
$\frg_\kappa$. The Jacobian in the Haar measure \eqref{eq:liHaar} is $\lambda(p_0,\vec p\,
)=\de^{\kappa^{-1} \, p_0}$,
and the $C^*$-algebra generated by
\beq
W(\tilde f\, ):= \int_{G_\kappa} \, \dd p_0\ \dd\vec p\ \de^{\kappa^{-1} \, p_0}\ \tilde f(p_0,\vec p\,)\,
W(p_0,\vec p\,) \ ,
\eeq
for $\tilde f\in\CC_c^\infty(G_\kappa)$, is the representation of
$\CC_r^*(G_\kappa)$ by $W$. The convolution product of two functions
$\tilde f,\tilde f'\in \CC_c^\infty(G_\kappa)$ takes the explicit form
\beq\label{eq:ConvAlgebraKappaMinkowski}
(\tilde f\circledast_\kappa \tilde f'\,)(p_0,\vec p\, )= \int_{G_\kappa}\,
\dd p_0'\ \dd \vec p\,'\ \de^{\kappa^{-1} \, p_0'}\ \tilde f(p_0',\vec p\,'\,)\,
\tilde f\big(p_0-p_0'\,,\,\de^{\kappa^{-1} \,
  p_0'}\, (\vec p-\vec p\,'\, )\big) \ .
\eeq

\paragraph{Example.} The Moyal product is also a special case of this quantization. Consider
the real vector space $V=\FR^d$ with constant Poisson bivector $\pi=\frac12\, \pi^{ij} \,
\partial_i\wedge \partial_j \in\bigwedge^2 V^*$, $\pi^{ij}=-\pi^{ji}\in\FR$. The
Heisenberg algebra is the 2-step nilpotent Lie algebra $\frg_\pi$ of dimension $d+1$ with basis
$g^1,\dots,g^d,h$ and relations
\beq
[g^i,g^j]=\di\, \pi^{ij}\, h \eand [h,g^i]=0 \ .
\label{Moyalrels}\eeq
In this case, one defines the usual Weyl
operators
\beq
W(p)=\exp\big(\di\, p_i\, g^i\big)
\eeq
for any $p\in V^*$, which satisfy
\beq
W(p)\, W(p'\,)=\exp\big(-\mbox{$\frac\di2$}\, p_i\, \pi^{ij}\, p_j'\,
h \big) \ W(p+p'\,) \ .
\label{Weylgp}\eeq
Here $\lambda=1$, so the associated $C^*$-algebra is generated by
\beq
W(\tilde f\, ):=\int_{V^*}\, \dd p\ W(p) \, \tilde f(p)
\eeq
for $\tilde f\in\CC_c^\infty(V^*)$. Because (\ref{Weylgp}) is not a
group representation, this $C^*$-algebra is not related to the group
$C^*$-algebra; in fact, this is a projective representation of the
abelian group structure on $V^*$. One can transform it into a representation of
the 2-step nilpotent Heisenberg group $H_\pi^{d+1}$ which is a central
extension of $V^*$ by a fixed element $\pi\in \bigwedge^2 V^*$, in which case the $C^*$-algebra
associated to the relations (\ref{Moyalrels}) is related to the
$C^*$-algebra of $H_\pi^{d+1}$. Since $h$ commutes with everything we can
set it to a scalar $h=\hbar$; this means that the representation
generated by $W$ is irreducible. Then the twisting factor in
(\ref{fstarfp}) is given by
\beq
H(-p',p)= p-p' - \mbox{$\frac\hbar2$}\, p_i\, \pi^{ij}\, p_j' \ ,
\eeq
and so in this case the convolution product on the Heisenberg group
$H_\pi^{d+1}$ induces the usual Moyal product on $\CS_c(V)$. This defines the \emph{twisted} convolution $C^*$-algebra
$\CC_r^*(V^*,\sigma_{\pi,\hbar})$ where the twisting group two-cocycle
$\sigma_{\pi,\hbar}:V^*\times V^*\to \sU(1)$ is given by
\beq
\sigma_{\pi,\hbar}(p,p'\,):=\de^{-\frac{\di\, \hbar}2
\, p_i\,
\pi^{ij}\, p_j' } \ .
\eeq
The
cocycle condition here is equivalent to antisymmetry of $\pi^{ij}$ and
it ensures that the algebra of Weyl operators (\ref{Weylgp}) is
associative.

\subsection{Poisson geometry, Lie algebroids and integrating Lie groupoids\label{sec:PMandLieAlgebroids}}

The quantization of duals of Lie algebras can be lifted to a quantization of Lie algebroids.\footnote{See Appendix~\ref{app:A} for the definition and examples of Lie algebroids.} As we briefly review below, Poisson manifolds come with a natural Lie algebroid structure. We can therefore generalize the above quantization procedure to this class of manifolds.

Recall that the Poisson bivector field $\pi$ which defines the Poisson bracket
\begin{equation}
 \{f,g\}_\pi:=\pi(\dd f, \dd g)
\label{bracketpidef}\end{equation}
has vanishing Schouten bracket $[\pi,\pi]_M=0$. This is equivalent to the the Leibniz rule and the Jacobi identity
\begin{equation}
 \{f\, g,h\}_\pi=f\, \{g,h\}_\pi+\{f,h\}_\pi\, g \eand
 \{f,\{g,h\}_\pi\}_\pi=\{\{f,g\}_\pi,h\}_\pi+\{g,\{f,h\}_\pi\}_\pi
\end{equation}
being satisfied for all $f,g,h\in \CC^\infty(M)$. The pair $\big(\CC^\infty(M)\,,\,\{-,-\}_\pi\big)$ is called a {\em Poisson
algebra}; it has in particular the structure of a Lie algebra. If $M$ is a symplectic manifold with symplectic structure given by a closed non-degenerate two-form $\omega\in\Omega^2(M)$, then the inverse of $\omega$ gives rise to a Poisson bivector field
$\pi=\omega^{-1}$ and thus any symplectic manifold is also a Poisson manifold.

On the other hand, a bivector field $\pi$ on $M$ induces a map
$\pi^\sharp:T^*M \rightarrow TM$ via contraction together with a bracket on $\CC^\infty(M,T^*M)=\Omega^1(M)$,
\begin{equation}
 [\alpha,\beta]_\pi:=\CL_{\pi^\sharp (\alpha)}\beta-\CL_{\pi^\sharp
   (\beta)}\alpha-\dd\pi(\alpha,\beta)
\label{Omega1brackets}\end{equation}
for $\alpha,\beta\in\Omega^1(M)$, where $\CL$ denotes the Lie derivative. The objects $(T^*M,\pi^\sharp,[-,-]_\pi)$ form a Lie algebroid if and only if the Schouten bracket of $\pi$ vanishes. Thus, the cotangent bundle of a Poisson manifold is naturally a Lie algebroid. See~\cite{Kosmann:2007aa} for further aspects of the connection between Poisson structures and Lie algebroids.

We would now like to generalize the quantization of Lie algebra duals to Lie algebroids $E=T^*M$ describing the Poisson manifolds $(M,\pi)$. The dual bundle $E^*$ of such a Lie algebroid $E$ is naturally a Poisson manifold with Poisson structure inherited from the Lie bracket $[-,-]_E$ of $E$~\cite{Coste:1987aa}; explicitly, the Poisson bracket $\{-,-\}_{E^*}$ on $\CC^\infty(E^*)$ is given by:
\begin{itemize}
\item[(i)] $\{f\circ p_{E^*},g\circ p_{E^*}\}_{E^*}=0$ for
  $f,g\in\CC^\infty(M)$, where $p_{E^*}: E^*\to M$ is the bundle
  projection.
\item[(ii)] $\{X,f\circ p_{E^*}\}_{E^*}= (\CL_{\rho(X)} f)\circ
  p_{E^*}$ for $f\in\CC^\infty(M)$ and fiberwise linear functions
  $X\in\CC^\infty(M,E)\subset \CC^\infty(E^*)$.
\item[(iii)] $\{X,Y\}_{E^*}=[X,Y]_E$ for fiberwise linear functions
  $X,Y\in\CC^\infty(M,E)\subset \CC^\infty(E^*)$.
\end{itemize}
Note that the Poisson structure on $E^*$ is completely determined by that
on $M$.

By attempting to quantize the cotangent Lie algebroid $E=T^*M$ instead of $M$, we have doubled the dimension of the space to be quantized. This has to be undone by introducing a {\em polarization}. The next step in the quantization procedure is therefore to derive a twisted polarized convolution algebra of an integrating Lie groupoid, generalizing that of a Lie group; see Appendix~\ref{app:A} for the relevant definitions. In this paper, we will follow Hawkins' approach~\cite{Hawkins:0612363}, in which the elements of geometric quantization are lifted to groupoids. Moreover, the notion of integration of the Lie algebroid $T^*M$ of a Poisson manifold $(M,\pi)$ is shifted to the equivalent one of finding an integrating symplectic groupoid $\Sigma$ of the Poisson manifold $(M,\pi)$. The essential steps of his procedure are:
\begin{enumerate}
 \item[{\em (1)}] Find an integrating symplectic groupoid $\Sigma$ of the Poisson manifold $(M,\pi)$.
 \item[{\em (2)}] Construct a prequantization of $\Sigma$ with data $(E,\nabla,\sigma)$, where $E$ is the prequantum line bundle with connection $\nabla$ and $\sigma$ is a cocycle twist.
 \item[{\em (3)}] Endow $\Sigma$ with a groupoid polarization $\CP$.
 \item[{\em (4)}] Construct the $C^*$-algebra of $\Sigma/\CP$ as a twisted polarized convolution algebra, either through half-densities or through a Haar system.
\end{enumerate}
One issue with this prescription is that not every Poisson manifold allows for an
integrating symplectic groupoid, see Appendix~\ref{app:A} for more details. This, however, is
merely a groupoid version of the statement that not every manifold is
quantizable. Note that integrability of $T^*M$ is equivalent to
integrability of the brackets (\ref{Omega1brackets}) on
$\Omega^1(M)$. In the classes of examples that we consider in this paper, the integrating symplectic groupoid will always be taken to be a suitable variant of
the pair groupoid $\Sigma=\Pair(M)=M\times M$, which integrates the tangent Lie
algebroid $TM$.

In the ensuing subsections, we will review Hawkins' approach and its
ingredients to the extent necessary for our further constructions. We
shall also give the groupoid quantization yielding $\kappa$-Minkowski space as a novel example.

\subsection{Geometric, Berezin and Berezin-Toeplitz quantization\label{sec:GBB-Tquant}}

Before coming to the groupoid formalism, let us briefly review the pertinent ingredients of
geometric quantization. In the formulation of Kostant, a {\em prequantization} of a symplectic
  manifold $(M,\omega)$ consists of a hermitian line bundle $E$ over $M$
with connection $\nabla$ and curvature two-form $F_\nabla=\nabla^2= -2\pi\,
\di\, \omega$; this imposes the prequantization condition that the
symplectic form must define an integer cohomology class $[\omega]\in
H^2(M,\RZ)$, i.e., $\omega\in \clidf^2(M)$, where generally
$\clidf^k(M)$ denotes the group of closed $k$-forms (i.e.,\ de Rham $k$-cocycles) on $M$ with
integer periods. The
prequantum Hilbert space is identified with the space of
$L^2$-sections of $E$.\footnote{The prequantization condition ensures that this space is sufficiently large.} It is well-known that this Hilbert space is too
large, and it is necessary to reduce it by choosing ``half a canonical coordinate system''~\cite{Woodhouse:1992de}. This can be done
by introducing a {\em polarization}, i.e., a foliation of $(M,\omega)$
by Lagrangian submanifolds or, equivalently, a smooth distribution
$\CP\subset T_\FC M$ which is integrable and Lagrangian. Locally, the
cotangent bundle $T^*N$ of a Lagrangian submanifold $N\subset M$ can be
naturally identified with the full symplectic manifold $M$; a real polarization then
corresponds to a choice of ``momentum space/position space''
representation. A section $\psi\in L^2(M,E)$ is \emph{polarized} if it
is covariantly constant along the leaves of the polarization $\CP$,
i.e., $\nabla_X\psi=0$ for all $X\in \CC^\infty(M,\CP)$.

The construction of the actual Hilbert space $\CCH$ from the prequantum
Hilbert space and a polarization is in general quite complicated. Here we are merely interested in the quantization of K\"ahler manifolds using {\em K\"ahler polarization}: We identify $\CCH$ with (a completion of) $H^0(M,E)$, the space of global holomorphic sections of $E$.
This construction of the Hilbert space is common to geometric
quantization \cite{Woodhouse:1992de} as well as to Berezin and
Berezin-Toeplitz quantization~\cite{Bordemann:1993zv,Schlichenmaier-1996aa}, see also~\cite{IuliuLazaroiu:2008pk}. The inner product on $\CCH$ is obtained
from the hermitian structure $h$ on $E$ together with a suitable measure
$\dd \mu$ on $M$ such as, e.g., the Liouville measure $\frac{\omega^n}{n!}$
for a compact K\"ahler manifold $(M,\omega)$ of dimension $\dim_\FC
M=n$; it is given by
\begin{equation}
 \langle \psi_1| \psi_2\rangle:=\int_M\, \dd \mu(x)\ h(\psi_1(x), \psi_2(x)) \efor \psi_1, \psi_2\in H^0(M,E) ~.
\end{equation}

To complete the quantization, we need a map from a quantizable subset
of functions $\CC^\infty_q(M)\subseteq \CC^\infty(M)$ to endomorphisms
on $\CCH$. In Berezin and Berezin-Toeplitz
quantization,\footnote{While geometric quantization employs the same
  Hilbert space, the quantization map differs from the ensuing ones.} one uses the Rawnsley coherent states \cite{Rawnsley:1976gb} $|x\rangle\in\CCH$, $x\in M$, to construct the coherent state projector
\begin{equation}
 \hat{P}_x:=\frac{|x\rangle\langle x|}{\langle x|x\rangle}~,
\end{equation}
which is simultaneously a function on $M$ and an operator on $\CCH$,
and hence bridges the classical and quantum worlds. It induces the maps
\begin{equation}
\begin{aligned}
 B\,: \ &\sEnd(\CCH) \ \longrightarrow \
 \CC^\infty_q(M)~,&B(\hat{f}\, )(x)&:=\tr_\CCH(\hat{f}\, \hat{P}_x)~,\\[4pt]
 T\,: \ &\CC^\infty_q(M) \ \longrightarrow \
 \sEnd(\CCH)~,&T(f)&:=\int_M\, \dd \mu(x)\ f(x)\, \hat{P}_x~.
\end{aligned}
\end{equation}
The image of the map $B$ defines the set of quantizable functions
$\CC^\infty_q(M)$. Since $B$ is injective, we can define its inverse
$B^{-1}:\CC^\infty_q(M)\rightarrow \sEnd(\CCH)$, and this is the
quantization map in {\em Berezin quantization}. The map $T$ is used in
{\em Berezin-Toeplitz quantization}. The classic example is the fuzzy
sphere, which is the Berezin or Berezin-Toeplitz quantization of the
K\"ahler manifold $\FC P^1$.

Instead of working with a prequantum line bundle, we can also follow Souriau and work with the corresponding principal $\sU(1)$-bundle $p:P\to M$ endowed with a connection $A\in\Omega^1(P)$ such that $\dd A=p^*\omega$. The line bundle $E$ is then recovered as the associated bundle $E=P\times_\rho\FC$, where
$\rho(\de^{\di\, \phi})\triangleright z=\de^{\di\, \phi}\, z$ for all
$z\in \FC$. The KKS prequantization gives a faithful unitary
representation of the Poisson algebra $\big(\CC^\infty(M)\,,\,
\{-,-\}_{\omega^{-1}}\big)$ on a Hilbert space, which can be
constructed by using the \emph{Atiyah Lie algebroid} associated to
$P$. The Atiyah Lie algebroid sequence is the exact sequence of vector
bundles
\beq
{\rm ad}(P) \ \longrightarrow \ E_{\rm
  At}(P):=TP\,\big/\, \sU(1) \ \longrightarrow \
TM \ ,
\label{Atiyahseq}\eeq
where $TM$ is the tangent Lie algebroid of $M$ and ${\rm ad}(P)$, being
a bundle of Lie algebras over $M$, is naturally a Lie algebroid with the zero
anchor map and fiberwise Lie bracket. The vector bundle $E_{\rm
  At}(P):=TP/\sU(1)\to M$ naturally inherits from $TP$ the structure
of a Lie algebroid, called the Atiyah Lie algebroid; its smooth
sections are the $\sU(1)$-invariant vector fields on $P$, and it
integrates to the Atiyah Lie groupoid $G_{\rm At}(P)=P\times_{\sU(1)}
P$ which is the quotient of the pair groupoid $\Pair(P)=
P\times P$ by $\sU(1)$. In this sense, consideration of the Atiyah Lie algebroid
over $M$ is equivalent to considering the cotangent bundle $T^*M$ as a
Lie algebroid.

A connection $A$ on $P$ is equivalent to a splitting of the exact
sequence (\ref{Atiyahseq}),\footnote{This splitting occurs in the
  category of \emph{vector bundles}, not of Lie algebroids, or else
  the corresponding curvature vanishes. Alternatively, we can get
  non-trivial Chern classes by passing to $L_\infty$-algebroids.}
which gives a monomorphism of Lie algebras
\beq
\big(\CC^\infty(M)\,,\,
\{-,-\}_{\omega^{-1}}\big) \ \longrightarrow \
\big(\CC^\infty(M,E_{\rm At}(P))\,,\, [-,-]_{TP/\sU(1)}\big) \ .
\eeq
This yields a faithful representation of the Poisson algebra of
$\CC^\infty(M)$. In particular, the
Poisson algebra of $\CC^\infty(M)$ acts as linear differential
operators on $\sU(1)$-equivariant functions $P\to\FC$, which
correspond to (global) $L^2$-sections of the hermitian line bundle $E$
associated to $P$. The Poisson algebra $\CC^\infty(M,T^*P/\sU(1))$ of
the Lie algebroid $E_{\rm At}(P)$ is quantized to the $C^*$-algebra of
the groupoid $G_{\rm At}(P)$, which is isomorphic to the $C^*$-algebra
$\CCK(L^2(M))\otimes\CC^*_r(\sU(1))$ of $\sU(1)$-invariant compact
operators on $L^2(P)$ (see \S\ref{sec:quantmap} below).

This prequantization is only the first step, and to complete the picture one should introduce a polarization and restrict to a polarized Hilbert space. There are however no faithful representations of the Poisson algebra on the polarized Hilbert space and one has to use approximate Lie algebra homomorphisms. As we will be working in the Kostant picture using prequantum line bundles, we refrain from going into further details.

\subsection{Prequantization and polarization of symplectic groupoids}\label{subsec:2.4}

Prequantization of groupoids dates back to work of Weinstein and Xu \cite{MR1103911}, see also
\cite{Crainic:0403269,Crainic:0403268} for more recent accounts. Consider a
symplectic groupoid $(\Sigma,\omega)$ integrating a Poisson manifold
$(M,\pi)$. This means that $\Sigma\rightrightarrows M$ is a Lie
groupoid, $(\Sigma,\omega)$ is a symplectic manifold, $\omega$ is a
multiplicative two-form and the target map\footnote{We use
  $(\sfs,\sft,\sfm,\unit)$ to denote the structure maps of the
  groupoid. Moreover, $\pr_1$ and $\pr_2$ denote the obvious
  projections from the ``set of composable arrows'' $\Sigma_{(2)}$,
  i.e.\ the 2-nerve of the simplicial manifold underlying $\Sigma$, to
  $\Sigma$.} of the groupoid $\sft:\Sigma\rightarrow M$ is a Poisson map, see Appendix \ref{app:A}. A \emph{prequantization} of $(\Sigma,\omega)$ is given by the prequantization of $\Sigma$ as a symplectic manifold, i.e., a hermitian line bundle
$E\rightarrow \Sigma$ endowed with a connection $\nabla$ such that
$F_\nabla=-2\pi\, \di\, \omega$, together with a
two-cocycle~$\sigma$.

The cocycle $\sigma$ provides an associative
multiplication on fibers of $E$ at different composable points of
$\Sigma$. It is a section of the hermitian coboundary line bundle
$\dpar^*E^*:=\pr_1^*E^*\otimes \sfm^*E\otimes \pr_2^*E^*$ over the 2-nerve
$\Sigma_{(2)}$, where $E^*$ denotes the line bundle dual to $E$. The fiber
multiplication is defined on $\psi_{g_1}\in E_{g_1}$ and
$\psi_{g_2}\in E_{g_2}$ by
\beq
\psi_{g_1}\bullet_\sigma \psi_{g_2}=\big\langle \sigma(g_1,g_2)\,,\,
\psi_{g_1}\otimes\psi_{g_2}\big\rangle \ ,
\eeq
which induces a multiplication
$\bullet_\sigma:\CC^\infty(\Sigma,E)\otimes\CC^\infty(\Sigma,E)\to \CC^\infty(\Sigma_{(2)},\sfm^*E)$ on sections
of $E$. Associativity is ensured by the multiplicative cocycle property
\beq
\sigma\big(g_1\,,\,\sfm(g_2,g_3)\big)\otimes\sigma(g_2,g_3)=
\sigma(g_1,g_2)\otimes \sigma\big(\sfm(g_1,g_2)\,,\,g_3\big)
\eeq
for $g_1,g_2,g_3\in\Sigma$. We
further demand that $\sigma$ has unit norm, and that it is covariantly
constant with respect to the (local) symplectic potential $\theta$ for
$\omega$,\footnote{This condition is necessary for it to be compatible with the
  polarization later on.}
\beq
\nabla\sigma= \dd\sigma-\di\, (\partial^*\theta)\, \sigma=0 \ .
\label{covconstsigma}\eeq

The condition on $\Sigma$ to be prequantizable is identical to that of
$M$ being prequantizable in the sense of Kostant
\cite{Crainic:0403269,Crainic:0403268}: All the periods of $M$ have to
be integer multiples of $2\pi$; in that case, the prequantization of
$\Sigma$ is unique (up to isomorphism).

A {\em polarization of a symplectic groupoid} $\Sigma$
\cite{Hawkins:0612363} is a polarization $\CP$ of $\Sigma$ as a symplectic
manifold which is {\em multiplicative}, i.e., the distribution $\CP\subset T_\FC \Sigma$ is compatible with and
closed under the groupoid multiplication:
$\sfm^*(\CP_{(2)})_{(g_1,g_2)}=\CP_{\sfm(g_1,g_2)}$, where
$\CP_{(2)}:=(\CP\times\CP)\cap T_\FC \Sigma_{(2)}$; see~\cite{Hawkins:0612363}
for further details.
The polarizations $\CP\subset T_\FC \Sigma$ that we will be mostly interested in
are given in terms of foliations of a fibration of
groupoids $\sfp: \Sigma\rightarrow \Sigma/\CP$,\footnote{See Appendix~\ref{app:A}.}
which give rise to strongly admissible polarizations. If the leaves of
$\CP$ are simply connected, then $E$ is trivial along these leaves and
there is a canonical identification $E\cong \sfp^* E_\CP$ where
$E_\CP$ is a line bundle over $\Sigma/\CP$. Abusing notation, we
denote the corresponding ``reduced'' cocycle with coefficients in
$E_\CP$ by the same symbol $\sigma$. Moreover, even when the
polarization $\CP$ is not the kernel foliation of a fibration of
groupoids, we continue to use the same notation $\Sigma/\CP$ to
indicate the polarized symplectic groupoid.

\subsection{Twisted polarized convolution algebra}\label{subsec:2.5}

To construct the twist element $\sigma$, we will always start from a
symplectic potential $\theta$ with $\dd \theta=\omega$ which is
\emph{adapted}, i.e., compatible with the polarization $\CP$ in the
sense that $\theta\in \CC^\infty(\Sigma,\CP^\perp)$. In these
cases, the line bundle $E_\CP$ is trivial on $\Sigma/\CP$, and
$\sigma\in \CC^\infty((\Sigma/\CP)_{(2)},\sU(1))$. To satisfy the
multiplicativity condition $0=\dpar^*\omega=\dpar^*\, \dd \theta=\dd
\, \dpar^*\theta$, we have to demand that $\dpar^* \theta$
is closed. The cocycle $\sigma$ is then constructed from the projection $\sfp:\Sigma\rightarrow \Sigma/\CP$ via
\begin{equation}
 \sfp^*(\sigma^{-1}\, \dd \sigma)=\di\, \dpar^*\theta~,
\end{equation}
where we have used (\ref{covconstsigma}).

The convolution product on $\CC_c^\infty(\Sigma/\CP)$ is now
defined as follows. Introduce a left Haar system of measures
$\{\dd\mu^x \ | \ x\in (\Sigma/\CP)_{(0)}\}$ on $(\Sigma/\CP)^x:=\{g\in\Sigma/\CP \ | \
\sft(g)=x \} =\sft^{-1}(x)$, i.e., $\dd\mu^{\sft(g)}(g^{-1}\,
h) = \dd\mu^{\sfs(g)}(h)$, and set
\beq
(\tilde f\circledast_\sigma \tilde f'\,)(g) &=& \int_{k\,k'=g}\,
\sigma(k,k'\,) \ \tilde f(k)\, \tilde f'(k'\,)\nonumber \\[4pt] &:=& \int_{(\Sigma/\CP)^{\sft(g)}}\,
  \dd\mu^{\sft(g)}(h)\ \sigma(g\, h, h^{-1})\ \tilde f(g\,h)\, \tilde
  f'(h^{-1})
\label{twistedconv}\eeq
for $\tilde f,\tilde f'\in\CC_c^\infty(\Sigma/\CP)$. Then the appropriate completion defines the $\sigma$-twisted reduced
convolution $C^*$-algebra $\CC_r^*(\Sigma/\CP,\sigma)$ of the
polarized symplectic groupoid $\Sigma/\CP$; it is the convolution
algebra of polarized sections of $E$.

A more canonical definition uses \emph{half-densities} and sections of
the associated complex line bundle $\Omega_{\Sigma/\CP}^{1/2}\to \Sigma/\CP$ to
define $\CC_r^*(\Sigma/\CP,\sigma)$, which ensures that the integrand
used in (\ref{twistedconv}) is always a density on each fiber
$(\Sigma/\CP)^x$ of the map $\sft$; for the most part the above definition
will suffice for the examples we consider.

\subsection{Generalized Bohr-Sommerfeld quantization condition\label{BohrSommerfeld}}

If the leaves of the polarization $\CP$ are not simply connected, then
the construction above requires some further truncation. In this case,
the leaves only admit parallel sections when the holonomy of the
connection $\nabla$ is trivial; if not,
then the symplectic potential $\theta$ is not adapted.
This condition on the holonomy can be regarded as a
generalized Bohr-Sommerfeld quantization condition. We define the
\emph{Bohr-Sommerfeld groupoid} as the subvariety
$\Sigma_0\subseteq\Sigma$ such that the holonomy\footnote{The formula (\ref{holgammatheta}) for the holonomy as written is symbolic in
general; for topologically non-trivial symplectic potentials $\theta$,
a precise expression is given in Appendix~\ref{app:B}.}
\beq
\hol_\gamma(\theta):=\exp\Big(2\pi\, \di\, \oint_\gamma\, \theta\Big)
\label{holgammatheta}\eeq
is equal to $1$ for all conjugacy classes of loops $[\gamma]\in
\pi_1(\Sigma_0)^\sim$, i.e., $\oint_\gamma\, \theta\in \RZ$. This in turn trivializes the line bundle $E_\CP$
over $\Sigma_0/\CP$, and so the previous construction can now be
applied to the reduced groupoid~$\Sigma_0/\CP$.

\subsection{Quantization map\label{sec:quantmap}}

Lagrangian submanifolds of $\Sigma$ with a section that is covariantly
constant along each leaf of the polarization $\CP$ (and which vanishes on
leaves not intersecting $E$) correspond to elements of $\CA:=\CC_r^*(\Sigma/\CP,\sigma)$.
From the definition of a symplectic groupoid, the graph of the
multiplication $\sfm$ is a Lagrangian submanifold of \footnote{For a
  symplectic manifold $(M,\omega)$, we denote $\overline{M}=(M,-\omega)$.}
$\overline{\Sigma}\times\overline{\Sigma}\times \Sigma$ and so
quantizes to an element of $\CA^*\otimes\CA^*\otimes\CA$, i.e., a map
$\CA\otimes\CA \to\CA$, the unit section
$\unit:M\hookrightarrow\Sigma$ is a Lagrangian submanifold which
quantizes to a unit element in $\CA$, and the graph of the inversion
$g\mapsto g^{-1}$ is a Lagrangian submanifold of
$\overline{\Sigma}\times \Sigma$ which quantizes to an element of
$\CA^*\otimes\CA$. These elements define the multiplication, identity
and inversion in the noncommutative algebra~$\CA$.

The procedure just outlined yields a
quantization of the dual $E^*$ of the associated Lie
algebroid $E=A(\Sigma/\CP)$. The quantization map $E^*\to \CC_r^*(\Sigma/\CP,\sigma)$ is given by the
composition of the extension of Fourier transform to
vector bundles $E\to E^*$, described in~\cite[\S7]{Landsman:2000aa}
using suitable left Haar systems on $E$ and $E^*$, with the extension of the exponential map
$E \to \Sigma/\CP$, described
in~\cite[\S2]{Landsman:2000aa} in the case that the Lie algebroid
$E$ is endowed with a connection. Here we regard the
twisted polarized convolution algebra $\CC_r^*(\Sigma/\CP,\sigma)$
as a quantization of the Poisson manifold $M$, in the sense that the
diagram
\begin{equation}
\vspace{4pt}
\begin{xy}
\xymatrix@C=20mm{
& \ \Sigma \ \ar@/^/[dl]^{\sfs} \ar@/_0.5pc/[dl]_{\sft}
\ar@//[dr]^{\sfp} & \\
M \ & &  \ \Sigma/\CP
}
\end{xy}
\vspace{4pt}
\end{equation}
can be used to ``pullback'' the groupoid multiplication on
$\Sigma/\CP$ to $\CC^\infty(M)$; however, even though the target map
$\sft:\Sigma\to M$ is a Poisson map and $\Sigma$ is $\sfs$-connected,
the pullback $\sft^* (f)$ of a function $f\in\CC^\infty(M)$ is
generically not quantizable, and in general the explicit quantization maps
$\CC^\infty(M)\to\CC_r^*(\Sigma/\CP,\sigma)$ are not known.

\paragraph{Example.} When $M=pt$ is a one-point space and the Lie groupoid $\Sigma$ is a Lie group $G$, the
left-invariant Haar measure on $G$ induces a left Haar system and the
convolution algebra $\CC^*_r(G)$ is the reduced group
$C^*$-algebra. The Lie algebra $\frg$ of $G$ is the associated Lie
algebroid $A(G)$, and the Poisson structure on the dual $A^*(G)=\frg^*$ is
the usual linear KKS Poisson structure. In this case the quantization
map $\CC_c^\infty(\frg^*)\to\CC_r^*(G)$ is essentially the composition of Fourier transformation on the
affine space $\frg^*\to\frg$ with the exponential map $\exp:\frg\to
G$, as constructed in \S\ref{sec:dualquant}.

\paragraph{Example.} When $\Sigma$ is the pair groupoid $\Pair(M)=M\times M$ of an oriented
manifold $M$, the 2-nerve is $\Sigma_{(2)}=M\times M\times M$ with
multiplication map $\sfm(x,y,z)=(x,z)$ and projections
$\pr_1(x,y,z)=(x,y)$, $\pr_2(x,y,z)= (y,z)$. The product in $\CC_r^*(M\times M)$ is the
convolution of kernels
\beq
(\tilde f\circledast \tilde f'\,)(x,z)=\int_M\, \dd\, {\rm vol}_M(y) \
\tilde f(x,y)\, \tilde f'(y,z) \ ,
\label{convkernel}\eeq
and the $C^*$-algebra $\CC_r^*(M\times M)$ acts as integral
kernels on $L^2(M)$, so that $\CC_r^*(M\times M)$ is isomorphic
to the algebra $\CCK(L^2(M))$ of compact operators. In this case the
associated Lie algebroid $A(\Pair(M))$ is the tangent bundle $TM$,
with the standard symplectic Poisson structure on the dual
$A^*(\Pair(M))=T^*M$; the quantization map is thus
$\CC_c^\infty(T^*M)\to\CCK(L^2(M))$. If $M$ is additionally a
symplectic manifold of dimension $2k$ with symplectic two-form $\omega_M=\dd\theta_M$, then it is
easy to check that the symplectic structure
$\omega:=\sfs^*\omega_M-\sft^*\omega_M$ on $\Sigma$ is multiplicative:
If we write $\omega(x,y)=\omega_M(x)-\omega_M(y)$ for $x,y\in
M$, then one has
\begin{equation}
 \begin{aligned}
    \pr_1^* \omega(x,y,z)&=\omega_M(x)-\omega_M(y)~,\\[4pt]
    \pr_2^* \omega(x,y,z)&=\omega_M(y)-\omega_M(z)~,\\[4pt]
    \sfm^*\omega(x,y,z)&=\omega_M(x)-\omega_M(z)~,
 \end{aligned}
\end{equation}
and hence $\partial^*\omega= \pr_1^*\omega-\sfm^*\omega+\pr_2^*\omega=0$.
If $E_M$ is a prequantum line bundle on $M$, then a prequantum line bundle on
$\Sigma=M\times \overline{M}$ is $E= E_M \boxtimes \overline{E_M}$; if
$\CP_M$ is a polarization on $M$, then a symplectic groupoid
polarization on $\Sigma$ is given by $\CP=\CP_M\times
\overline{\CP_M}$. The local symplectic potential
$\theta:=\sfs^*\theta_M-\sft^*\theta_M$ is also multiplicative,
$\partial^*\theta=0$, whence the prequantization cocycle is
$\sigma=1$. The ensuing restriction to the algebra
$\CC_r^*(\Sigma/\CP)$ with the convolution product (\ref{convkernel})
may be constructed from polarized sections of the line bundle
$E\otimes\Omega_{\CP}^{1/2}$, where
$\Omega_\CP:=\det(\CP_M^\perp)\boxtimes
\det(\,\overline{\CP_M}\,^\perp)$~\cite{Hawkins:0612363}. Then
$\CC_r^*(\Sigma/\CP)$ is isomorphic to the algebra of compact
operators on the space of $\CP_M$-polarized sections of the complex
line bundle $E_M\to M$. In (\ref{convkernel}) the product of polarized
sections $\tilde f(x,y)\, \tilde f'(y,z)$ lives in the square root of
$(\Omega_\CP)_{(x,y)}\otimes (\Omega_\CP)_{(y,z)}$. In order to
integrate this over $\sft^{-1}(y)=(\Sigma/\CP)^y$, we have to tensor
with non-vanishing sections of the square root of $\det
T_{(x,y)}^*\sft^{-1}(y)$ and of $\det T_{(y,z)}^*\sfs^{-1}(y)$ to get
a top form. By~\cite[Thm.~5.3]{Hawkins:0612363}, this can be done by
contraction with the Liouville form
$\dd\mu_M=\frac{(\omega_M)^k}{k!}=\det\omega_M\big|_{\sft^{-1}(y)}$ on
$M$. See~\cite{deM.Rios:2002ux} for a derivation of the Moyal product
on $M=\FR^2$ within this framework.

\subsection{Groupoid quantization and $\kappa$-Minkowski space}

Let us now study a non-trivial example for the groupoid quantization
of the dual of a Lie algebra in detail: the $\kappa$-Minkowski algebra
from \S\ref{sec:dualquant}. We start from the real vector space $V=\FR^d$ with the
$+$-KKS Poisson structure induced by the $\kappa$-Minkowski Lie
algebra \eqref{kappaminkalg}. For $d=2$ the corresponding
$\kappa$-Minkowski group $G_\kappa \cong \FR \rtimes\FR_{>0}^*$ is isomorphic to a
connected affine group on the real line, also known as an {\em $a\,
x+b$-group} (here with $a=\de^{-\kappa^{-1}\, p_0}$, $b=p_1$). A
symplectic groupoid over an affine group on $\FR$ has been constructed
in \cite{Mikami:1991aa}; in the following we work in a different
parameterization for arbitrary $d\geq2$ and with a different affine group. This example is a
special case of that studied in~\cite[\S6.3]{Hawkins:0612363}.

As a manifold, the symplectic groupoid over $V$ is the
cotangent bundle $\Sigma=T^*G_\kappa$; we will use
coordinates $x=(x^0,\vec x\, )$ on the affine space $\frg_\kappa^*\cong
T^*_{g}G_\kappa$ and coordinates $p=(p_0,\vec p\, )$ on the group manifold
$G_\kappa$. We will follow now the general construction of groupoid structure maps on the cotangent bundle of a Lie group, see Appendix \ref{app:A}.
The unit embedding is trivially given by $\unit:\frg_\kappa^*\rightarrow
T^*_eG_\kappa$, or explicitly by $\unit_{(x^0,\vec x\, )}=\big((x^0,\vec
x\, )\,,\,(0,\vec 0\,)\big)$.

To derive the source and target maps, note that the left and
right actions of $G_\kappa$ on itself read
\begin{equation}
 L_{p}(q) =(p_0+q_0,\vec p+\de^{-\kappa^{-1}\, p_0}\, \vec q\, ) \eand
 R_{p} (q)=( q_0+p_0,\vec q+\de^{-\kappa^{-1}\,  q_0}\, \vec p\,
 )
\end{equation}
with inverses
\begin{equation}
 L_{p^{-1}} (q) =\big( q_0-p_0\,,\,\de^{\kappa^{-1}\, p_0}\,
 (\vec q-\vec p\, )\big)\eand
 R_{p^{-1}} (q) =\big( q_0-p_0\,,\,\vec q-\de^{-\kappa^{-1}\,
   ( q_0-p_0)}\, \vec p \, \big)
\end{equation}
where $p^{-1}=(-p_0,\de^{\kappa^{-1}\, p_0}\,\vec p\, )$ for $p, q\in
G_\kappa$. These actions induce derivative maps $\dd L_p=L_{p*}:
T_ q G_\kappa\rightarrow T_{p\,  q}G_\kappa$ and $\dd R_p=R_{p*}:
T_ q G_\kappa\rightarrow T_{ q\, p}G_\kappa$. The source and
target maps $\sfs,\sft:T^*G_\kappa\rightarrow \frg_\kappa^*\cong T^*_e
G_\kappa$ are then given by
\begin{equation}
\big\langle \sfs(x,p)\,,\,( y^0, \vec y\,)\big\rangle :=( x^0, \vec
 x\, )\circ
 \dd R_{p}( y^0, \vec y\, ) = \big( x^0-\kappa^{-1}\, p_i\, x^i \quad
 \vec x\,\big)\, \begin{pmatrix}  y^0 \\ \vec y\,^\top \end{pmatrix}
\end{equation}
and
\begin{equation}
\big\langle \sft( x,p)\,,\,( y^0, \vec y\, )\big\rangle :=( x^0, \vec
 x\, )\circ \dd L_{p}( y^0, \vec y\, ) =\big( x^0 \quad
 \de^{-\kappa^{-1}\, p_0}\, \vec x\, \big)\, \begin{pmatrix} y^0 \\ \vec
   y\,^\top \end{pmatrix}
\end{equation}
where we identify $V$ with its dual $V^*$.

Multiplication of two points $( x,p)$ and $( y, q)$ in $T^*G_\kappa$
is defined if $\sfs( y, q)= \sft( x,p)$ or
\begin{equation}
  y^0= x^0+\kappa^{-1} \, \de^{-\kappa^{-1}\, p_0}\,  q_i\, x^i \eand  \vec y=\de^{-\kappa^{-1}\, p_0}\, \vec x~.
\end{equation}
As $ y\in T^*_ q G_\kappa$ is hence determined by $( x,p)$ and the
fiber point $ q$, the 2-nerve of the groupoid $\Sigma$ is
$\Sigma_{(2)}=T^*G_\kappa \times G_\kappa$. We will use coordinates
$\big( (x^0, \vec x\, )\,,\, (p_0,\vec p\, )\,,\,( q_0,\vec  q\, ) \big)$
on $\Sigma_{(2)}$. In these coordinates the product $\sfm$ and the projections $\pr_{1},\pr_{2}$ read
\begin{equation}
\begin{aligned}
\sfm( x,p, q)&:=\big( (x^0+\kappa^{-1}\, \de^{-\kappa^{-1}\, p_0}\,
 q_i\, x^i, \vec x\, )\,,\,(p_0+ q_0,\vec
p+\de^{-\kappa^{-1}\, p_0}\, \vec q\, )\big)~, \\[4pt]
\pr_1( x,p, q)&:=\big( (x^0, \vec x\, )\,,\, (p_0,\vec p\,) \big)~,\\[4pt]
\pr_2( x,p, q)&:=\big( (x^0+\kappa^{-1}\, \de^{-\kappa^{-1}\, p_0}\,
 q_i\, x^i,\de^{-\kappa^{-1}\, p_0}\,\vec x\, )\,,\,( q_0,\vec
 q\, )\big)~.
\end{aligned}
\end{equation}

A polarization $\CP$ of the groupoid $\Sigma$ is given by the kernel
of the tangent map to the bundle projection $\sfp:T^*G_\kappa\to
G_\kappa$, regarded as a fibration of groupoids, i.e., $\Sigma/\CP= G_\kappa$.

Define the adapted symplectic potential $\theta= x^\mu\, \dd p_\mu$,
which gives rise to the canonical symplectic structure $\omega=\dd
\theta =\dd x^\mu\wedge\dd p_\mu$ on
$T^*G_\kappa$. Note that $\unit^*\theta=0$ and $\theta$ is conormal to
the groupoid polarization $T^*G_\kappa\rightarrow G_\kappa$. Moreover,
one has
\beq
\dpar^*\theta&=& x^\mu\, \dd p_\mu+\big( x^0+\kappa^{-1}\,
 \de^{-\kappa^{-1}\, p_0}\, q_i\, x^i\big)\, \dd
  q_0+\de^{-\kappa^{-1}\,  q_0}\, x^i\, \dd  q_i \nonumber \\ && -\, \big(
 x^0+\kappa^{-1}\, \de^{-\kappa^{-1}\, p_0}\,  q_i\,
 x^i\big)\, \big(\dd p_0+\dd q_0 \big) \nonumber \\ && -\, x^i\,\big(\dd
 p_i+\de^{-\kappa^{-1}\, p_0}\, \dd  q_i-\kappa^{-1}\,
 \de^{-\kappa^{-1}\, p_0}\,  q_i\, \dd p_0\big) \ = \ 0~.
\eeq
Thus the twist element $\sigma=1$ is trivial and we obtain the
untwisted convolution algebra
\eqref{eq:ConvAlgebraKappaMinkowski}. This is a more general feature
of the groupoid quantization of duals of Lie algebras~\cite[\S6.3]{Hawkins:0612363}.

\section{Quantization of 2-plectic manifolds and loop spaces}

In this section we will discuss the use of loop spaces and knot spaces in
the quantization of 2-plectic manifolds. We start with an outline of
our approach, and then review all necessary notions from loop space geometry before discussing the loop space extension of the groupoid approach to quantization.

\subsection{Overview}

Just as a symplectic structure on a manifold $M$ is defined in terms
of a closed, non-degenerate two-form, a 2-plectic structure is given
by a closed, non-degenerate three-form. As we saw in
\S\ref{sec:GBB-Tquant}, a symplectic form with integer periods
represents the first Chern class of a prequantum line bundle in
geometric quantization, fixing it up to isomorphism. Correspondingly,
a 2-plectic form with integer periods specifies the Dixmier-Douady
class of an abelian prequantum gerbe. Pushing the analogy further, we
recall that the Hilbert space in geometric quantization is derived
from the space of polarized global sections of the prequantum line
bundle. The corresponding notions for a gerbe, however, seem to be
still unclear. One can in principle work with the realization of gerbes as principal $\sPU(\CH)$-bundles
$P\to M$ where $\CH$ is an infinite-dimensional separable Hilbert space (see e.g.~\cite{Bouwknegt:2000qt,Bouwknegt:2001vu}), and consider the 
space of sections thereof; indeed, $H^3(M,\RZ)$ is the group of isomorphism
classes of projective bundles $P$ (with infinite-dimensional separable
fibers) while the automorphism group of $P$ is the group $H^2(M,\RZ)$ of complex
line bundles on $M$, see e.g.~\cite{Atiyah:2004jv}. 
 However, the appropriate definition of a polarization still remains
 to be found. But the lesson here is that attempting to
resort to conventional geometric settings thus inevitably leads us into the
realm of infinite-dimensional analysis, which will be a recurrent
theme below.

We can circumvent these quantization issues by using a trick that goes back to
Gaw\k{e}dz\-ki~\cite{Gawedzki:1987ak}, see also~\cite{0817647309}. The
Dixmier-Douady class of the abelian prequantum gerbe on $M$ gives rise
to the first Chern class of a prequantum line bundle on the loop space
$\CL M$ of $M$ via the cohomological transgression homomorphism. A representative of the
first Chern class yields a symplectic structure on $\CL M$. We will
focus on three-dimensional Riemannian 2-plectic manifolds. Their knot
spaces, i.e.\ the loop spaces factored by the group of
reparameterizations of the loops, comes with a natural complex
structure. This complex structure can be combined with the symplectic
structure obtained from the transgressed volume form to a K\"ahler
structure. In principle, we can then follow the usual recipe of
geometric quantization to construct a Hilbert space as the space of global sections of the corresponding prequantum line bundle on knot space. The problem with the definition of a polarization is solved e.g.\ by using ordinary 
K\"ahler polarization on the global sections.

This approach is also well-motivated from a different perspective:
While a symplectic form on $M$ induces a Lie algebra provided by a
Poisson bracket on smooth functions on $M$, a 2-plectic form induces a
Lie 2-algebra consisting of smooth functions together with a subspace of
one-forms on $M$. The only non-trivial binary bracket in this Lie
2-algebra is a Poisson-like bracket on the one-forms. As we will see
below, the prequantization of this Lie 2-algebra is rather clear. One
can construct a representation of this Lie 2-algebra, just as the KKS
prequantization gave a representation of the usual Poisson
algebra. However, it is not known how to implement a polarization into
this picture. If we apply the transgression map to switch to the loop
space setting, the situation becomes much nicer: The transgression map
embeds the subspace of one-forms on $M$ into the space of smooth functions on the loop space of $M$ and it maps the 2-plectic form on $M$ to a symplectic form on $\CL M$. On loop space, the transgression of the Poisson-like brackets 
of two one-forms is precisely the loop space Poisson bracket of the transgression of the one-forms.

To actually quantize loop space, we then apply Hawkins' groupoid approach. The reason for this is the following: Ultimately, we hope to establish a more direct quantization procedure for multisymplectic manifolds including polarization. It seems that the latter will involve nonassociative structures, and a model based on a Hilbert space and its endomorphisms seems no longer suitable. Hawkins' approach, however, circumvents the construction of a Hilbert space. Moreover, it is much more suitable for categorification. It thus seems more likely that the groupoid approach will allow us to compare the loop space quantization with a yet to be developed direct quantization of 2-plectic manifolds.

\subsection{Nambu-Poisson geometry}

Consider a manifold $M$ of dimension $d$ together with its algebra of
smooth functions $\CA=\CC^\infty(M)$. A {\em Nambu-Poisson structure
  of order $n$} on $M$ \cite{Nambu:1973qe,Takhtajan:1993vr}\footnote{see also \cite{deAzcarraga:2010mr} for a very detailed review} is an
$n$-ary, totally antisymmetric and multi-linear map $\{-,\cdots, -\}:\CA^{\wedge n}\rightarrow \CA$ which satisfies the Leibniz rule
\begin{equation}
 \{f_1 \, f_2,f_3,\ldots,f_{n+1}\}=f_1\,
 \{f_2,\ldots,f_{n+1}\}+\{f_1,\ldots,f_{n+1}\}\, f_2~,
\end{equation}
and the {\em fundamental identity}
\begin{equation}
\begin{aligned}
\{f_1,\ldots,f_{n-1},\{g_1,\ldots,g_n\}\} =&\ \{\{f_1,\ldots,f_{n-1},g_1\},g_2,\ldots,g_n\}\\
&\, +\ldots+\{g_1,\ldots,g_{n-1},\{f_1,\ldots,f_{n-1},g_n\}\}
\end{aligned}
\end{equation}
for $f_i,g_i\in \CC^\infty(M)$. As special cases, we have the usual
Poisson structures for $n=2$ and a derivation $\{- \}$ on $\CA$ for $n=1$.

Nambu-Poisson manifolds can be used as multiphase spaces in Nambu
mechanics, akin to the way in which Poisson structures on manifolds
define phase spaces in Hamiltonian mechanics; such multiphase spaces
are a starting point for \emph{higher quantization}. A Nambu-Poisson
bracket gives the vector space $\CC^\infty(M)$ of smooth functions on
$M$ the structure of an \emph{$n$-Lie algebra} \cite{Filippov:1985aa}
called a {\em Nambu-Poisson algebra}. An $n$-Lie algebra is a vector
space $\CA$ with an $n$-ary, totally antisymmetric multilinear map
which satisfies the fundamental identity. It has often been suggested
that a Nambu-Poisson structure of order $n$ should turn into an $n$-Lie
algebra under quantization, just as a Poisson algebra turns into a
Lie algebra of endomorphisms on a Hilbert space,
see~e.g.~\cite{Axenides:2008rn,DeBellis:2010pf} and references therein. This suggestion,
however, does not seem to yield quantum spaces with all the desired features.

Recall that a Poisson structure on $M$ can be defined in terms of a bivector field $\pi\in \CC^\infty(M,TM\wedge TM)$ via (\ref{bracketpidef}) for which the Schouten bracket vanishes. For general Nambu-Poisson structures, we consider a multivector field $\pi\in \CC^\infty(M,\bigwedge^n TM)$, which reads in local coordinates $x^i$, $i=1,\ldots,d$, as
\begin{equation}
 \pi=\pi^{i_1\ldots i_n}\, \der{x^{i_1}}\wedge \cdots\wedge \der{x^{i_n}}~.
\end{equation}
The higher analogue of the vanishing of the Schouten bracket reads as
\begin{equation}\label{eq:higherSchouten}
 \pi^{i_1\ldots i_{n-1} l}\, \der{x^l}\pi^{j_1\ldots j_n}-\pi^{l
   j_2\ldots j_{n}}\, \der{x^l}\pi^{i_1\ldots i_{n-1}j_1}-\cdots-\pi^{
   j_1\ldots j_{n-1}l}\, \der{x^l}\pi^{i_1\ldots i_{n-1}j_n}=0
\end{equation}
together with
\begin{equation}\label{eq:quadraticRelation}
 N_{i_1i_2\ldots i_nj_1j_2\ldots j_n}+N_{j_1i_2\ldots i_ni_1j_2\ldots j_n}=0~,
\end{equation}
where
\begin{equation}
\begin{aligned}
N_{i_1\ldots i_nj_1\ldots j_n}:=& \ \pi_{i_1\ldots i_n}\, \pi_{j_1\ldots
  j_n}+\pi_{j_ni_1i_3\ldots i_n}\, \pi_{j_1\ldots j_n} \\ &\, +\cdots
+ \pi_{j_ni_2\ldots i_{n-1}i_1}\, \pi_{j_1\ldots j_{n-1}i_n} -
\pi_{j_ni_2\ldots i_n}\, \pi_{j_1\ldots j_{n-1}i_1} ~.
\end{aligned}
\end{equation}
Now the $n$-vector field $\pi$ defines a Nambu-Poisson structure on
$M$ via 
\beq
\{f_1,\ldots,f_n\}_\pi:=\pi(\dd f_1,\ldots,\dd f_n)
\eeq
if and only if $\pi$ satisfies both \eqref{eq:higherSchouten} and \eqref{eq:quadraticRelation}, cf.~\cite{Takhtajan:1993vr}. Note that \eqref{eq:quadraticRelation} is automatically satisfied for $n=2$.
For further details on Nambu-Poisson structures, see \cite{springerlink:10.1007/BF00400143,Vaisman:1999aa}.

\subsection{Multisymplectic geometry}\label{subsec:MultisymplecticGeometry}

A Poisson structure is often derived from a symplectic structure on
$M$: A symplectic form $\omega$ defines a map $TM\rightarrow T^*M$. As
$\omega$ is non-degenerate, this map is invertible and its inverse is
a bivector field $\pi=\omega^{-1} :T^*M\rightarrow TM$. In particular,
there is a {\em Hamiltonian vector field} $X_f$ for every function
$f\in \CC^\infty(M)$ which is defined through $\dd
f=\iota_{X_f}\omega:=\omega(X_f,-)$. The Poisson structure is then
given by $\{f,g\}_\pi =\omega(X_f,X_g)=\CL_{X_g}f=\dd f(X_g)$. The Hamiltonian vector fields $X_f$ satisfy $\CL_{X_f}\omega=0$
and therefore generate symplectomorphisms of the manifold $M$. Since
$[X_f,X_g]=X_{\{f,g\}}$, the map $f\mapsto X_f$ provides an embedding
of the Poisson algebra $(\CC^\infty(M),\{-,-\}_\pi )$ into the Lie
algebra of symplectomorphisms. On two-dimensional manifolds these are
the area-preserving diffeomorphisms, and in the cases $M=\FR^2$ and
$M=S^2$ the Lie algebra of Hamiltonian vector fields coincides with the Lie algebra of symplectomorphisms.

To a certain extent, we can generalize this construction to
Nambu-Poisson structures by introducing multisymplectic forms. An
$n$-form $\varpi$ is called a {\em multisymplectic $n$-form} or an
{\em $n-1$-plectic form}\footnote{The number $n-1$ here indicates the
  degree of categorification of the notion of symplectic structure;
  in particular 1-plectic amounts to symplectic.} on $M$ if it is
closed and non-degenerate, i.e.,\ $\iota_X\varpi=0$ if and only if $X=0$ for $X\in \CC^\infty(M,TM)$. Simple examples of $n$-plectic forms are volume forms on orientable $n$-dimensional manifolds.

A multisymplectic $n$-form $\varpi$ defines a map $TM\rightarrow \bigwedge^{n-1}T^*M$. This map is only invertible for $d=n$, in which case the multisymplectic structure is given by a volume form on $M$. Volume forms $\varpi$ define a Nambu-Poisson structure via
\begin{equation}
 \{f_1,\ldots,f_n\}_{\varpi^{-1}} =\varpi^{-1}(\dd f_1, \ldots, \dd
 f_n) ~.
\end{equation}
In this paper, we will be mostly interested in the case $n=d=3$, and
in particular we will consider the 2-plectic Nambu-Poisson manifolds
$\FR^3$, $\FT^3$ and $S^3$.

Multisymplectic $p+1$-forms also naturally give rise to Poisson-like
brackets on certain $p-1$-forms: Assume that the manifold $M$ is
endowed with a $p$-plectic structure $\varpi$. The space of {\em
  Hamiltonian $p-1$-forms} on $M$, $\frH^{p-1}(M,\varpi)$, is the
space of $p-1$-forms $\alpha$ for which there is an associated
Hamiltonian vector field $X_\alpha\in \CC^\infty(M,TM)$ such that $\dd
\alpha=\iota_{X_\alpha}\varpi$. Again, these Hamiltonian vector fields
form a Lie algebra and
satisfy $\CL_{X_\alpha}\varpi=0$; hence they generate multisymplectomorphisms
of $M$. There are now two obvious generalizations of the Poisson
bracket on $\frH^{p-1}(M,\varpi)$~\cite{Baez:2008bu}: The {\em
  hemi-bracket} and the {\em semi-bracket} are defined on
$\alpha,\beta\in\frH^{p-1}(M, \varpi)$ as
\begin{equation}\label{Poissonlike}
 \{\alpha,\beta\}_{h,\varpi}:=\CL_{X_\alpha}\beta\eand
 \{\alpha,\beta\}_{s,\varpi} :=\iota_{X_\alpha}\, \iota_{X_\beta}\varpi~,
\end{equation}
respectively. Both brackets yield maps $\frH^{p-1}(M,\varpi)\times\frH^{p-1}(M,\varpi)\rightarrow \frH^{p-1}(M,\varpi)$. Note that
\begin{equation}\label{eq:diffHemiSemi}
 \{\alpha,\beta\}_{h,\varpi} -\{\alpha,\beta\}_{s,\varpi} =\dd \, \iota_{X_\alpha} \beta~.
\end{equation}
Furthermore, the hemi-bracket satisfies the Jacobi identity but is
not antisymmetric, while the semi-bracket is antisymmetric but does
not satisfy the Jacobi identity.\footnote{For ``exact'' multisymplectic manifolds, a further bracket can be constructed that is both antisymmetric and satisfies the Jacobi identity \cite{Forger:2002aa}.} Due to \eqref{eq:diffHemiSemi}, the failure of the brackets to be antisymmetric or to fulfill the Jacobi identity is always an exact form.

\subsection{Prequantization of 2-plectic manifolds}

We will now briefly review the prequantization of 2-plectic manifolds
following \cite{Rogers:2010sc}, see also~\cite{Baez:2008bu}. We will
see that the Poisson Lie algebra $\Pi_\omega= \CC^\infty(M)$ of smooth
functions on a symplectic manifold $(M,\omega)$ is replaced by the Lie
2-algebra $\Pi_\varpi=\CC^\infty(M)\oplus \frH^1(M,\varpi)$ on a 2-plectic
manifold $(M,\varpi)$. Moreover, just as the Atiyah Lie algebroid
allowed us to construct a faithful representation of the Poisson
algebra on its space of sections, i.e.,\ a prequantization, an exact Courant algebroid\footnote{See Appendix \ref{app:C} for the definitions related to Courant algebroids.} allows us to define a representation of the Lie 2-algebra. As argued in \cite{Baez:2008bu,Rogers:2010sc}, this should be regarded as a prequantization of $(M,\varpi)$.

The semi-bracket on a $p$-plectic manifold is well-defined on the
quotient of the space $\frH^{p-1}(M,\varpi)$ by the group of closed
$p-1$-forms on $M$; in particular, the quotient by exact $p-1$-forms
is a Lie algebra and if $M$ is contractible then it can be resolved by
the augmented de Rham complex
\begin{equation}
 \FR~\hookrightarrow~\CC^\infty(M)~\xrightarrow{ \ \dd \ }
 ~\Omega^1(M) ~ \xrightarrow{ \ \dd \ } ~\cdots~
 \xrightarrow{ \ \dd \ } ~ \Omega^{p-2}(M)~ \xrightarrow { \ \dd \ } ~\frH^{p-1}(M, \varpi)~.
\end{equation}
As shown in \cite{Barnich:1997ij}, such a resolution gives rise to
an $L_\infty$-algebra. Here, however, we will use a truncated form of
this homotopy Lie algebra.

Recall that a Lie 2-algebra\footnote{In this paper, we are interested in {\em semistrict} Lie 2-algebras. More general, {\em weak} Lie 2-algebras have been defined in \cite{Roytenberg:0712.3461}. Semistrict Lie 2-algebras (as well as the hemistrict ones derived from the hemi-bracket) are special cases of these.} is a two-term $L_\infty$-algebra
$L_{-1}\xrightarrow{~\mu_1~}L_0$ endowed with unary, binary and ternary
maps $\mu_1$, $\mu_2$, $\mu_3$ with grading $1,0,-1$ that satisfy
homotopy Jacobi identities, see~\cite{Baez:2003aa}. On a 2-plectic manifold $(M,\varpi)$ we define the Lie 2-algebra $\Pi_\varpi$ as the vector space
\begin{equation}
 \Pi_\varpi=L_{-1}\oplus L_0\ewith L_0=\frH^1(M,\varpi)\eand L_{-1}=\CC^\infty(M)
\end{equation}
together with the maps
\begin{equation}
\begin{aligned}
\mu_1(f+\alpha)=& \ \dd f~, \\[4pt] \mu_2(f+\alpha,g+\beta)=& \
\{\alpha,\beta\}_{s ,\varpi}=\varpi(X_\alpha,X_\beta,-)~, \\[4pt]
\mu_3(f+\alpha,g+\beta,h+\gamma)=& \ \varpi(X_\gamma,X_\beta,X_\alpha)~,
\end{aligned}
\end{equation}
where $f,g,h\in L_{-1}$, $\alpha,\beta,\gamma\in L_0$ and $X_\alpha$
denotes again the Hamiltonian vector field corresponding to the
one-form $\alpha$.

Consider now the exact Courant algebroid $C$ associated with the
2-plectic manifold $(M,\varpi)$ together with its associated strong
homotopy Lie algebra $L_\infty(C)$ as reviewed in Appendix
\ref{app:C}. By~\cite[\S 7]{Rogers:2010sc} there exists a morphism of
Lie 2-algebras, given by $\alpha\mapsto \lambda(X_\alpha)+\alpha$, which embeds $\Pi_\varpi$ into $L_\infty(C)$. In this
sense, the Courant algebroid $C$ allows us to construct a
representation of the Lie 2-algebra $\Pi_\varpi$, just as the Atiyah
Lie algebroid enabled us to construct a representation of the Poisson algebra on a symplectic manifold.

Altogether, the problem of prequantizing 2-plectic manifolds seems to
be solved. To obtain a complete quantization, however, we have to
restrict to polarized sections. A possible generalization of
polarizations to multisymplectic manifolds is explored in
\cite{Cantrijn:1999aa}: For a $p$-plectic vector space $(V,\varpi)$
the {\em $k$-orthogonal complement} of a subspace $W\subseteq V$ is the subspace
\begin{equation}
 W^{\perp,k}=\big\{v\in V \ \big| \
 \varpi(v,w_1,\ldots,w_k)=0~~\mbox{for all}~~w_1,\ldots,w_k\in W \big\}~.
\end{equation}
The subspace $W$ is called {\em $k$-Lagrangian} if
$W=W^{\perp,k}$. Correspondingly, a submanifold $N$ of a $p$-plectic
manifold $(M,\varpi)$ is called {\em $k$-Lagrangian} if the tangent
space $T_xN$ is a
$k$-Lagrangian subspace of $T_xM$ for all $x\in N$. Despite this
notion, it seems still
unclear how to properly incorporate polarization into the quantization
of multisymplectic manifolds. This is why we will now turn our
attention to loop spaces, where this problem is under control.

\subsection{Differential geometry of loop spaces}\label{subsec:LoopsAndKnots}

The {\em free loop space} $\CL
M$ of a $d$-dimensional manifold $M$ is the space of smooth maps $S^1\rightarrow M$, $\CL
M:=\CC^\infty(S^1,M)$; it is the configuration space of a 
bosonic string sigma-model on $S^1\times\FR$ with target space $M$. We can turn $\CL M$ into a smooth manifold
modeled on the loop space of $\FR^d$ \cite{Pressley:1988aa}, see also
\cite{Stacey:2006aa}. In particular, there are open covers of
$\CL M$ given by $\CU=(\CL U_a)$ where $U=(U_a)$ is a cover of
$M$. For example, the loop space $\CL S^d$ of the $d$-dimensional
sphere can be covered by the patches $\CU_a=\CL U_a=\CL(S^d \,
\backslash \, \{a\})$, $a\in S^d$.

We will describe a loop by a map $x:S^1\rightarrow M$, and use local
coordinates $(x^ i)$ on some patch $U_a$ of $M$ yielding parametrized loops
$\big(x^ i(\tau)\big) :S^1\rightarrow U_a$. To streamline notation, we will
usually combine the continuous loop parameter $\tau\in S^1$ with the
discrete index $ i=1,\ldots,d$ into a multi-index and write $x^{ i\tau}:=x^ i(\tau)$.

The infinite-dimensional tangent bundle $T \CL M$ over the free loop space $\CL M$ has
fibers given by $T_x \CL M:=\CC^\infty(S^1,x^*TM)$, i.e., a tangent vector to
a loop $x\in\CL M$ is a vector field along the map $x(\tau)$. In local coordinates
we can describe a section $X$ of $T\CL M$ as a map
$\big(X^ i(\tau)\big) :S^1\rightarrow x^*TM$. Together with the functional
derivatives $\delder{x^ i(\tau)}$, this map combines into a
derivation acting on functionals on $\CL M$ given by
\begin{equation}
 X=\oint \, \dd \tau~X^ i(\tau)\, \delder{x^ i(\tau)}=\oint \, \dd
 \tau~X^{ i\tau}\, \delder{x^{ i\tau}}~ ,
\end{equation}
where throughout we denote integration over the circle $S^1$ by $\oint$.
There is a natural diffeomorphism of the manifolds $T\CL M$ and $\CL T
M$. Note that each loop $x(\tau): S^1\rightarrow M$ comes with a natural tangent vector
$\xd(\tau):=\frac{\dd x(\tau)}{\dd\tau} \in \CL TM\cong T \CL M$.

We define the cotangent bundle $T^*\CL M$ as the vector bundle dual to
$T\CL M$; it is the phase space of a closed string sigma-model on the cylinder
$S^1\times\FR$ with target space $M$. Unlike $T\CL M$, this
definition is more subtle: While $\CL T^*M$ is modeled on the vector
space $\CL\FR^d$, the model space for $T^*\CL M$ is the dual space
$(\CL\FR^d)^*$ consisting of $\FR^d$-valued distributions on the
circle. In analogy with the tangent bundle, we hence define a
restricted cotangent bundle $\hat{T}^*\CL M$ by the union of pullbacks
of the cotangent spaces of $M$, i.e., its fibers are $\hat{T}^*_x\CL
M:=\CC^\infty(S^1,x^* T^*M)$. Clearly $T^*\CL M\supset \hat{T}^*\CL
M$ by the map $f\mapsto\oint\,\dd\tau \ f(\tau)$ on $\CL\FR\to\FR$,
and we have $\hat{T}^*\CL M\cong \CL T^* M$. In either case, we
write a section of the (restricted) cotangent bundle $\alpha$ as
\begin{equation}
 \alpha=\oint\, \dd \tau~\alpha_{ i\tau}~\delta x^{ i\tau}~,
\end{equation}
where
\begin{equation}
 \Big\langle \delta x^{ i\tau}\,,\, \delder{x^{ j\sigma}}\Big\rangle:=\delta^ i{}_ j~\delta(\tau-\sigma)~.
\end{equation}
For $\alpha\in \CC^\infty(\CL M,\hat{T}^*\CL M)$ the quantity
$\alpha_{ i\tau}$ is a function of $\tau$, while for $\alpha\in
\CC^\infty(\CL M,T^*\CL M)$ the map
\begin{equation}
 \alpha_{i\tau}\, :\, T\CL M \ \longrightarrow \ \FR~,~~~X=\oint\, \dd
 \tau \ X^{i\tau}\, \delder{x^{i\tau}}~~\longmapsto~~\oint \, \dd \tau \
\alpha_{ i\tau} \, X^{i\tau}
\end{equation}
is generally a distribution. 

The definition of $T^*\CL M$ gives rise to sections of the
antisymmetric tensor products of this bundle, i.e., the loop space $n$-forms
$\Omega^n(\CL M)$. We will call such differential forms {\em local} if they can be
written in local coordinates as a single integral over the loop as
\begin{equation}
 \alpha=\oint\, \dd \tau~\frac{1}{n!}\,\alpha_{ i_1\ldots i_n\tau}~\delta x^{ i_1\tau}\wedge \cdots \wedge \delta x^{ i_n\tau}~.
\end{equation}
Note that all one-forms on loop space are therefore local. Using the total differential $\delta$, which in local coordinates reads as
\begin{equation}
 \delta=\oint\, \dd \tau~\delta x^{ i\tau}\,\delder{x^{ i\tau}}~,
\end{equation}
we can define closed and exact forms, and therefore the de Rham cohomology groups $H^n(\CL M,\FR)$.

The differential geometry of $\CL M$ can be made rigorous by regarding loop
spaces e.g.\ as diffeological spaces, where the diffeology on
$\CL M$ is equivalent to a smooth Fr\'echet manifold structure. If
$(M,g)$ is a Riemannian manifold, then there is also
a natural
Riemannian structure on $\CL M$ induced from that on $M$ by
\beq\label{eq:RiemannianStructLS}
\langle X_1,X_2\rangle := \oint \, \dd\tau\ |\xd(\tau)|~
g_{x(\tau)}\big(X_1(\tau)\,,\, X_2(\tau)\big)
\eeq
for $X_1,X_2\in T_x\CL M$, together with a Banach
manifold completion of $T\CL M$ with respect to the $L^\infty$-norm on
$T_{x}\CL M$ given by
\beq
\|X \| :=\sup_{\tau\in S^1}\, \big|X(\tau)\big| \ .
\eeq

\subsection{Knot spaces}

For our purposes, the free loop space $\CL M$ is too big and we
have to impose various restrictions. First of all, we can demand that
the maps $x:S^1\rightarrow M$ have nice properties. We thus introduce
the loop space of immersions $\CL_\di M\subset \CL M$ with the
additional condition that up to isolated points, the map $x$ is an
embedding. The images of the maps contained in $\CL_\di M$ are called
{\em singular knots} \cite{0817647309}. Moreover, we can restrict to
the space of actual embeddings $\CL_\de M\subset \CL_\di M \subset \CL
M$; here the images of the maps $x\in \CL_\de M$ are knots, and the maps themselves give rise to parametrizations of knots.
We would like our description to be invariant under
reparameterizations of these knots, which are given by the smooth action of
the group of orientation-preserving diffeomorphisms of the circle
$\CR=\sDiff^+(S^1)$ via precomposition. Under such a coordinate change $\tau\mapsto\tilde \tau=R(\tau)$, we have e.g.\ the relations
\begin{equation}\label{dotxrepar}
\tilde{x}(\tilde{\tau})=x(\tau)~,~~~
\dot{\tilde{x}}(\tilde{\tau})=\frac{\dd \tau}{\dd \tilde{\tau}}\,
\dder{\tau}x(\tau)= \dot R(\tau)^{-1}\, \dot x(\tau)\eand \dd\tilde\tau=\dot R(\tau)\, \dd\tau~.
\end{equation}
The $\CR$-fixed points define a natural embedding $M\embd\CL M$
given by the constant paths (zero modes) $x(\tau)=x_0\in M$.
The quotient of $\CL M$ by $\CR$ is singular. But if we divide $\CL_\di M$ and $\CL_\de M$ by the action of $\CR$, we obtain the space of singular oriented knots, $\CK_{\di}M$, and the space of oriented knots, $\CK_{\de} M$ \cite{0817647309}. Our constructions work on both types of knot spaces, and we will therefore simply write $\CK M$, leaving the choice of the type of knots to the reader. However, when referring to ``knots'' and ``knot spaces'' we will always mean {\em oriented} knots.

In our description of knot spaces, we will still use the local coordinates from the description of loop space. To do this, we will have to make sure that all formulas are invariant under reparameterizations. At infinitesimal level, reparameterizations are generated by the vector fields
\begin{equation}
 R=\oint\, \dd \tau~R(\tau)\,\xd^{i\tau} \,\delder{x^{ i\tau}}~.
\end{equation}
A smooth functional $f$ on $\CK M$ therefore satisfies
\begin{equation}
 \xd^{ i\tau}\,\delder{x^{ i\tau}}\,f=0~.
\end{equation}
This restricts the tangent bundles of the knot spaces, if their local sections are regarded as derivations acting on functionals. In particular, we impose the relations
\begin{equation}\label{eq:restrictionsTangentCotangent}
 \xd^{ i\tau}\,\delder{x^{ i\tau}}=0\eand \delta x^{ i\tau}\,\xd_{ i\tau}=0~.
\end{equation}
For the second relation, we assumed a metric on knot space induced
from the target manifold $M$ to raise and lower indices. Because of
these relations, differential forms on knot spaces are of a special
form. For example, a one-form $\alpha$ can be written as
\begin{equation}\label{eq:OneFormsOnKnots}
\alpha(x)=\oint\, \dd \tau~\alpha_{[ i_1 i_2]\tau}(x)\,\xd^{ i_1\tau}~\delta x^{ i_2\tau}~,
\end{equation}
where $\alpha_{[ i_1 i_2]\tau}(x)=-\alpha_{[ i_2 i_1]\tau}(x)$. Note
that the identification of $T\CK M$ with $\CK T M$ does not hold for
knot spaces, as reparametrization transformations act differently on both spaces.

\subsection{Transgression}\label{sec:transgression}

The {\em transgression map}
\beq
\CT \,:\, \Omega^{k+1}(M) \ \longrightarrow \ \Omega^k(\CL M)
\eeq
is defined via the correspondence
\begin{equation}
\xymatrix{
 & \CL M\times S^1 \ar[dl]_{ev} \ar[dr]^{pr} & \\
M &  & \CL M
}
\end{equation}
between $M$ and its free loop space $\CL M$. Here $ev$ is the
evaluation map of the loop $x(\tau_0)\in M$ at the given angle
$\tau_0\in S^1$, and
$pr$ is the obvious projection. The transgression map is the
composition of the pullback along $ev$
with the pushforward $pr_!$ given by integration over the fiber $S^1$ of $pr$. It produces
the loop space $k$-form
\begin{equation}
(\CT\alpha)_{x} = \oint\, \dd\tau\ \iota_{\dot x}(ev^*\alpha) \efor \alpha\in\Omega^{k+1}(M) \ ,
\label{CTalpha}\end{equation}
where $\iota_{\dot x}$ is contraction with the natural tangent vector
$\xd(\tau)$ to a loop $x: S^1\rightarrow M$. Explicitly, in local coordinates $x^{ i\tau}$ we have
\begin{equation}\label{eq:Transgression}
\begin{aligned}
 \alpha=& \ \frac{1}{(k+1)!}\,\alpha_{ i_1\ldots i_{k+1}}(x)\ \dd
 x^{ i_1}\wedge \cdots \wedge \dd x^{ i_{k+1}} \\& \qquad \qquad \ \longmapsto \
 \CT \alpha :=\oint\, \dd \tau~\frac{1}{k!}\,\alpha_{ i_1\ldots  i_{k+1}}(x(\tau))\, \xd^{ i_{k+1}\tau}~\delta x^{ i_1\tau}\wedge \cdots \wedge \delta x^{ i_k\tau}~.
\end{aligned}
\end{equation}
Differential forms on loop space which are in the image of the transgression map and therefore of the above form are called {\em ultra-local}.

The image of a transgression is invariant under the group of
loop reparameterizations $\CR$. Moreover, the transgression map commutes with the exterior derivative, i.e., $\delta(\CT\alpha)=\CT(\dd\alpha)$. In fact, it descends on integer cohomology to a homomorphism
\beq
\CT\,:\, H^{k+1}(M,\RZ) \ \longrightarrow \ H^k(\CL M,\RZ) \ .
\eeq
For $k=1$ there is a natural transgression map which
computes the holonomy of a connection on a line bundle over $M$. For $k=2$ transgression takes a Dixmier-Douady
class on a manifold $M$ to a first Chern class on the free
loop space $\CL M$; put differently, a bundle gerbe $\CG$ on $M$ gives
rise to a line bundle $\CT\CG$ on $\CL M$.\footnote{See
  Appendix~\ref{app:B} for the pertinent definitions regarding gerbes.} In particular,
transgression can also be refined to a homomorphism on differential cohomology $\hat H^3(M)\to \hat
H^2(\CL M)$. We can thus take a prequantum gerbe on $M$ and map it to a prequantum line bundle on $\CL M$, which can be quantized in the usual way.

We can restrict the image of the transgression to the open subsets
$\CL_\di M$ and $\CL_\de M$ of $\CL M$ consisting of restricted
immersions and embeddings. As the transgression map is invariant under actions of the group of reparameterizations $\CR=\sDiff^+(S^1)$, the restricted image of a transgression descends to a differential form on the corresponding knot spaces.

It should be stressed, however, that the transgression map is {\em
  not} surjective. In particular, one-forms on knot space are of the
form \eqref{eq:OneFormsOnKnots} but this does not imply they are in
the image of the transgression map: $\alpha_{[ i_1 i_2] \tau}(x)$ is not necessarily of the form $\alpha_{[ i_1 i_2]}(x(\tau))$.

\subsection{Poisson algebras on knot spaces\label{subsec:Poissonknots}}

Every loop space $\CL M$ is naturally a symplectic manifold with local
symplectic potential $\vartheta=\oint \, \dd \tau\ \xd_{i\tau}\,
\delta x^{i\tau}$. The corresponding two-form $\omega=\delta
\vartheta$ is closed and it is weakly symplectic in
the sense that it defines an injection $T \CL M\rightarrow T^*\CL M$ with dense image.

On the knot space $\CK M$, however, the symplectic potential
$\vartheta$ vanishes and we have to find a different source for a
symplectic structure. The most natural way is to assume a 2-plectic
structure on $M$, that we can transgress to a symplectic structure on
the knot space. Given a closed non-degenerate three-form $\varpi$ on
$M$, we define a symplectic structure $\omega$ on $\CK M$ via the transgression
\begin{equation}
 \omega=\CT \varpi=\oint\, \dd\tau\ \iota_{\dot x}(ev^*\varpi)~.
\end{equation}
As the transgression map commutes with the exterior derivative, this form is closed. Moreover, its kernel consists of vector fields
\begin{equation}
 X=\oint \, \dd \tau\ X(\tau)\,\xd^{i\tau}\, \delder{x^{i\tau}}
\end{equation}
which generate reparameterizations and are therefore excluded from the tangent bundle on knot space. On $T\CK M$, the two-form $\omega$ is therefore non-degenerate and defines a symplectic structure.

We can now invert $\CT \varpi$ to define a Poisson bracket on knot space $\CK M$. The result is a global bivector field $\Pi\in
\CC^\infty(\CL M,\bigwedge^2\CL TM)$ with vanishing Schouten bracket. Locally, we have
\begin{equation}
 \oint \, \dd
 \tau~\Pi^{i\sigma,j\tau}\,\omega_{j\tau,k\rho}=\delta^i{}_k\, \delta(\sigma-\rho)~,
\end{equation}
where $\Pi^{i\sigma,j\tau}$ and $\omega_{j\tau,k\rho}$ are the
components of the bivector field $\Pi$ and of the two-form $\omega=\CT\varpi$. In these components, the Poisson bracket reads as
\begin{equation}
 \{f,g\}_\Pi:=\oint \, \dd \sigma \ \oint \, \dd \tau\ \Pi^{i\sigma,j\tau}
 \, \Big(\, \delder{x^{i\sigma}}f\, \Big)\, \Big(\,
 \delder{x^{j\tau}}g\, \Big)
\end{equation}
for loop space functionals $f,g\in\CC^\infty(\CL M)$.

The hemi-bracket and semi-bracket that we defined in
\S\ref{subsec:MultisymplecticGeometry} are mapped to ordinary Poisson
brackets on loop space under transgression. Recall that the 2-plectic
structure yields Poisson-like brackets (\ref{Poissonlike}) of
one-forms on $M$. The transgression map now converts these to the
Poisson bracket on loop space as
\begin{equation}
 \CT\big(\{\alpha,\beta\}_{h,\varpi}\big)=\CT\big(\{\alpha,\beta\}_{s,\varpi}\big)=\{\CT\alpha,\CT\beta\}_{(\CT\varpi)^{-1}}~,
\end{equation}
which follows from the observation $\CT (\iota_{X_\alpha}\varpi)
=\iota_{X_{\CT\alpha}}(\CT\varpi)$ and that $\CT (\dd f) =\delta (\CT
f)=0$ as $\CT f=0$ for $f\in\CC^\infty(M)$. This feature can be
regarded as further motivation for using loop spaces or knot spaces.

With these structures inherited on the loop space $\CL M$, we may
formally follow the recipes of \S\ref{QuantPoisson}. Alternatively, we could have started from a Nambu-Poisson structure induced by a trivector field $\pi$ on $M$. The key feature here is that the local quantity $\Pi^{ij}= \pi^{ijk}\,\dot x_k$ behaves like an
ordinary Poisson structure.

\subsection{K\"ahler geometry of knot spaces of three-dimensional 2-plectic manifolds}

Let us now restrict to Riemannian three-dimensional manifolds $M$. In
these cases there is a natural complex structure on the corresponding
knot space $\CK M$~\cite{Marsden:1983aa}: The tangent space $T_x\CK M$
restricted to a point $x(\tau)$ on the loop $x$ is the orthogonal
complement of the tangent vector $\xd(\tau)$ in $T_{x(\tau)}M$ and
therefore isomorphic to the plane $\FR^2$. The tangent vector to the loop
provides an axis of rotation on this plane, and we define the action of
an almost complex structure $\CJ$ as the rotation of the restricted
vector around this axis by angle $\frac{\pi}{2}$. This action clearly extends smoothly to the whole of $T_x\CK M$ and we have $\CJ^2=-\id$.

In local coordinates $x^{i\tau}$, we have
\begin{equation}
 \CJ X=\CJ\oint \, \dd \tau~X^{i\tau}\,\delder{x^{i\tau}}=\oint\, \dd
 \tau~\tilde{\eps}^{\, ijk}\,X_{j\tau}\,\frac{\xd_{k\tau}}{|\xd^\tau |}~\delder{x^{i\tau}}~,
\end{equation}
where indices are raised and lowered with the metric $g$ on $M$,
e.g.,\ $X_{j\tau}:=g_{jl}(x(\tau))\, X^{l\tau}$, and $\tilde{\eps}^{\,
  ijk}$ is the ``metric-corrected'' Levi-Civita symbol given by
\begin{equation}
 \tilde{\eps}_{ijk}=\det(g)^{{1}/{2}}\, \eps_{ijk}\eand
 \tilde{\eps}^{\, ijk}=\det(g)^{-{1}/{2}}\, \eps^{ijk}=\det(g)^{-1}\, \tilde{\eps}_{ijk}~.
\end{equation}
A straightforward calculation confirms $\CJ^2=-\id$. It should be
stressed that this almost complex structure is not integrable, in the
sense that no non-empty subset of a complex Fr\'echet space is
biholomorphic to an open subset. However, it is {\em formally
  integrable} in the sense that the corresponding Nijenhuis tensor vanishes. Formal integrability is in fact sufficient for the introduction of a Dolbeault differential and the definition of holomorphic functions.

One can now show that $\omega=\CT \varpi$ is of type $(1,1)$ with
respect to the complex structure $\CJ$, i.e.,
$\omega(\CJ X_1,\CJ X_2)=\omega(X_1,X_2)$. Furthermore, the induced inner product $\langle-,-\rangle$ on $T \CK M$ defined in \eqref{eq:RiemannianStructLS} satisfies
\begin{equation}
 \langle X_1,X_2 \rangle=\omega(X_1,\CJ X_2)~,
\end{equation}
because the volume form $\varpi$ on $M$ originates from the Riemannian
metric $g$ on $M$. Altogether, we have turned the knot space $\CK M$
into a K\"ahler manifold, up to technical issues associated with the notions of integrability of the complex structure.
For further details on symplectic and complex structures on loop spaces and knot spaces, see~\cite{0817647309} and references therein.

\subsection{Prequantization of symplectic groupoids on knot spaces}\label{ssec:PrequantKnotSpace}

Our aim is to map the problem of quantizing a 2-plectic manifold
$(M,\varpi)$ onto a geometric quantization problem on the knot
space $\CK M$ of $M$. Thus far we have established the transgression map \cite{0817647309} that takes a prequantum gerbe
over a manifold $M$ to a prequantum line bundle over its knot space
$\CK M$. Let us discuss this point in some more detail.

The 2-plectic structure on $M$ is specified by a closed,
non-degenerate globally defined three-form $\varpi$; if
$\varpi\in\clidf^3(M)$, then $\varpi$ arises as the curvature of the connective structure of a
gerbe $\CG$ represented by a smooth Deligne two-cocycle $(g_{abc},A_{ab},B_a)$ with
respect to an open cover $U= (U_a)$ of $M$ (see Appendix~\ref{app:B}). We assume that the cover $U$ is fine enough for $\CU=(\CU_a):=(\CK U_a)$ to cover the free knot space $\CK M$. The transgression map then yields the closed symplectic two-form $\omega:=\CT\varpi$ on knot
space $\CK M$. The corresponding smooth hermitian line bundle $\CT\CG$ over $\CL M$ has fiber
$\CT\CG_{x}$ over a
circle $x:S^1\hookrightarrow M$ equal to the set of isomorphism
classes of flat
trivializations of the pullback $x^*\CG$ of the gerbe $\CG$ to
$S^1$. It comes with a product, inherited from the bundle gerbe product: If two
knots $x_1$ and $x_2$ are composable, then
$\CT\CG_{x_1\circ x_2}\cong \CT\CG_{x_1}\otimes
\CT\CG_{x_2}$. The connection $\vartheta_\CG \in\Omega^1(\CT\CG)$ whose curvature is
$\CT\varpi$ can be written in the
patch $\CK U_a$ as
\beq
\vartheta_\CG =\delta z_a+\CT B_a \eon \CK U_a \ ,
\label{varthetaalpha}\eeq
where $z_a\in\FC$ are coordinates on the fibers. Since
$B_a-B_b=\dd A_{ab}$ on the intersection
$U_{ab}=U_a\cap U_b$, one has
\beq
z_a-z_b=\CT A_{ab} \eon \CK U_{ab} \ .
\eeq
The $\sU(1)$-action on the fibers is given by the gauge group of the gerbe, described
in Appendix~\ref{app:B}, i.e., the Cheeger-Simons cohomology group $\hat H^2(M)$ of line bundles with
connection $(E,a)$ modulo gauge transformations; from
(\ref{varthetaalpha}) it follows that the fiber coordinate $z\in\FC$
shifts under this action as $z\mapsto z+\CT a$. Since $\hat
H^2(S^1)\cong\sU(1)$, the pullback $x^*E$ of the line bundle $E\to M$
to a knot $x\cong S^1$
is a trivial line bundle with flat connection, i.e., the pullback
$x^*\CG$ of
the gerbe $\CG$ over $M$ to a knot is a principal homogeneous space for $\sU(1)$, and
$\CT\CG$ becomes a principal $\sU(1)$-bundle on $\CK M$.

Instead of directly quantizing the prequantum gerbe $\CG$ over $M$, we will use the Poisson structure on $\CK M$ induced by $\CT \varpi$ and apply Hawkins' groupoid approach to quantization. We will make no attempt here to develop a proper rigorous theory of symplectic
groupoids over knot spaces and their quantization, paralleling the
description of the previous section; this is a highly involved technical task
that could presumably invoke the setting of diffeological spaces. Our
main concern here will be to understand what sort of noncommutative
knot spaces arise, and how they capture the quantum geometry of closed strings and M-branes in background fluxes. In
all examples that we treat later on the groupoids over knot space $\CK
M$ will be of the form $\CK\Sigma$ for a (not necessarily symplectic)
groupoid $\Sigma\rightrightarrows M$; this will simplify the
development somewhat as the appropriate groupoid structures on
$\CK\Sigma\rightrightarrows \CK M$ can be obtained by using
functoriality of the
transgression map. But even then, the complete construction of appropriate
analogs of entities like a convolution $C^*$-algebra appear to be
prohibitively complicated. In particular, it is well-known that there are
no reparametrization invariant measures on the loop space $\CL M$, and
suitable convolution products should presumably be constructed in some
way from
the quasi-invariant \emph{Wiener measure} $\CCD x(\tau)\ \exp\big(-\frac12\,
\oint \, \dd\tau \ |\dot x(\tau)|^2\big)$ for the quantum
mechanics of free geodesic motion on $M$.

Our choice of symplectic groupoid will be
partly dictated by the existence of suitable co-isotropic restrictions of the velocity vectors
$\dot x(\tau)$ to one-dimensional sub-bundles of $T M\subset\CK TM$
for which our quantization will reduce to a known quantization of the
induced Poisson structure on a corresponding codimension one
submanifold $N$ of $M$. We may then mimick the symplectic groupoid
quantization of $N$ by using transgression techniques.

\subsection{Regression}

Let $(\CK \Sigma,\omega)$ be a symplectic groupoid integrating the
Poisson manifold $(\CK M,\Pi)$. We will investigate what
the symplectic groupoid quantization corresponds to on the original
groupoid $\Sigma\rightrightarrows M$. As a start, one can ask to what
corresponds the prequantization data of $(\CK\Sigma,\omega)$;
naturally, the answer would be a particular prequantum bundle gerbe on
$\Sigma$ with connective structure. For this, one requires an inverse to the
transgression map $\CT$, called a
\emph{regression map}, whose details have been recently worked out
by Waldorf~\cite{Waldorf:2010aa,Waldorf:2011aa} (see
also~\cite{Mackaay:2000ac}). Note that because the map $\CT$ is not surjective, the inverse of the transgression exists only on the image of $\CT$. We will briefly sketch the construction
of the connective structure $\CT^{-1}\omega$ on the bundle gerbe, glossing over many of the intricate
technical details. For simplicity, we perform the inversion on loop space.

For this, we embed loop space
$\CL\Sigma$ into path space $\CP\Sigma$; after factoring out
reparameterizations and thin homotopy, the path space is naturally a
groupoid $\CP\Sigma\rightrightarrows\Sigma$ with source and target
fibrations $ev_0, ev_1:\CP\Sigma\to\Sigma$ which evaluate a
path $g(\tau)\in\CC^\infty([0,1],\Sigma)$ at its initial and final endpoints $g(0), g(1)\in\Sigma$, and groupoid
multiplication given by concatenation of paths. In what follows we take
$\CP\Sigma$ to be the space of smooth oriented paths with the same initial points
$g(0)=g_0\in\Sigma$. It can be shown that the resulting regression map is independent of the choice of $g_0$. The fibered
product $\CP\Sigma^{[2]}$ over $g\in\Sigma$ consists of pairs of paths
between $g_0$ and $g$ in $\Sigma$. By reversing the orientation of one
of the paths, such a pair may be identified with a loop based at
$g_0$; to simplify notation, in the
ensuing construction we will take $\CL \Sigma$ to be the space of
smooth loops in $\Sigma$ based at $g_0$. Then we can identify
$\CP\Sigma^{[2]}$ with $\CL\Sigma$. A technical subtlety here
  is that the loop resulting from concatenation of two paths may not
  be smooth at the two points where the paths are composed. This problem
  is rectified by reparametrizing the paths $g(\tau)$ in
  $\CP\Sigma$ such that they have ``sitting instants'' at these points, i.e., they are locally
  constant in neighborhoods of $\tau=0$ and $\tau=1$. On $\CL \Sigma$ this
  structure is defined in terms of an integral over $[0,1]$ which is invariant
  under such reparameterizations. In particular, differential forms
  $\eta\in\Omega^k(\CL \Sigma)$ which are compatible with the
  multiplication in the smooth path groupoid $\CP\Sigma\rightrightarrows
  \Sigma$ respect composability of paths; if in addition they are
  reparametrization invariant then we may define the
 regression form
  $\CT^{-1}\eta\in \Omega^{k+1}(\Sigma)$ at the point $g_0\in\Sigma$ by
\beq
(\CT^{-1}\eta)_{g_0}(v_0,v_1,\dots,v_k):=\lim_{\sigma\to0^+}\,\frac1\sigma\,
\eta_{g(\tau)|_{[-1,\sigma]}} (\tilde v_1,\dots,\tilde
v_k)\big|_{g(0)=g_0,\dot g(0)=v_0}
\label{regreta}\eeq
where $v_i\in T_{g_0}\Sigma$, $g(\tau)\in\CP\Sigma$ is any smooth path
such that $g(\tau)\big|_{[-1,0]}$ is thin-homotopic to the constant path $g_0$, and $\tilde v_i\in
\CL T_{g_0}\Sigma$ are extensions of the tangent vectors $v_i$ along the path
$g(\tau)$ such that $\eta_{g(\tau)|_{[-1,0]}} (\tilde v_1,\dots,\tilde
v_k)=0$, see e.g.~\cite{Saemann:2010cp}. This map also commutes with
the exterior derivative, i.e.,
$\dd(\CT^{-1}\eta)=\CT^{-1}(\delta\eta)$, where the exterior
derivative $\delta$ on $\CP\Sigma$ respects the groupoid structure by
including boundary contributions from evaluation at the endpoints.

Let $\CCL$ be a hermitian line bundle over $\CL\Sigma$ with
an associative
fiberwise product~\cite{Waldorf:2010aa}. Regression then maps the line bundle
$\CCL\to\CL\Sigma$ to the bundle gerbe $\CT^{-1}\CCL$ on $\Sigma$ given by
\beq
\xymatrix@C=10mm{
\CCL \ar[d] & \\
\CL \Sigma \ \ar@< 2pt>[r]^{\pr_1} \ar@< -2pt>[r]_{\pr_2}
 & \ \CP \Sigma \ar[d]^{ev_1} \\
 & \Sigma
}
\eeq
Note that this construction holds without the need of introducing a
connection~\cite{Waldorf:2011aa}; however, in this case the gerbe is
unique only up to thin homotopies of paths and there is no distinguished inverse
map in general.

Let $\nabla=\delta+\vartheta$ be a reparametrization invariant connection on $\CCL\to\CL \Sigma$ of
curvature $F_\nabla=-2\pi\, \di\, \omega$ which commutes with the
multiplication on $\CCL$. Then we may apply (\ref{regreta}) to
$F_\nabla\in\Omega^2(\CL\Sigma)$ to get a three-form
$H:=\CT^{-1}F_\nabla \in \Omega^3(\Sigma)$. We define a curving
$B\in\Omega^2(\CP\Sigma)$ by pulling $H$ back along the evaluation map
$ev:\CP\Sigma \times [0,1]\to \Sigma$ and integrating over the fiber
to get
\beq
B= \int_0^1\, \dd\tau \ \iota_{\dot x}\,ev^*\big(\CT^{-1}F_\nabla
\big) \ .
\eeq
One readily checks the requisite compatibility equations
\beq
\pr_1^*(B)-\pr_2^*(B)= F_\nabla \eand \delta
B= ev_1^*\big( \CT^{-1} F_\nabla\big)
\eeq
over $\CL\Sigma$ and $\CP\Sigma$, respectively.
Note that the two-form $B$ depends only on the curvature of the line
bundle $\CCL$; see~\cite{Waldorf:2010aa} for further technical details and
requirements of this construction.

The regression map $\CT^{-1}$ provides a functorial isomorphism between the
differential cohomology group $\hat H^3(\Sigma)$ and the subgroup of
$\hat H^2(\CL \Sigma)$ consisting of classes of line bundles on $\CL
\Sigma$ with fiberwise product and reparametrization invariant compatible connection. It gives a functorial equivalence between
flat bundles on $\CL \Sigma$ and flat gerbes on $\Sigma$. It also restricts to a bijection between
stable isomorphism classes of trivial gerbes with connection and isomorphism
classes of trivial bundles with fiberwise product and
reparametrization invariant compatible
connection; the transgression map (\ref{CTalpha}) in this case induces a group
isomorphism between the gauge equivalence classes
$\Omega^2(\Sigma)/\clidf^2(\Sigma)$ of topologically trivial
$B$-fields $b$ and the subgroup of $\Omega^1(\CL
\Sigma)/\delta\CC^\infty(\CL \Sigma,\FR)$ consisting of reparametrization invariant
compatible topologically
trivial gauge fields $a$ modulo compatible gauge transformations, with
the inverse map given by (\ref{regreta}). Note that here we implement
the distinction stable isomorphism (as opposed to actual isomorphism)
in order to account for the fact that gauge transformations of gerbes consist themselves
of line bundles and so are naturally compared via bundle morphisms,
see Appendix~\ref{app:B}; put differently, bundle gerbes are
objects of a 2-category, in contrast to line
bundles which are objects of a category.

\subsection{A generic example}

A representative class of symplectic groupoids that we shall frequently encounter are cotangent
groupoids with arrow set the cotangent bundle $\Sigma=T^*M$ over a
manifold $M$ of dimension $d$, the canonical symplectic structure, and various choices for the structure maps
$(\sfs,\sft,\unit,m)$. The cotangent bundle $T^*\CK M$ is the phase space of
a closed string sigma-model on $S^1\times\FR$, but in the following we
will restrict ourselves to the sub-bundle $\hat{T}^*\CK M\subset
T^*\CK M$ which is defined
analogously to the tangent bundle, see\
\S\ref{subsec:LoopsAndKnots}. The cotangent bundle over the knot space
of $M$ is naturally an infinite-dimensional symplectic manifold with
symplectic two-form written in local canonical coordinates $(x,p)$
on $\Sigma=T^*M$ as
\beq
\omega=\oint \, \dd\tau\ \delta x^{i\tau} \wedge\delta p_{i\tau}
\label{loopomegagen}\eeq
at a given knot $x:S^1\hookrightarrow M$. The corresponding Poisson brackets should
be evaluated on smooth global sections of the (trivializable) infinite jet
bundle $J^\infty (\Sigma\times S^1) \to S^1$, i.e., equivalence
classes (or ``jets'')
of smooth sections of the trivial fibration $\Sigma\times S^1\to S^1$ whose Taylor
coefficients coincide to all orders, with induced coordinates
$(\tau,x,p,\dot x,\dot p,\dots)$; if we wish to consider only ultra-local
functionals that do not depend explicitly on the loop parameter $\tau$,
then we should restrict to functionals on the infinite jet space
$J^\infty(\CL\Sigma)\subset J^\infty(\Sigma\times S^1)$ consisting
of jets of smooth maps $S^1\to\Sigma$. There is a symplectic potential
$\vartheta\in\Omega^1(\CL\Sigma)$, with $\omega=\delta\vartheta\in
\Omega^2_{{\rm cl},\RZ}(\CL\Sigma)$, given by
\beq
\vartheta=\oint \, \dd\tau \ p(\tau) \ ,
\label{loopthetagen}\eeq
where the globally defined momentum
\beq
p(\tau)= p_{i\tau} \, \delta x^{i\tau}
\eeq
is a normal covector field on the knot $x$, i.e., a section $p\in \hat
T^*_x\CL M= 
\CC^\infty(S^1,x^*T^*M)$. Since $\omega$ is exact, it is the curvature of the trivial prequantum
line bundle $\CCL=\hat T^*\CK M\times\FC$, which naturally has a trivial
fiberwise product (induced by multiplication in $\FC$); the
wave functionals of the quantum sigma-model are sections of this bundle.

The
corresponding bundle gerbe is also trivial and up to thin homotopy
it is given by
\beq
\xymatrix@C=10mm{
\hat T^*\CK M\times \FC \ar[d] & \hat T^*\CP M\times \FC \ar[d] \\
\hat T^* \CK M \ \ar@< 2pt>[r]^{\pr_1} \ar@< -2pt>[r]_{\pr_2}
 & \ \hat T^* \CP M \ar[d]^{ev_1} \\
 & T^*M
}
\eeq
Even though the symplectic potential \eqref{loopthetagen} is not in the image of the transgression map, let us apply the regression formula (\ref{regreta}) to it. This yields a quadratic form
$\tilde b:=\CT^{-1}\vartheta$ given at a point $(x,p)\in T^*M$ by
\beq
\tilde b_{(x,p)}\big(v\,,\,\mbox{$\frac\partial{\partial p_i}$}\big)=0 \eand \tilde b_{(x,p)}\big(v\,,\,\mbox{$\frac\partial{\partial
    x^i}$}\big)= p_i
\eeq
for all tangent vectors $v\in T_{(x,p)}\Sigma$. Note that $\tilde b$ only defines a
section of $T^*\Sigma\otimes T^*\Sigma$: Although $\vartheta$
commutes with the multiplication on $\CCL$, it is
\emph{not} a reparametrization invariant one-form on $\hat T^*\CK M$ and
so its regression $\CT^{-1}\vartheta$ does not automatically antisymmetrize to
give an element of $\Omega^2(\Sigma)$. On the other hand, as a section
of $\CK\Sigma$ the momentum $p(\tau)$ transforms as in
(\ref{dotxrepar}) and the symplectic potential (\ref{loopthetagen}) is
reparametrization invariant; when viewed in this way the antisymmetrization of
the section $\tilde b$ yields the two-form
\beq
b=-\sum_{i,j=1}^d\,\big(\mbox{$\frac12$}\,( p_i-p_j) \, \dd x^i\wedge
\dd x^j + p_i\, \dd x^i\wedge \dd p_j \big)
\label{bfieldgen}\eeq
with curvature
\beq
H=\dd b= \sum_{i,j=1}^d \,\big(\dd x^i\wedge \dd x^j\wedge \dd p_i+ \dd
x^i\wedge \dd p_i\wedge \dd p_j \big) \ .
\eeq
Note that we may shift $b$ by
any closed two-form $\beta\in\Omega^2(\Sigma)$, $\dd\beta=0$, which
transgresses to a gauge transformation of the symplectic potential
\beq
\vartheta_\beta=\oint \, \dd\tau \ \big(p(\tau)+\iota_{\dot x}\,ev^*\beta\big)
\ .
\label{thetabetagen}\eeq
This modification generally affects the quantization of
$\hat T^*\CK M$;
it depends on the
cohomology class of a flat gerbe 1-connection in $H^2(M,\sU(1))$.

\subsection{Winding sectors and Bohr-Sommerfeld 2-leaves\label{BS2leaves}}

There are two main modifications which must be made when the 2-plectic manifold $M$
is multiply connected. They can be deduced by noting that the homotopy
groups of the loop space are given by
\beq
\pi_n(\CL M)\cong\pi_{n+1}(M)\times\pi_n(M)
\label{loophomotopy}\eeq
whenever $M$ is connected.

Firstly, setting $n=0$ in (\ref{loophomotopy}) we note that in this case
the knot space $\CK M$ (and the underlying loop space $\CL M$) is
disconnected with connected components $\CK_w M$, called ``winding sectors'', parametrized by the set $w\in \pi_1(M)^\sim$
of conjugacy classes of the fundamental group of $M$. These winding
modes appear as zero
modes for the velocity vectors such that $\oint \, \dd\tau\ \dot
x(\tau) = w$. Formally, the twisted
convolution $C^*$-algebra of the polarized symplectic groupoid on $\CK
M$ then decomposes as
a graded vector space
\beq
\CC_r^*(\CK\Sigma/\CP,\sigma)=\bigoplus_{w\in\pi_1(M)^\sim} \,
\CC_{r,(w)}^*(\CK_w\Sigma/\CP,\sigma) \ ,
\eeq
where $\CC_{r,(w)}^*(\CK_w\Sigma/\CP,\sigma)$ is the space of
polarized sections of a line bundle $\CCL_w\to\CK_w\Sigma$ in the
topological sector $w\in\pi_1(M)^\sim$. The collection of these line
bundles $\CCL=(\CCL_w)_{w\in\pi_1(M)^\sim}$ is regarded as a single
line bundle on $\CK\Sigma$. Hence on loop space one quantizes not just a functional $f$ but
rather a pair $(f,w)\in\CC^\infty(\CK M)\times\pi_1(M)^\sim$. In general
the multiplication in $\CC_r^*(\CK\Sigma/\CP,\sigma)$ does \emph{not}
preserve the grading; instead, the group
multiplication in $\pi_1(M)$ gives $\CC_r^*(\CK\Sigma/\CP,\sigma)$ the
structure of an algebra over
$\CC_{r,(e)}^*(\CK_e\Sigma/\CP,\sigma)$.

Secondly, setting $n=1$ in (\ref{loophomotopy}) we see that
 we must impose a Bohr-Sommerfeld quantization condition over
the leaves of the polarization as we did in
\S\ref{BohrSommerfeld}, but now parametrized by conjugacy classes
$[S]\in\pi_2(M)^\sim$ of the second homotopy group of $M$ in each
winding sector $w\in\pi_1(M)^\sim$. For this, we construct a subvariety
$\Sigma_0\subseteq \Sigma$ by demanding that the holonomy
$\hol_{\gamma_w}(\vartheta)=1$ for all loops $\gamma_w$ on
$\CK_w\Sigma_0$ and for all $w\in\pi_1(M)^\sim$. If
the symplectic potential $\vartheta$ happens to lie in the image of the
transgression map, then this condition should be compatible with the regressed picture: Under regression,
the trivialization of $\CCL_w$ over $\CK_w\Sigma_0/\CP$ determines a gauge
equivalence class of trivial gerbes with connective structure over
$\Sigma_0/\CP$. To fix a trivialization, we note that if
$\gamma_w$ is the image of the smooth map $\ell_w(\rho) :S^1\to \CK_w\Sigma_0$,
then the image of the adjoint map $\ell_w^\vee(\tau,\rho) :=
\ell_w(\rho)(\tau)$ is a torus $\FT_{\gamma_w}^2\cong S^1\times S^1$, or ``2-loop'', in
$\Sigma_0\subseteq \Sigma$; conversely, any smooth 2-loop
$\FT^2\subset\Sigma_0$ defines a loop on $\CK_w\Sigma_0$ for each
winding mode $w\in\pi_1(M)^\sim$. Then the holonomy of the symplectic
potential around the loop $\gamma_w$ is equal to the 2-holonomy
of the regressed connective structure on the gerbe $\CT^{-1}\CCL_w$ around the
associated 2-loop~\cite{Waldorf:2010aa},\footnote{See Appendix~\ref{app:B} for the precise
expression for the 2-holonomy of a topologically non-trivial $B$-field.}
\beq
\hol_{\gamma_w}(\vartheta)=\hol_{\FT_{\gamma_w}^2}(B_w):= \exp\Big(2\pi\, \di\,
\oint_{\FT_{\gamma_w}^2}\, B_w\Big) \ ,
\eeq
and hence Bohr-Sommerfeld quantization on $\CK_w\Sigma$ requires that
$\oint_{\FT^2}\,B_w \in\RZ$ for all 2-loops $\FT^2\subset\Sigma_0$. 

\section{Quantization of vector spaces\label{sec:Moyal}}

In these remaining sections we will consider three explicit examples
of the formalism described in the preceding sections. Here we will derive the Moyal product on the Euclidean
space $\FR^d$, which was reviewed in \S\ref{sec:dualquant}, using the groupoid
approach to quantization. We shall then extend it to the 2-plectic manifold
$\FR^3$ endowed with its natural volume form.

\subsection{Symplectic geometry\label{subsec:vectorspaces-sympl}}

The following construction is an extended version of that presented in
\cite[\S6.2]{Hawkins:0612363}. As we will draw parallels to our new
construction in the next subsection, we will be very explicit in our
discussion. We consider the Poisson manifold given by the real vector
space $V=\FR^d$ with constant
Poisson structure $\pi^{ij}= -\pi^{ji}$, $i,j=1,\dots,d$. As discussed
in \S\ref{sec:PMandLieAlgebroids}, the Poisson structure on $V$
turns the cotangent bundle $T^*V$ into a Lie algebroid. This Lie
algebroid can be integrated to a Lie groupoid if and only if there is
an integrating symplectic groupoid for $V$ as defined in Appendix \ref{app:A} \cite{Crainic:2002aa}.

The preferred choice for such an integrating symplectic groupoid is
$(\Sigma,\omega)$, where $\Sigma$ is the cotangent bundle $T^*V \cong
V\times V^*$. We will use dual Cartesian coordinates $x^i$ on $V$ and $p_i$
on $V^*$. In these coordinates, the symplectic two-form $\omega$ is
given by $\omega=-\langle \dd p,\dd x\rangle= \dd x^i\wedge \dd
p_i$. The groupoid structure on $\Sigma$ depends on the matrix
$\pi=(\pi^{ij})$, regarded as a linear map $\pi:V^*\to V$. The object inclusion map of
the groupoid $\Sigma$ is trivially given by
$\unit_x:(x^i)\mapsto(x^i,x^i)$, where we used the isomorphism between
$V$ and its dual $V^*$. We can encode an arrow $\alpha\in\Sigma$ by a pair $(x^i,p_i)$ using the Bopp shifts
\begin{equation}
\xymatrix{
   (x^i+\tfrac{1}{2}\, \pi^{ij}\, p_j) \ar@/^2pc/[rr]^{\alpha}
&& (x^i-\tfrac{1}{2}\, \pi^{ij}\, p_j)
}~.
\end{equation}
The source and target maps on the groupoid $\Sigma$ are thus
\begin{equation}
 \sfs(\alpha)=\sfs(x,p)=(x^i+\tfrac{1}{2}\,\pi^{ij}\, p_j)\eand
 \sft(\alpha)=\sft(x,p)=(x^i-\tfrac{1}{2}\,\pi^{ij}\,p_j)~.
\end{equation}
Recall that one condition for $\Sigma$ to be an integrating groupoid
is that $\sft:\Sigma\to V$ is a Poisson map. That this is indeed the
case here follows from a straightforward computation
\begin{equation}
\{\sft^*f,\sft^*g\}_{\omega^{-1}}= \Big(\, \derr{\, \sft^i}{x^k}\,\derr{f(y)}{y^i}\,
    \derr{\, \sft^j}{p_k}\, \derr{g(y)}{y^j}-\derr{\, \sft^i}{x^k}\,
    \derr{g(y)}{y^i}\, \derr{\, \sft^j}{p_k}\,
    \derr{f(y)}{y^j}\, \Big)\bigg|_{y=\sft(x,p)}= \sft^*\{f,g\}_\pi~.
\end{equation}

The set of composable arrows (2-nerve) $\Sigma_{(2)}$ can be identified
with $V\times V^*\times V^*$: An element $(x,p,p'\, )$
represents the concatenation $\alpha_1\circ\alpha_2$ of two arrows
$\alpha_1$ and $\alpha_2$ as
\begin{equation}
\xymatrix{
   x^i+\pi^{ij}\,(p_j+p'_j\, ) \ar@/^2pc/[rr]^{\alpha_1}
&& x^i+\pi^{ij}\,(p_j-p'_j\, ) \ar@/^2pc/[rr]^{\alpha_2}
&& x^i-\pi^{ij}\,(p_j+p'_j\, )
}~.
\end{equation}
In coordinates, the projections $\pr_1(\alpha_1\circ\alpha_2)=\alpha_1$ and $\pr_2(\alpha_1\circ\alpha_2)=\alpha_2$ are given by
\begin{equation}
 \pr_1(x,p,p'\, )=(x^i+\tfrac{1}{2}\,\pi^{ij}\,p_j,p'_i\, )\eand
 \pr_2(x,p,p'\, )=(x^i-\tfrac{1}{2}\,\pi^{ij}\,p_j',p_i)~,
\end{equation}
and the groupoid multiplication $\sfm(\alpha_1,\alpha_2)=\alpha_1\circ\alpha_2$ reads
\begin{equation}
\sfm(x,p,p'\, )=(x^i,p_i+p'_i\, )~.
\end{equation}

This completes the construction of the relevant structures on the groupoid $\Sigma$. To
verify that $\Sigma$ integrates $V$, we have to check that $\omega=\dd
x^i\wedge\dd p_i$ is multiplicative for this groupoid structure, i.e.,
$\dpar^*\omega=0$. One readily computes
\begin{equation}
 \begin{aligned}
    \pr_1^* \omega&=\dd x^i\wedge \dd p'_i+\tfrac{1}{2}\,\pi^{ij}\,\dd p_i\wedge \dd p'_j~,\\[4pt]
    \pr_2^* \omega&=\dd x^i\wedge \dd p_i-\tfrac{1}{2}\,\pi^{ij}\,\dd p'_i\wedge \dd p_j~,\\[4pt]
    \sfm^*\omega&=\dd x^i\wedge \dd p_i+\dd x^i\wedge \dd p'_i~,
 \end{aligned}
\end{equation}
and thus $\dpar^*\omega:=\pr_1^*\omega-\sfm^*\omega+\pr_2^*\omega=0$.

The prequantization and polarization of $\Sigma$ are straightforward:
Since $\omega$ is exact, the prequantum line bundle is the trivial bundle $\Sigma\times \FC$,
which we endow with a connection $\nabla$ of curvature
$F_\nabla=-2 \pi\, \di \, \omega$. The projection $\sfp:\Sigma\rightarrow
\Sigma/\CP$, $\sfp=\sfs-\sft$ with $\CP={\rm span}\big\{\der{x^i}\big\}$ maps
$\Sigma$ to $V^*$. Note that $\sfp$ is a fibration of groupoids if we
regard $V^*$ as an additive group.

The twist element $\sigma=\sigma_\pi$ used in
the construction of the twisted convolution algebra can be obtained
from an adapted symplectic potential $\theta$ as described in \S\ref{subsec:2.4} and \S\ref{subsec:2.5}. The simplest choice is
$\theta= x^i\, \dd p_i$, for which $\dpar^*\theta$ is
closed. Moreover, this symplectic potential is perpendicular to the polarization $\CP$ and
therefore adapted. The twist element
$\sigma_\pi$ on $(\Sigma/\CP)_{(2)}=V^*\times V^*$ is obtained from the equation
\begin{equation}
 \sigma_\pi^{-1}\, \dd \sigma_\pi
 =\di\, \dpar^*\theta=\dd\big(-\tfrac{1}{2}\, p_i\,\pi^{ij}\,  p_j'\, \big)~,
\end{equation}
and up to an irrelevant constant phase we arrive at the group cocycle
\begin{equation}
 \sigma_\pi(p,p'\,)=\de^{-\frac{\di}{2}\, p_i\,\pi^{ij}\,  p_j'\,
 } ~.
\label{Vcocycle}\end{equation}

A Haar system is given by the translationally invariant Haar measure
$\dd p$ on $V^*$. Alternatively, the relevant half-form bundles are
\begin{equation}
 \Omega_\Sigma^{1/2}=\sqrt{\mbox{$\bigwedge^d$} T^*(V\times V^*)} \eand
 \Omega^{1/2}_\CP=\sqrt{\mbox{$\bigwedge^d$}T^*V } \ .
\end{equation}
Altogether, the twisted polarized convolution algebra on $\Sigma/\CP\cong V^*$ is then given by functions on $V^*$ with the convolution product
\begin{equation}
 (\tilde f\circledast_\pi \tilde g)(p)=\int_{V^*}\, \dd p' \
 \sigma_\pi(p',p-p'\, )\ \tilde f(p'\, )\, \tilde g(p-p'\,) \ ,
\end{equation}
since in this case $(\Sigma/\CP)^{pt}= V^*= \sft^{-1}(\sft(p))$ for all
$p\in V^*$.
We therefore recover the algebra of functions on the Moyal plane
$V_\pi$ as a
quantization of the associated dual Lie algebroid $A^*(\Sigma/\CP)=V$.

\subsection{2-plectic geometry}\label{subsec:vectorspaces-2pl}

We will now generalize the above discussion to the case of the
2-plectic manifold $(\FR^3,\varpi)$. In usual Cartesian coordinates,
the 2-plectic form reads as $\varpi=\pi^{-1}\, \dd x^1\wedge \dd
x^2\wedge \dd x^3$. To quantize this space, we switch to its knot
space $\CK \FR^3$ which is a symplectic manifold with symplectic form
obtained by transgressing the 2-plectic form to get
\begin{equation}
  \omega:=\CT\varpi=\oint\, \dd \tau \ \varpi_{ijk}\, \xd^{k\tau}\, \delta x^{i\tau}\wedge \delta x^{j\tau}=\oint\, \dd \rho\ \oint\, \dd \tau \ \varpi_{ijk}\, \xd^{k\tau}\, \delta(\tau-\rho)\, \delta x^{i\rho}\wedge \delta x^{j\tau}~,
\end{equation}
as explained in \S\ref{subsec:Poissonknots}. We define a
Poisson structure from the symplectic structure via the formula
$\{f,g\}_{\omega^{-1}} :=\iota_{X_f}\, \iota_{X_g}\omega$ for $f,g\in
\CC^\infty(\CK \FR^3)$, where $X_f$ denotes the Hamiltonian vector
field defined through $\iota_{X_f}\omega=\dd f$. Here we find
\begin{equation}
 \{f,g\}_\pi :=\oint\, \dd \rho \ \oint\, \dd \tau \ \delta(\tau-\rho)
 \, \pi^{ijk}\, \frac{\xd_{k\tau}}{|\xd^\tau |^2}\, \Big(\,
 \delder{x^{i\rho}}f\, \Big)\, \Big(\, \delder{x^{j\tau}}g\, \Big)~,
\end{equation}
where $\pi^{ijk}=\pi\, \eps^{ijk}$. One readily checks that this
Poisson bracket indeed satisfies the Jacobi identity. We shall now use the
groupoid approach to quantize the underlying Poisson algebra on
knot space.

First, we need an integrating symplectic groupoid. The groupoid quantization of $V=\FR^3$ suggests using
the (restricted) cotangent groupoid $\hat{T}^*\CK\FR^3$. As before, we describe this space by the standard coordinates
$(x^{i\tau},p_{i\tau})$ on the (restricted) loop space
$\hat{T}^*\CL_{\di}\FR^3\cong \CL_{\di} T^*\FR^3=\CL_\di\Sigma$. This is possible as
long as all expressions obtained are invariant under the appropriate
action of the
group of reparametrization transformations $\CR= \sDiff^+(S^1)$. In
particular, the loop space coordinates $x^{i\tau}$ are invariant under actions of $\CR$ and the $p_{i\tau}$ transform in the
same representation of $\CR$ as $\xd^{i\tau}$. In these coordinates the canonical
symplectic two-form $\omega$ on $\CL_\di\Sigma$ is given by
\begin{equation}
 \omega=\oint\, \dd \tau\ \oint\, \dd \rho\ \delta(\tau-\rho)\, \delta x^{i\tau}\wedge \delta p_{i\rho}~,
\end{equation}
and it can be derived from the symplectic potential
\begin{equation}
 \vartheta=\oint \, \dd \tau\ x^{i\tau}\, \delta p_{i\tau}~.
\end{equation}
Both of these differential forms are invariant under the group of reparametrization transformations~$\CR$. 

On $\CL_\di\Sigma$, we choose source and target maps invariant under the
action of $\CR$ which are given by
\begin{equation}
 \begin{aligned}
 \sfs\big(x^{i\tau}\,,\,p_{i\tau}\big)&=x^{i\tau}+\frac{1}{2}\,
 \pi^{ijk}\, p_{j\tau}\, \frac{\xd_{k\tau}}{|\xd^\tau |^2}~,\\[4pt]
 \sft\big(x^{i\tau}\,,\,p_{i\tau}\big)&=x^{i\tau}-\frac{1}{2}\,
 \pi^{ijk}\, p_{j\tau}\, \frac{\xd_{k\tau}}{|\xd^\tau |^2}~,
 \end{aligned}
\end{equation}
where we used the Euclidean metric on $\FR^3$ to lower indices as
$\xd_{i\tau}:=\delta_{ij}\, \xd^{j\tau}$. Here the target map $\sft$ is a Poisson map only to lowest order in $\pi$, i.e.,\
\begin{equation}
\begin{aligned}
 \{\sft^*f,\sft^*g\}_{\omega^{-1}}:=&\oint\, \dd\tau \ \bigg( \Big(\,
 \delder{x^{i\tau}}\sft^*f\, \Big)\, \Big(\,
 \delder{p_{i\tau}}\sft^*g\, \Big)-\Big(\, \delder{x^{i\tau}}\sft^*g\,
 \Big)\, \Big(\, \delder{p_{i\tau}}\sft^*f\, \Big) \bigg) \\[4pt]
=& \ \sft^*\{f,g\}_\pi+\CO(\pi^2) \efor f,g\in \CC^\infty(\CL_\di \FR^3)~.
\end{aligned}
\end{equation}
For the moment, let us nevertheless continue and complete the groupoid
structure first to lowest order in $\pi$. The 2-nerve of $\CL_\di\Sigma$ is
$\CK_\di\Sigma_{(2)} =\CL_\di(\FR^3\times \FR^3\times \FR^3)/\CR$, and
comparison with the case $V= \FR^3$ suggests using the projections
and multiplication maps given by
\begin{equation}
\begin{aligned}
 \pr_1(x^{i\tau},p_{i\tau},p'_{i\tau})&:=\Big(x^{i\tau}+\frac{1}{2}\,
 \pi^{ijk}\, p_{j\tau}\, \frac{\xd_{k\tau}}{|\xd^\tau |^2}\,,
 \,p'_{i\tau}\Big)~, \\[4pt]
 \pr_2(x^{i\tau},p_{i\tau},p'_{i\tau})&:=\Big(x^{i\tau}-\frac{1}{2}\,
 \pi^{ijk}\, p'_{j\tau}\, \frac{\xd_{k\tau}}{|\xd^\tau
   |^2}\,,\,p_{i\tau}\Big)~, \\[4pt]
 \sfm(x^{i\tau},p_{i\tau},p'_{i\tau})&:=
 \big(x^{i\tau}\,,\,p_{i\tau}+p'_{i\tau} \big)~.
\end{aligned}
\end{equation}
The canonical symplectic structure $\omega$ on $\CL_\di\Sigma$ is also
multiplicative only to lowest order in $\pi$, i.e.,
$\dpar^*\omega=\CO(\pi)$. We have therefore arrived at an integrating
symplectic groupoid for $(\CL_\di \FR^3,\pi)$ to lowest order in
$\pi$, and we can now attempt to correct these structures order by
order. We will demonstrate this procedure to the first non-trivial order in $\pi$.

The first step is to ensure that $\sft$ is a Poisson map to order
$\CO(\pi^2)$. To preserve the groupoid structures, we will attempt to
do so by replacing $p_{i\tau}$ in all groupoid maps and the symplectic
form by a more general expression $\tilde{p}_{i\tau}=p_{i\tau}+\pi \, p^{(1)}_{i\tau}$. Such an expression can indeed be found and it reads as
\begin{equation}
 \begin{aligned}
  \tilde{p}_{i\tau}&=p_{i\tau}+\pi\, \bigg(\frac{p_{i\tau}\,
    \xdd^{j\tau}\,
    \xd_{j\tau}}{3|\xd^\tau|^4}+\frac{\eps_{ijk}}{|\xd^\tau|^4}\,
  \Big(\, \frac{1}{2}\, \xdd^{j\tau}\, p^{k\tau}\, \xd^{l\tau} \,
  p_{l\tau}+\frac{4}{3}\, \xd^{j\tau}\, \xdd^{k\tau}\, p_{l\tau}\,
  p^{l\tau} \\
&\hspace{6cm} +\, \frac{1}{2}\, \xd^{j\tau}\, p_{k\tau}\,
\xd^{l\tau}\, \pd_{l\tau}+\pd^{j\tau}\, \xd^{k\tau}\, p_{i\tau}\,
\xd^{i\tau}\, \Big)\bigg)~.
 \end{aligned}
\end{equation}
To lowest order in $\pi$, we can invert this formula to get
\begin{equation}
  p_{i\tau}=\tilde{p}_{i\tau}-\pi\, \Big(\,
  \frac{\tilde{p}_{i\tau}\, \xdd^{j\tau}\, \xd_{j\tau}}{3|\xd^\tau|^4}+\ldots\,
  \Big)~,
\end{equation}
and therefore replacing $p_{i\tau}$ by $\tilde{p}_{i\tau}$ corresponds to a mere coordinate change.

The next step requires us to modify the groupoid structure so that $\omega$ is multiplicative to lowest order in $\pi$. We have
\begin{equation}
\begin{aligned}
 \dpar^*\omega=& \ \pr_1^*\omega+\pr_2^*\omega-\sfm^*\omega \\[4pt] =&
 \ \frac{1}{2}\, \oint \, \dd \tau\ \pi^{ijk}\, \delta\Big(\,
 \frac{\xd_{k\tau}}{|\xd^\tau|^2}\, \Big)\wedge
 \delta(\tilde{p}^{\,\prime}_{i\tau}\, \tilde{p}_{j\tau})=\oint\, \dd
 \tau\ \delta x^{i\tau}\wedge R_{i\tau}~,
\end{aligned}
\end{equation}
where $R_{i\tau}$ can be chosen to be a closed one-form of order
$\CO(\pi)$ which transforms like $p_{i\tau}$ under actions of the
group $\CR$. For closed $R_{i\tau}$, there is a potential $R_{i\tau}=\delta r_{i\tau}$ and we can use this to modify the groupoid multiplication according to
\begin{equation}
 \sfm(x,\tilde{p},\tilde{p}\,'\, )=(x^{i\tau},\tilde{p}_{i\tau}+\tilde{p}^{\,\prime}_{i\tau}+r_{i\tau})~.
\end{equation}
Note that both $\sfs(\sfm(x,\tilde{p},\tilde{p}\, '\, ))$ and
$\sft(\sfm(x,\tilde{p},\tilde{p}\,'\, ))$ are modified to second order
in $\pi$ and therefore the groupoid structure is still preserved to
first order in $\pi$. Moreover, this modification ensures that
$\omega$ is multiplicative, and therefore there is a symplectic potential with
\begin{equation}\label{eq:pretwist}
 \dpar^*\vartheta=\delta\Big(\, -\frac{1}{2}\, \oint\, \dd \tau\
 \pi^{ijk}\, \tilde{p}_{i\tau}\, \tilde{p}^{\,\prime}_{j\tau}\,
 \frac{\xd_{k\tau}}{|\xd^\tau|^2}\, \Big)~.
\end{equation}
This form is clear to first order in $\pi$ by computing
$\dpar^*\vartheta$ for $r_{i\tau}=0$. The contributions from
$r_{i\tau}$ add the missing terms of the form $-\frac{1}{2}\, \oint \,
\dd \tau\ \pi^{ijk}\, \tilde{p}_{i\tau}\,
\tilde{p}^{\,\prime}_{j\tau}\,
\delta\big(\frac{\xd_{k\tau}}{|\xd^\tau|^2}\big)$. This algorithm can presumably be iterated to compute the corrections to arbitrary order in $\pi$.

The group cocycle corresponding to \eqref{eq:pretwist} reads as
\begin{equation}\label{eq:twist}
 \sigma_\pi(x,\tilde{p},\tilde{p}\,'\, )=\exp\Big(\, -\frac{\di}{2}\,
 \oint\, \dd \tau\ \pi^{ijk}\, \tilde{p}_{i\tau}\,
 \tilde{p}^{\,\prime}_{j\tau}\, \frac{\xd_{k\tau}}{|\xd^\tau|^2}\, \Big)~.
\end{equation}
As discussed in \S\ref{ssec:PrequantKnotSpace}, we will not attempt a
precise definition of the corresponding convolution algebra, as e.g.\
the choice of measure on loop space is unclear to us. We can
nevertheless glean a few features of the quantization from the form of
the twisting cocycle \eqref{eq:twist}. Recall that we worked with the
symplectic two-form $\omega=\oint \, \dd\tau\ \delta x^{i\tau}\wedge
\delta \tilde{p}_{i\tau}$. Therefore the variable $\tilde{p}$ is
canonically conjugate to $x$ and the form of $\sigma_\pi(x,\tilde
p,\tilde p\,'\, )$ implies a relation between quantized coordinate functions
given by
\begin{equation}\label{quantcoordrel}
 {}\big[\hat{x}^{i\tau}\,,\,\hat{x}^{j\rho}\big]=\di \,
 \pi^{ijk}\, \widehat{\frac{\xd_{k\tau}}{|\xd^\tau|^2}} \
 \delta(\tau-\rho )~.
\end{equation}
To first order in $\pi$, this relation is identical to those of the
noncommutative loop spaces found in
\cite{Bergshoeff:2000jn,Kawamoto:2000zt,Berman:2004jv} describing the quantum
geometry of open membranes ending on M5-branes in a constant
supergravity $C$-field
background. The higher order terms in $\pi$ computed
by~\cite{Kawamoto:2000zt,Berman:2004jv} could presumably come from
higher order corrections to the groupoid structure above.

This loop space construction can be reduced to that of the symplectic space
$(\FR^2,\omega)$ by applying a simple dimensional reduction
prescription from M-theory to string theory: We choose a particular
M-theory direction, say $x^3$, and compactify $\FR^3\rightarrow
\FR^2\times S^1$ along this direction on a circle $S^1$ of radius $r$. Imposing the condition that loops extend around this direction yields
\begin{equation}
 x^{i}(\tau)=x^i_0+\frac{\tau}{r}\, \delta^{i,3} \ , \qquad
 \xd^i(\tau) =\frac{1}{r}\, \delta^{i,3} \eand p^{3}(\tau) =0
\end{equation}
with $x_0^i\in\FR$ constant zero modes. The Nambu-Poisson and multisymplectic
structures on $\FR^3$ reduce to Poisson and symplectic structures on
$\FR^2$ given by $\pi^{ij}=r\,\pi^{ij3}$ and $\omega_{ij}= r^{-1} \,
\varpi_{ij3}$. In general, the loop space groupoid structure at a
given $\tau_0\in S^1$ reduces to that of $\FR^2$ after integrating over the loop parameter.

\section{Noncommutative tori}

In this section we extend the groupoid quantization of
\S\ref{sec:Moyal} to tori. The underlying groupoid structures turn out
to be the essentially the same; the novelty in this class of examples
is that the generalized Bohr-Sommerfeld quantizations of
\S\ref{BohrSommerfeld} and \S\ref{BS2leaves} come into play. Hence we
shall only highlight the topological aspects which crucially affect
the quantization.

\subsection{Symplectic geometry}

Geometric
quantization on the symplectic groupoid associated to the Poisson
$d$-torus leads rather naturally to the noncommutative torus. The groupoid quantization of the torus $\FT^2$ with its standard
constant symplectic structure was originally done by
Weinstein~\cite{Weinstein:1991aa}, who was the first to apply
Bohr-Sommerfeld quantization to a symplectic groupoid with
polarization having multiply connected leaves. Here we follow
again~\cite[\S6.5]{Hawkins:0612363} and work on tori of arbitrary
dimension. Let $V=\FR^d$ be a real vector space of dimension $d$, and
let $\Lambda=\RZ^d\subset V$ be a lattice of rank $d$. Then the
quotient $\FT^d=V/\Lambda$ is a $d$-dimensional torus. We endow
$\FT^d$ with a constant Poisson structure given as before by an
antisymmetric $d\times d$ matrix $\pi$ on~$V$.

As in the case of vector spaces from
\S\ref{subsec:vectorspaces-sympl}, we take the symplectic groupoid
$(\Sigma,\omega)$ to be the cotangent groupoid $\Sigma=T^*\FT^d\cong
\FT^d\times V^*$ with the canonical symplectic structure $\omega=\dd
x^i\wedge\dd p_i$ for $x=(x^i)\in\FT^d$ and $p=(p_i)\in V^*$. The
structure maps of the groupoid are exactly those described in
\S\ref{subsec:vectorspaces-sympl}. Likewise, the prequantization is
given by the trivial line bundle $E=\Sigma\times\FC$ with connection
$\nabla=\dd +\theta$, where now the global symplectic
potential is
\beq
\theta=p_i\, \dd x^i \ .
\label{thetatorus}\eeq
The polarization $\CP={\rm span}\big\{\der{x^i} \big\}$ is the kernel
of the fibration of groupoids $\FT^d\times V^*\to V^*$, so that again
$\Sigma/\CP\cong V^*$.

The crucial distinction now from the quantization of the universal
covering space $V$ is that the
symplectic potential (\ref{thetatorus}) is \emph{not} adapted to
$\CP$; this is also reflected in the fact that the leaves of $\CP$,
being $d$-tori, are not simply connected. There are natural
identifications $H_1(\FT^d,\RZ)=\Lambda$ and
$H^1(\FT^d,\RZ)=\Lambda^*$. Let $\gamma_i\subset\FT^d$ for
$i=1,\dots,d$ be a basis of one-cycles dual to the one-forms $\dd x^i$
in this sense. Given an arbitrary cycle $\gamma=m^i\,\gamma_i$,
with winding numbers $m=(m^i)\in\pi_1(\FT^d)=\pi_0(\Lambda) = \Lambda$, the holonomy of
the connection given by (\ref{thetatorus}) is
\beq
\hol_\gamma(\theta)= \de^{2\pi\,\di\,\langle p,m\rangle} \ .
\eeq
The Bohr-Sommerfeld quantization condition $\hol_\gamma(\theta)=1$
thus holds if and only if $\langle p,m\rangle\in\RZ$ for all
$m\in\Lambda$, i.e., $p\in\Lambda^*\subset V^*$. The Bohr-Sommerfeld
groupoid is then $\Sigma_0=\FT^d\times\Lambda^*$, which is a
disconnected union
\beq
\Sigma_0=\bigsqcup_{p\in\Lambda^*}\, \FT_p^d
\eeq
of the Bohr-Sommerfeld leaves $\FT_p^d\cong \FT^d$; its quantization
is given by the twisted convolution algebra of
$\Sigma_0/\CP=\Lambda^*$. Note that while the locations of the
Bohr-Sommerfeld leaves depend on the particular choice of symplectic
potential $\theta$, any two such choices are intertwined by the action
of the group ${\sf SL}(d,\RZ)$ of automorphisms of the lattice
$\Lambda$, which preserve the symplectic two-form $\omega$.

The computation of the group two-cocycle $\sigma_\pi:\Lambda^*\times
\Lambda^*\to \sU(1)$ proceeds exactly as before and results in the skew
bicharacter (\ref{Vcocycle}), but now with $p,p'\in\Lambda^*$. A left
Haar system is given by discrete measure on the abelian group
$\Lambda^*$, and the twisted convolution product is
\begin{equation}
 (\tilde f\circledast_\pi \tilde g)(p)=\sum_{p'\in\Lambda^*}\,
 \sigma_\pi(p',p-p'\, )\ \tilde f(p'\, )\, \tilde g(p-p'\,) \ .
\end{equation}
Thus for $\tilde f,\tilde g\in\CS(\Lambda^*)$, the twisted group
convolution algebra $\CC_r^*(\Lambda^*,\sigma_\pi)$ is the usual
algebra of functions on the noncommutative torus $\FT^d_\pi$, regarded as a
deformation of the algebra of Fourier series on $\FT^d$;
the quantization map is achieved via pullback along the Fourier
transformation $\CC^\infty(\FT^d)\to \CS(\Lambda^*)$.

\subsection{2-plectic geometry\label{subsec:torus-2pl}}

Now let us endow $\FT^3$ with a constant Nambu-Poisson structure
specified by an antisymmetric tensor $\pi^{ijk}$ of rank~$3$.
Equivalently, we can consider the three-torus $\FT^3$ with its standard 2-plectic structure
$\varpi=\pi^{-1}\, \dd x^1\wedge \dd x^2\wedge \dd x^2$; the standard
prequantum bundle gerbe with connective structure associated to this
2-plectic form is described in Appendix~\ref{app:B}. We
transgress the quantization problem to the knot space $\CK\FT^3$,
using the cotangent groupoid $\CK\Sigma= \CK T^*\FT^3$, which we may identify with $\hat T^*\CK
\FT^3$. A loop $x(\tau):S^1\to\FT^3$ induces a homomorphism of homology
groups $H_1(S^1,\RZ)\to H_1(\FT^3,\RZ)$. Since $H_1(S^1,\RZ)=\RZ$ and
$H_1(\FT^3,\RZ)=\Lambda$, it follows that to any map $x(\tau)\in\CK
\FT^3$ we may assign an element $w=w\big(x(\tau)\big)= (w^i)\in\Lambda$ whose
components are the winding numbers of the loop. The Fourier mode
expansion of any knot $x$ is therefore given by
\beq
x(\tau)= x_0+\tau\, w+ \sum_{n\neq0}\, \de^{2\pi\, \di\, n\, \tau} \,
\xi_n
\eeq
with $x_0\in\FT^3$ and $\xi_{-n}=\overline{\xi_n}$ complex
numbers, so that the tangent vectors $\dot x(\tau)$ now have zero modes, i.e.,
$\oint \, \dd\tau\ \dot x(\tau)=w$. This defines a decomposition of the knot space
\beq
\CK\FT^3\cong \FT^3\times\Lambda\times \CV
\eeq
where $\CV$ is the infinite-dimensional real vector space of ``oscillators''; in particular
the knot space is Pontryagin self-dual, i.e.,
$\widehat{\CK\FT^3}\cong \CK\FT^3$, and hence respects the T-duality
symmetry of closed string theory with target space $\FT^3$.
Thus the cotangent groupoid
is a disconnected union
\beq
\CK\Sigma=\bigsqcup_{w\in\Lambda}\, \CK_w\Sigma \ ,
\label{CKwindingsum}\eeq
where each connected component is an identical copy $\CK_w\Sigma\cong
\FT^3\times V^*\times \CV$; this decomposition nicely parallels the
decomposition of the prequantum bundle gerbe on $\FT^3$ given in Appendix~\ref{app:B}. On each component we construct the analogous
groupoid structures to the ones in the case of $\CK\FR^3$ from
\S\ref{subsec:vectorspaces-2pl}, and then take the direct sum over all
winding sectors $w\in\Lambda$.

To lowest order in $\pi$, the symplectic two-form on $\CK_w\Sigma$ is
given by (\ref{loopomegagen}). The global symplectic
potential is given by (\ref{loopthetagen}), and its holonomy can be
computed as the 2-holonomy around the 2-loop $\FT^2_{w,\gamma}\subset \FT^3$ associated to a
cycle $\gamma=m^i\, \gamma_i$ with $m=(m^i)\in\Lambda$ in
the topological sector $w\in\Lambda$. The 2-loop is parametrized by
the adjoint map
\beq
\ell^\vee_w(\tau,\rho) = \ell_0+m\,
\rho+ w\, \tau +\sum_{l,n\neq0}\, \de^{2\pi\,\di\,(l\,
  \rho+ n\, \tau)}\, \lambda_{l,n}
\eeq
with $\ell_0\in\FT^3$, while the global momentum $p(\tau)\in
\CC^\infty(S^1,x^*T^*M)$ has mode expansion
\beq
p(\tau)= p_{0,i}\, \dd x_0^i+\sum_{n\neq0}\, \de^{2\pi\,\di\,
  n\, \tau}\, \rho_{n,i}\, \dd\xi_n^i
\eeq
with $p_0=(p_{0,i}) \in V^*$. The 2-holonomy is then given by
\beq
\hol_{w,\gamma}(\vartheta)=\exp\Big(2\pi\, \di\, \oint\, \dd\rho \
\oint \, \dd\tau \
\frac{\partial\ell_w^{\vee,i}(\tau,\rho)}{\partial\rho} \, p_i(\tau)
\Big) = \de^{2\pi\,\di\, \langle
  p_0, m\rangle} \ .
\eeq
The Bohr-Sommerfeld 2-leaves are thus parametrized by quantized
momentum zero modes $p_0\in\Lambda^*\subset V^*$, and hence to lowest
order the Bohr-Sommerfeld variety is
\beq
\CK_w\Sigma_0 = \FT^3\times\Lambda^*\times \CV~,
\eeq
independently of the winding mode $w\in\Lambda$. The locations of the
Bohr-Sommerfeld 2-leaves do however depend on the gauge orbits of the
symplectic potential through the winding sectors. Given a constant
two-form $\beta$ on $\FT^3$, representing the cohomology class of a
flat gerbe 1-connection in $H^2(\FT^3,\sU(1))$, consider a gauge
transformation (\ref{thetabetagen}) of the symplectic potential. Then
the Bohr-Sommerfeld quantization condition is generalized to
\beq
p_0+\iota_w\beta \ \in \ \Lambda^* \ ,
\eeq
which defines a lattice $\Lambda^*_{w,\beta}\cong\Lambda^*$ with
$\Lambda^*_{0,\beta}=\Lambda^* =\Lambda^*_{w,0}$.

At higher order in $\pi$, we can redefine the momentum $p(\tau)$ to
$\tilde p(\tau)$ as described in \S\ref{subsec:vectorspaces-2pl}, and
the Bohr-Sommerfeld quantization condition now reads $\tilde
p_0\in\Lambda^*$; as above, this modification only affects the precise
locations of the Bohr-Sommerfeld 2-leaves. It is easy to see that, in each fixed topological
sector $w\in\Lambda$, the corrected groupoid structures follow exactly
as described in \S\ref{subsec:vectorspaces-2pl}; although additional
terms do arise through $\oint\, \dd\tau\ \xd(\tau)=w$ from integrating
by parts over the loop, they are always hit by the loop space exterior
derivative $\delta$ and therefore vanish. We thus arrive again at the
twisting cocycle (\ref{eq:twist}), but now with the quantized zero
mode condition $\tilde p_0, \tilde p^{\,\prime}_0\in\Lambda^*$.

We can see the effect of the winding sectors $w\in\Lambda$ on the
quantization of loop space explicitly at the
level of zero modes: Integrating the commutation relations
(\ref{quantcoordrel}) over the loop parameters $\tau,\rho$ we arrive
at
\beq
{}\big[\hat{x}_0^{i}\,,\,\hat{x}_0^{j}\big]=\di \,
 \pi^{ijk}\, w_k \ .
\eeq
This relation reproduces the expected noncommutative geometry of
closed strings in $\FT^3$ with constant $H$-flux in the non-geometric
$Q$-space duality
frame~\cite{Lust:2010iy,Condeescu:2012sp}: Only closed strings which
extend around cycles of the torus can probe noncommutativity, in which
case the geometry is that of the noncommutative torus
$\FT^3_{\pi_w}$ where $\pi_w^{ij}=\pi^{ijk}\, w_k$. On the other hand,
after summing over
all winding sectors $w\in\Lambda$ as in (\ref{CKwindingsum}) we
obtain the anticipated quantum geometry of the corresponding
T-fold obtained via T-duality from the torus $\FT^3$ with
$H$-flux~\cite{Bouwknegt:2004ap}:\footnote{See~\cite{Mylonas:2012pg} for a discussion of the
  relation between these two perspectives on the noncommutative
  $Q$-space geometry.} Since $\pi_w^{ij}$ has rank~$2$, the
full quantized loop space contains a fibration of noncommutative
two-tori.

\section{Fuzzy spheres}

For our final example, we demonstrate how to recover the standard
fuzzy two-sphere from the groupoid quantization of $S^2$; this is a
special instance of the class of examples given in
\S\ref{sec:quantmap}. Then we describe how to extend this quantization
to the loop space of the 2-plectic manifold $S^3$.

\subsection{Symplectic geometry\label{subsec:fuzzy-sympl}}

Let us start with a detailed discussion of the groupoid quantization of $S^2$; we will refer to it later
on when discussing the groupoid quantization of $\CL_\di S^3$.
The canonical symplectic structure on $S^2$ is given by its
area form
\begin{equation}\label{eq:volS2}
 \omega_{S^2}=\tfrac{1}{4\pi}\, \sin \theta^1 \ \dd \theta^1\wedge \dd \theta^2
\end{equation}
written in standard angular coordinates $0\leq \theta^1\leq \pi$,
$0\leq \theta^2\leq 2\pi$. The sphere has unit area with respect to
this area form, $\int_{S^2}\, \omega_{S^2}=1$.
Over $S^2$, we can define a family of prequantum line bundles $E_k$,
$k\in\RZ$, with connection $\nabla$ and associated curvature
$F_\nabla$ such that $F_\nabla :=\nabla^2=-2\pi\, \di\, k\, \omega$. The first Chern number of $E_k$ is computed as
\begin{equation}
 c_1(E_k) =\int_{S^2}\, \frac{\di\, F}{2\pi}=k~.
\end{equation}

To construct a K\"ahler polarization on the sections of $E_k$, we introduce
local gauge potentials for the connection $\nabla$ and consider the
natural complex structure on $S^2$. We will cover $S^2$ by the two
patches $U_+$, for which $\theta^1\neq \pi$, and $U_-$, for which
$\theta^1\neq 0$. On these patches we define the respective potentials
\begin{equation}
 A_+= \frac{\di\, k}{2} \, \big(\cos\theta^1-1 \big)\, \dd \theta^2 \eand
 A_-=\frac{\di \, k}{2}\, \big(\cos\theta^1+1 \big)\, \dd \theta^2 ~.
\end{equation}
The complex structure on $S^2$ is given by the linear map $J$ acting on the vector fields $\der{\theta^i}$ that span $\CC^\infty(S^2,TS^2)$ as
\begin{equation}
 J\der{\theta^1}=\frac{1}{\sin\theta^1}\, \der{\theta^2} \eand
 J\der{\theta^2}=-\sin\theta^1\, \der{\theta^1}
\end{equation}
with $J^2=-\id$. The usual projection of a one-form onto its
antiholomorphic part is given by
\begin{equation}\label{eq:antiHolomProj}
 \Pi^{0,1}\, :\, \xi \ \longmapsto \ \xi^{0,1}=\tfrac{1}{2}\, (\xi+\di\, \xi \circ J)~,
\end{equation}
and thus the antiholomorphic parts of the connections read as
\begin{equation}
\begin{aligned}
 \nabla^{0,1}_\pm&=\Pi^{0,1}\circ(\dd_\pm+A_\pm)\\[4pt]
 &=\frac{1}{2}\, \Big(\, \der{\theta^1}+\frac{\di}{\sin
     \theta^1}\, \der{\theta^2}-\frac{k\,
     (\cos\theta^1\pm1)}{2\sin\theta^1}\, \Big) ~ \dd \theta^1\\
 &~~~~~~~~+\, \frac{1}{2}\, \Big(\, \der{\theta^2}-\di\, 
   \sin\theta^1\, \der{\theta^1}+\frac{\di \, k}{2}\,
   (\cos\theta^1\pm1)\, \Big)
 ~ \dd \theta^2
\end{aligned}
\end{equation}
on $U_\pm$. We will now construct covariantly constant $L^2$-sections
$\psi$ of the line bundle $E_k$. For example, on $U_+$ we look for
solutions to $\nabla^{0,1}_+\psi_+=0$ or equivalently
\begin{equation}
\Big(\, \der{\theta^2}-\di \, \sin\theta^1\,
\der{\theta^1}+\frac{\di\, k}{2}\, (\cos\theta^1-1)\, \Big) \psi_+=0~.
\end{equation}
The general solution is the product of a particular solution and
solutions $s_+$ to $\overline{\partial}_+ s_+:=\Pi^{0,1}\circ \dd_+
s_+=0$, giving
\begin{equation}
 \psi_+(\theta^1,\theta^2) =\Big(\,
 \frac{1}{1+\tan^2\big(\frac{\theta^1}{2}\big)}\,\Big)^{k/2}\,
 s_+\Big(\tan\big(\mbox{$\frac{\theta^1}{2}$} \big)\, \de^{\di\, \theta^2}\Big)~.
\end{equation}
Square integrability with respect to the area form \eqref{eq:volS2},
\begin{equation}
 \frac1{4\pi}\, \int_0^\pi\, \dd\theta^1 \ \int_0^{2\pi}\, \dd\theta^2 \ \sin\theta^1
 \ 
 ~\overline{\psi}_+~\psi_+~<~\infty~,
\end{equation}
implies that $s_+$ is a polynomial of degree $p\leq k$.
By introducing complex stereographic coordinates $z_+=\tan\big(\frac{\theta^1}{2} \big)\,
\de^{\di\, \theta^2}$, we arrive at a more familiar formulation: Then
$s_+$ are polynomials of maximal degree $k$ in $z_+$ and therefore
determine sections of the holomorphic line bundle $\CO_{\CPP^1}(k)$ of degree
$k$ over the complex projective line $\CPP^1\cong S^2$, which is a K\"ahler manifold with K\"ahler two-form
\begin{equation}
 \omega_{\CPP^1}=\omega_{S^2}~,~~~
 \omega_{\CPP^1}\big|_{U_+}=\frac{\di}{2\pi} \,\dpar_+ \,
 \overline{\dpar}_+ \log(1+z_+\, \overline{z}_+)=\frac{\di}{2\pi}\,
 \frac{\dd z_+\wedge \dd \overline{z}_+}{(1+z_+\, \overline{z}_+)^2}~.
\end{equation}

For the integrating symplectic groupoid we take the pair groupoid
$\Sigma= \Pair(\CPP^1)=\CPP^1\times\CPP^1$ with multiplicative symplectic
two-form $\omega=\sfs^*\omega_{\CPP^1}-\sft^*\omega_{\CPP^1}$. The
prequantum line bundle on $\Sigma$ is
$E=\CO_\Sigma(k,0)\otimes\overline{\CO_\Sigma(0,k)}$, endowed with
connection of curvature $-2\pi\,\di\,k\, \omega$, while the
prequantization cocycle is trivial, $\sigma=1$. The polarization on
$\Sigma$ is given by the generalized K\"ahler polarization $\CP=T_{\FC}^{0,1}\CPP^1\times
T_{\FC}^{1,0}\CPP^1$; in local complex coordinates
$(z,\overline{z};w,\overline{w}\,)\in\FC\times\FC$ one has $\CP={\rm
  span}\big\{\frac{\partial}{\partial\overline{z}},
  \frac\partial{\partial w} \big\}$, and a basis of covariantly constant
sections is given by $\underline{\varepsilon}=(\dd
z,\dd\overline{w}\,)$.

The algebra $\CC_r^*(\Sigma/\CP)$ is constructed from polarized
sections $\psi\otimes\sqrt{\,\underline{\varepsilon}}\in
\CC^\infty\big(\Sigma, E\otimes\Omega_\CP^{1/2} \, \big)$, where
$\Omega_{\CP}=\det\big(T_{\FC}^{0,1}\CPP^1\big)\boxtimes
\det\big(T_{\FC}^{1,0}\CPP^1\big)$. This identifies
\beq
\CC_r^*(\Sigma/\CP)\cong H^0\big(\CPP^1\,,\,\CO_{\CPP^1}(k)\big)
\otimes H^0\big(\CPP^1\,,\,\overline{\CO_{\CPP^1}(k)}\,\big) \ ,
\eeq
which is the operator algebra
$\CCK\big(H^0(\CPP^1,\CO_{\CPP^1}(k))\big)$ anticipated from
\S\ref{sec:GBB-Tquant}. The Hilbert space
$\CCH_k:=H^0(\CPP^1,\CO_{\CPP^1}(k))$ is finite-dimensional, and in terms of homogeneous coordinates $(z_0,z_1)\in\FC^2$ on $\CPP^1$ it is spanned by
homogeneous polynomials of degree $k$. It follows that a basis of
$\CC_r^*(\Sigma/\CP)$ is provided by the sections
\beq
\psi_{p,q}(z,\overline{z};w,\overline{w}\,) = \frac{z_0^{p}\, z_1^{k-p}\,
  \overline{w}_0^{\,q}\, \overline{w}_1^{\,{k-q}}}{|z|^k\, |w|^k} \ ,
\qquad p,q=0,1,\dots, k
\eeq
where $|z|^2:= \overline{z}_0\, z_0+\overline{z}_1\, z_1$. Hence any
section $\tilde f\in\CC_r^*(\Sigma/\CP)$ can be expanded as a finite
linear combination
\beq
\tilde f(z,\overline{z};w,\overline{w}\,) = \sum_{p,q=0}^k\, \bfF_{pq}
\ \psi_{p,q}(z,\overline{z};w,\overline{w}\,) \ , \qquad
\bfF_{pq}\in\FC \ ,
\eeq
which under the mapping $\tilde f\mapsto \bfF:=(\bfF_{pq})$ identifies
$\CC_r^*(\Sigma/\CP)$ with the linear space $\sMat_{\FC}(k+1)$ of
$(k+1)\times (k+1)$ complex matrices. The convolution product
(\ref{convkernel}) is given by
\beq
(\tilde f\circledast \tilde f'\,)(z,\overline{z};w,\overline{w}\,) &=&
\int_{\CPP^1}\, \omega_{\CPP^1}(y, \overline{y}\,) \ \tilde
f(z,\overline{z};y, \overline{y}\,)\, \tilde f'(y, \overline{y};
w,\overline{w}\,) \nonumber \\[4pt]
&=& \sum_{p,q,p',q'=0}^k\, \bfF_{pq}\, \bfF'_{p'q'} \
\psi_{p,q'}(z,\overline{z};w,\overline{w}\,) \nonumber \\ && \qquad
\qquad \times\, \int_{\CPP^1}\, \omega_{\CPP^1}(y, \overline{y}\,) \
\psi_{q,p'}(y, \overline{y}; y, \overline{y}\,)
\eeq
for polarized sections $\tilde f=\sum_{p,q}\, \bfF_{pq}\ \psi_{p,q}$
and $\tilde f'=\sum_{p,q}\, \bfF'_{pq}\ \psi_{p,q}$. That this formula
induces the matrix product $\bfF\cdot \bfF'$ now follows from the identity
\beq
\int_{\CPP^1}\, \omega_{\CPP^1}(y, \overline{y}\,) \
\psi_{p,q}(y, \overline{y}; y, \overline{y}\,) = \delta_{p,q}
\label{spqortho}\eeq
expressing orthonormality of the basis sections of
$\CCH_k=H^0\big(\CPP^1\,,\,\CO_{\CPP^1}(k)\big)$, and therefore
\beq
(\tilde f\circledast \tilde f'\,)(z,\overline{z};w,\overline{w}\,) =
\sum_{p,q=0}^k\, (\bfF\cdot \bfF'\,)_{pq} \
\psi_{p,q}(z,\overline{z};w,\overline{w}\,)
\eeq
with $(\bfF\cdot \bfF'\,)_{pq}= \sum_r\, \bfF_{pr}\, \bfF'_{rq}$. Hence
$\CC_r^*(\Sigma/\CP)\cong \sMat_{\FC}(k+1)$ as $C^*$-algebras.

Note that the ``diagonal'' subalgebra of $\CC_r^*(\Sigma/\CP)$
consists of functions on $\CPP^1$. The explicit quantization map can
then be given by identifying
the space of homogeneous polynomials of degree $k$ with the
$k$-particle Hilbert space in the Fock space of two harmonic
oscillators as
\beq
H^0\big(\CPP^1\,,\,\CO_{\CPP^1}(k)\big)\cong {\rm
  span}\big\{|p\rangle\ \big| \ p=0,1,\dots k\big\} \ , \qquad
|p\rangle:=\frac{\big(\hat a_0^\dag\,\big)^p\, \big(\hat
  a_1^\dag\,\big)^{k-p}}{\sqrt{p!\, (k-p)!}} |0\rangle
\eeq
where $|p\rangle$ are orthonormal number basis states in the Fock
space generated by creation and annihilation operators obeying $[\hat
a_\alpha,\hat a_\beta^\dag\,]=\delta_{\alpha\beta}$ and $\hat
a_\alpha|0\rangle=0$ for $\alpha,\beta=0,1$. The Berezin quantization
map of \S\ref{sec:GBB-Tquant} in this case is given
by~\cite{IuliuLazaroiu:2008pk}
\beq
B^{-1}\,:\, \CC_q^\infty(\CPP^1) \ \longrightarrow \
\CC_r^*(\Sigma/\CP) \ , \qquad
B^{-1}\big(\psi_{p,q}(z,\overline{z};z,\overline{z}\,) \big) =
|p\rangle \langle q| \ .
\eeq
The orthonormality relation (\ref{spqortho}) then follows from
\beq
\int_{\CPP^1}\, \omega_{\CPP^1}(z, \overline{z}\,) \
\psi_{p,q}(z, \overline{z}; z, \overline{z}\,) &=&
\tr_{\CCH_k}\big( B^{-1}(\psi_{p,q})\big)
\nonumber \\[4pt] &=& \sum_{r=0}^k\, \langle r|p\rangle\, \langle
q|r\rangle \ = \ \delta_{p, q} \ .
\eeq

This construction can be straightforwardly generalized to all complex
projective spaces $\CPP^n$ with $n\geq1$, and in fact to arbitrary
compact K\"ahler manifolds with integral symplectic two-forms. See~\cite{Bonechi:2010yh} for the symplectic
groupoid quantization of $S^2$ as a coset space, which yields the standard
Podle\`s quantum sphere.

\subsection{2-plectic geometry}

Let us now come to the quantization of $S^3$; as previously we closely
parallel the quantization of $S^2$. Following our strategy, the notion
of a prequantum line bundle is replaced by that of a prequantum
gerbe. To simplify the analysis involved, we then transgress the gerbe
to a line bundle over loop space and from this derive a Hilbert space.

Just as the line bundles over $S^2$ are characterized by their Chern
class $k\in H^2(S^2,\RZ)\cong\RZ$ up to isomorphism, the gerbes over
$S^3$ are characterized by their Dixmier-Douady class $k\in
H^3(S^3,\RZ)\cong\RZ$. The canonical 2-plectic form on $S^3$ is given
by its volume form which reads as
\begin{equation}
 \varpi_{S^3}=\mbox{$\frac{1}{2\pi^2}$}\, \sin^2\theta^1\, \sin
 \theta^2 \ \dd \theta^1\wedge \dd \theta^2\wedge \dd \theta^3
\end{equation}
in standard angular coordinates $0\leq \theta^{1},\theta^{2}\leq \pi$,
$0\leq \theta^3\leq 2\pi$, and the volume of the three-sphere with
respect to $\varpi_{S^3}$ is $\int_{S^3}\, \varpi_{S^3}=1$. This
volume form is determined by a contact structure on $S^3$
which is specified by a connection one-form $\kappa$ on the Hopf fibration $\pi:S^3\to S^2$ defined
via the pullback $\dd\kappa=\pi^*(\omega_{S^2})$. The global contact
form $\kappa$ determines a splitting $T^*S^3=E^*\oplus E^0$ of the
co-oriented contact distribution $E\to S^3$, a trivialization
$E^0=S^3\times\FR$, and an orientation on $S^3$ with volume form
$\varpi_{S^3}=\kappa\wedge\dd\kappa$. Moreover, it makes $S^3$ locally
isomorphic to the first jet bundle $T^*N\times\FR$ of real-valued functions on a
Legendrian submanifold $N$, i.e., a one-dimensional submanifold whose
tangent bundle lies in the hyperplane field $\ker\kappa$. The pushforward
$\pi_!:\frH^{1}(S^3, \varpi_{S^3})\to \CC^\infty(S^2)$ of Hamiltonian one-forms on $S^3$
by integration over the $S^1$-fibers of the Hopf fibration then yields
the Poisson algebra of functions on $S^2$. 

The prequantum gerbes $\CG_k$, $k\in \RZ$, have Dixmier-Douady class
$[k\, \varpi_{S^3}]$, curvature
$H=-2\pi^2\, \di\, k\, \varpi_{S^3}$ and topological charge
\begin{equation}
 dd(\CG_k) =\int_{S^3} \, \frac{\di \, H}{2\pi^2}=k~.
\end{equation}
We cover $S^3$ by the two patches $U_+$ and $U_-$, for which $\theta^1\neq \pi$ and $\theta^1\neq 0$ respectively. The connective structure with respect to these patches reads as
\begin{equation}
 \begin{aligned}
  B_+&=- \frac{\di\, k}{2}\, \left(\theta^1-\tfrac{1}{2}\,
    \sin(2\theta^1)\right)\, \sin\theta^2~\dd \theta^2\wedge \dd \theta^3~,\\[4pt]
  B_-&=- \frac{\di\, k}{2}\,
  \left(\theta^1-\tfrac{1}{2}\sin(2\theta^1)-\pi\right)\, \sin\theta^2~\dd \theta^2\wedge \dd \theta^3~,\\[4pt]
  A_{+-}&=\frac{\pi\, \di\, k}{2}\, \cos\theta^2 \ \dd \theta^3~.
 \end{aligned}
\end{equation}
Note that the intersection $U_{+-}=U_+\cap U_-$ is homotopic to $S^2$
and the gauge potential $A_{+-}$ coincides (up to an exact term) with
the standard gauge potential on the line bundle $E_k\to S^2$ from \S\ref{subsec:fuzzy-sympl}.

We now transgress the prequantum gerbe $\CG_k$ to a line bundle
$\CT\CG_k$ on knot space $\CK S^3$. As we want a local description in
terms of components of a connective structure, we have to introduce a
cover of $\CK S^3$. For this, we start from the cover $(U_a)$, $a\in
S^3$, with $U_a:=S^3\, \backslash\, \{a\}$ and consider the knot
spaces of the open sets $\CK U_a$. We thus cover $\CK S^3$ by the patches $\CU_a:=\CK U_a$. Working with angular coordinates has the advantage that different patches $U_a$ merely correspond to different ranges of the angles.

The gauge potentials of the prequantum line bundle on the patches
$\CU_a$ can be obtained from a transgression of the $B$-fields on the
patch $U_a$. For example, over $U_+$ we have
\begin{equation}
 \CA_+:=\CT B_+=- \frac{\di\, k}{2}\, \oint\, \dd\tau \ 
 \Big(\theta^{1\tau}-\frac{1}{2}\, \sin(2\theta^{1\tau})\Big)\, \sin\theta^{2\tau}~
 \big(\dot{\theta}^{2\tau}\, \delta
 \theta^{3\tau}-\dot{\theta}^{3\tau} \, \delta \theta^{2\tau} \big)~.
\end{equation}
The canonical complex structure $\CJ_{S^3}$ on $\CK S^3$ acts in local coordinates on the vector fields $\delder{\theta^{i\tau}}$ as
\begin{equation}
 \CJ_{S^3}\delder{\theta^{i\tau}}=\tilde{\eps}_{ijk}\,
 \frac{\thetad^{k\tau}}{\big|\thetad^\tau\big|}\, g^{jl}\, \delder{\theta^{l\tau}}~,
\end{equation}
where $g$ is the metric on $S^3$ defined by the line element
\begin{equation}
\dd s^2=(\dd \theta^1)^2+\sin^2 \theta^1\, \big((\dd
\theta^2)^2+\sin^2\theta^2\, (\dd \theta^3)^2 \big)~,
\end{equation}
while
\begin{equation}
 \tilde{\eps}_{ijk}:=\sqrt{g}\,
 \eps_{ijk}~,~~~\tilde{\eps}^{\,ijk}:=\frac{1}{\sqrt{g}}\, \eps_{ijk}
 \eand \sqrt{g}=\sin^2\theta^1\, \sin\theta^2~.
\end{equation}
One easily verifies that $\CJ_{S^3}^2=-\id$.

Let us now analyze the covariantly constant sections with respect to
the antiholomorphic part of the connection $\nabla^{0,1}$ on the line bundle $\CT\CG_k$ defined by
the gauge potentials $\CA_a$ on the patches $\CU_a$. Using
\eqref{eq:antiHolomProj}, we have for example on $\CU_+$ the formula
\begin{equation}
\nabla^{0,1}_{+}=\frac{1}{2}\, \oint\,  \dd
\tau~\delta\theta^{i\tau}\, \Big(\,
\delder{\theta^{i\tau}}+\di\,\tilde{\eps}_{ijk}\,
\frac{\thetad^{k\tau}}{\big|\thetad^\tau \big|}\, g^{jl}\,
\delder{\theta^{l\tau}}\, \Big) +\CA_+^{0,1} ~,
\end{equation}
and a corresponding formula involving $\CA_+^{1,0}$. From comparison with the
analysis of covariantly constant sections of the prequantum line
bundles $E_k$ on $S^2$, the interesting part of covariantly constant
sections $\psi_a$ of $\CT\CG_k$ over $\CU_a$ are the holomorphic
functionals $s_a$ satisfying $\overline{\delta}_a s_a =0$. The construction of these sections is surprisingly complicated, and we will use results of~\cite{0792.58008} for this. Instead of reviewing these constructions in generality, we present the concrete application to the case at hand.

Let us restrict to the patch $\CU_a \cong \CK \FR^3$. In Cartesian
coordinates $x^i$, $i=1,2 ,3$, the metric $g$ can be read off from the
line element
\begin{equation}\label{eq:metricS3}
 \dd s^2=\Big(\, \frac{2}{1+(x^1)^2+(x^2)^2+(x^3)^2}\, \Big)^2\,
 \big((\dd x^1)^2+(\dd x^2)^2+(\dd x^3)^2 \big)~.
\end{equation}
The Dolbeault operator on loop space is rewritten as
\begin{equation}
 \overline{\delta}_{a}=\frac{1}{2}\, \oint \, \dd \tau~\delta
 x^{i\tau}\, \Big(\, \delder{x^{i\tau}}+\di\,\tilde{\eps}_{ijk}\,
 \frac{\xd^{k\tau}}{|\xd^\tau|}\, g^{jl}\, \delder{x^{l\tau}}\, \Big)~.
\end{equation}
Plugging in \eqref{eq:metricS3}, the Dolbeault operator reduces to
that on the knot space $\CK\FR^3$ of Euclidean space $\FR^3$ given by
\begin{equation}\label{eq:DolbeaultS3}
 \overline{\delta}_{a}=\frac{1}{2}\, \oint \, \dd \tau~\delta
 x^{i\tau}\, \Big(\, \delder{x^{i\tau}}+\di\,\eps_{ijk}\,
 \frac{\xd^{k\tau}}{|\xd^\tau|}\, \delder{x^{j\tau}}\, \Big)~,
\end{equation}
where in this formula $|\xd^\tau|$ denotes the norm of the vector
$\xd^\tau$ with respect to the Euclidean metric on $\FR^3$. Constructing the holomorphic local sections $s_a$ thus amounts to finding holomorphic functionals on $\CK \FR^3$.

Recall that the tangent space at a loop $x\in\CK\FR^3$ consists of the two-dimensional orthogonal complements to the tangent vector to the loop at every point $x(\tau)$. To describe these complements, it is helpful to switch to a twistor formulation as in \cite{0792.58008}. This idea goes back to Drinfeld and LeBrun. The twistor description we will use was developed in full generality in \cite{LeBrun:1984-02-01Z:601}; see also \cite{Popov:2005uv} for a detailed discussion in a notation closely related to ours here.
The connection between loop spaces and twistor geometry arises because
oriented lines in a Riemannian three-manifold, like the tangent space to a loop at a point $x(\tau)$, are parametrized by LeBrun's twistor Cauchy-Riemann manifolds. At the heart of any twistor geometry is the \emph{double fibration}, which for $\FR^3$ reads as 
\begin{equation}\label{eq:DoubleFibration}
 \xymatrix{
 & F \ar[dl]_{\Pi_1} \ar[dr]^{\Pi_2} & \\
P &  & \FR^3
}
\end{equation}
The space $F=\FR^3\times \CPP^1$ is called the {\em correspondence
  space}, while $P$ is the {\em (mini)twistor space} which can be identified with the total space of the holomorphic line bundle $\CO_{\CPP^1}(2)$ or equivalently with the holomorphic tangent bundle over $\CPP^1$. The map $\Pi_1$ is given by 
\begin{equation}
 \Pi_1(x,\lambda)=x^{\ald\bed}\, \lambda_\ald\, \lambda_\bed~, 
\end{equation}
where $\lambda_\ald$, $\ald=1,2$, are homogeneous coordinates on $\CPP^1$ and 
\begin{equation}
 \big(x^{\ald\bed}\big) = \big( x^{\bed\ald}\big) =\left(\begin{array}{cc} 
 x^1+\di\, x^2 & -x^3 \\
 -x^3 & -x^1+\di \, x^2
\end{array}\right)~.
\end{equation}
The map $\Pi_2$ is the trivial projection from $F\cong \FR^3\times
\CPP^1$ to $\FR^3$. It is easy to see that this double fibration
establishes two correspondences: Between points in $\FR^3$ and
sections of $P\cong \CO_{\CPP^1}(2)$, and between points in $P$ and oriented lines in $\FR^3$.

The correspondence space $F$ is a {\em Cauchy-Riemann manifold}: it
comes with an integrable distribution $\CD\subset TF$ that is almost Lagrangian,
i.e., $[\CD,\CD]\subseteq\CD$ and $\CD\cap
\overline{\CD}=\varnothing$. As with many Cauchy-Riemann manifolds,
the distribution $\CD$ is induced by an embedding into a complex
manifold. Here we can embed $F$ into Penrose's twistor space $T\cong
\FR^4\times S^2$, which is biholomorphic to the total space of the
rank two holomorphic vector bundle $\CO_{\CPP^1}(1)\oplus\CO_{\CPP^1}(1)$, cf.~\cite{Popov:2005uv}. Explicitly, we have
\begin{equation}
 \CD={\rm span}\Big\{\, \lambda^\ald\, \lambda^\bed \,
 \mbox{$\der{x^{\ald\bed}}$}\, \Big\} \ \oplus \ T_\FC^{0,1}\CPP^1~.
\end{equation}
Using $\CD$, we can define the smooth sub-bundle $HF=TF\cap (\CD\oplus
\overline{\CD}\, )$. On this sub-bundle, there is an endomorphism
$J_F$ mapping ${\rm Re}(X) \in HF$ to $-{\rm Im}( X) \in HF$ for every
$X\in \overline{\CD}$; it satisfies $J^2_F=-\id$ and the bundle $HF\to
F$ has a canonical orientation.

At every point $x(\tau)$ of the loop, we can now write the tangent vector as
\begin{equation}
 \xd^{\ald\bed\tau}=\lambda^{\ald\tau}\,
 \lambdah^{\bed\tau}+\lambdah^{\ald\tau}\, \lambda^{\bed\tau}~,
\end{equation}
where $\lambda^{\ald\tau}:=\eps^{\ald\bed}\, \lambda_{\bed\tau}$ and
$\lambdah^{\ald\tau}:=\eps^{\ald\bed}\,
\overline{\lambda}\,^{\bed\tau}$. The ambiguity in the choice of
$\lambda^{\ald\tau}$ is a phase, which is factored out because we consider
them as homogeneous coordinates on $\CPP^1$. We thus obtain a map
$x(\tau)\mapsto \lambda^{\ald\tau}\in \CPP^1$, and we can lift a loop
$x\in\CK\FR^3$ to a loop $\tilde{x}=(x,\lambda)\in \CK_T F$; the loop
space $\CK_T F$ is the space of {\em transverse} loops in $F$
\cite{0792.58008}, i.e., the loops form equivalence classes of smooth
immersions $\tilde x:S^1\rightarrow F$ under reparametrization
transformations such that for every $\tau\in S^1$, $\tilde x(\tau)$ is
transverse to $H_{\tilde x(\tau)} F$. One can show that the tangent
space to $\CK_T F$ at a loop $\tilde{x}$ is isomorphic to
$\CC^\infty(S^1, \tilde{x}^*HF)$, and the endomorphism $J_F$ on $HF$ induces a complex structure $\CJ_F$ on $T\CK_T F$ at each point of the loop~\cite{0792.58008}.

The interesting point now is that the map $x\mapsto \tilde{x}$, which embeds $\CK \FR^3$ into $\CK_T F$, intertwines the complex structure $\CJ_F$ with the complex structure $\CJ_{\FR^3}$ on $\CK \FR^3$. The (complexified) orthogonal complement to $\xd^{\ald\bed\tau}$, which is the tangent space at $x$ restricted to the point $x(\tau)$, is spanned by the vectors
\begin{equation}
 v^{\ald\bed\tau}:=\lambda^{\ald\tau}\, \lambda^{\bed\tau}\eand
 \overline{v}\,^{\ald\bed\tau}:=\lambdah^{\ald\tau}\, \lambdah^{\bed\tau}~,
\end{equation}
because the Euclidean inner product reads explicitly as 
\begin{equation}
(v,\xd)=-\tfrac{1}{2}\,
\eps_{\ald\gad}\, \eps_{\bed\ded}\, v^{\ald\bed\tau}\, \xd^{\gad\ded\tau} 
\end{equation}
and $\lambda^{\ald\tau}\, \lambda_{\ald\tau} =\lambdah^{\ald\tau}\,
\lambdah_{\ald\tau}=0$. We rewrite the loop space Dolbeault operator as
\begin{equation}
 \overline{\delta}_{a}=\frac{1}{2}\, \oint\, \dd \tau~ \delta
 x^{\ald\bed\tau}\, \overline{\delta}_{\ald\bed\tau}
\end{equation}
and examine the vector fields
\begin{equation}
 \lambda^{\ald\tau}\, \lambda^{\bed\tau}\, \overline{\delta}_{\ald\bed\tau}~,~~~
 \lambda^{\ald\tau}\, \lambdah^{\bed\tau}\,
 \delb_{\ald\bed\tau}\eand\lambdah^{\ald\tau}\, \lambdah^{\bed\tau}\, \delb_{\ald\bed\tau}~.
\end{equation}
We find that $\lambda^{\ald\tau}\, \lambda^{\bed\tau}\,
\delb_{\ald\bed\tau}f=0$ is equivalent to $\lambda^{\ald\tau}\,
\lambda^{\bed\tau}\, \frac\delta{\delta x^{\ald\bed\tau}} f=0$ for any
$f\in \CC^\infty(\CK \FR^3)$. Moreover, we have $\lambdah^{\ald\tau}\,
\lambdah^{\bed\tau}\, \delb_{\ald\bed\tau}=0$, and
$\lambda^{\ald\tau}\lambdah^{\bed\tau}\delb_{\ald\bed\tau}$ contains
the vector field $\xd^{\ald\bed\tau}\, \frac\delta{\delta
  x^{\ald\bed\tau}}$ which generates loop reparameterizations and therefore vanishes. We thus saw the intertwining of the complex structures explicitly.

It follows that we can pullback any holomorphic function on $\CK_T F$
along the map $x\mapsto \tilde{x}$ to obtain a holomorphic function on $\CK\FR^3$. Lempert gives a nice way of constructing holomorphic functions on $\CK_T F$ in~\cite{0792.58008}: Take a $(1,0)$-form $\alpha$ on $T\cong \CO_{\CPP^1}(1)\oplus\CO_{\CPP^1}(1)\supset F$. Then the map
\begin{equation}
 y \ \longmapsto \ f(y)=(\CT\alpha)_y= \oint \, \dd\tau \ \iota_{\dot
   y}(ev^* \alpha)~,~~~y\in \CK_T F
\end{equation}
defines a holomorphic function $\CK_T F\rightarrow \FC$. This formula
can be generalized to expressions of multiple integrals, as products
of holomorphic functions are holomorphic. Lempert explains this
generalization, which follows in the spirit of Chen's iterated
integrals, and conjectures that the space of functions thus obtained is locally dense in the space of holomorphic functions on $\CK_T F$. 

However, it seems rather clear to us that the corresponding set of
holomorphic functions obtained via pullback is not even dense in the
space of holomorphic functions on $\CK \FR^3$ (although we have no
proof). Let us nevertheless push the analysis a little further. For
example, consider the $(1,0)$-form
\begin{equation}
 \alpha=\frac{x^{1\ald}\, \lambda_{\ald}
\,\dd\big(x^{2\bed}\, \lambda_{\bed}\big)}{(\lambda_1)^2}
\end{equation}
on $T$, yielding the function
\begin{equation}
 f(x)=\oint\, \dd\tau~\frac{x^{1\ald\tau}\, \lambda_{\ald\tau} \,
 \big(\xd^{2\bed\tau}\, \lambda_{\bed\tau}+x^{2\bed\tau}\,
 \dot{\lambda}_{\bed\tau}
 \big)}{(\lambda_{1\tau})^2} \ewith \xd^{\ald\bed\tau}=\lambda^{\ald\tau}\, \lambdah^{\bed\tau}~.
\end{equation}
If we restrict ourselves to loops with $\xd^{1\tau}=\xd^{2\tau}=0\neq
\xd^{3\tau}$ and $\lambda^1=\lambdah^2=0=\dot{\lambda}^\ald$, then this function restricts to
\begin{equation}
 f(x)=\big(x^1+\di\, x^2 \big)~\oint\, \dd \tau~ \,
 \lambdah^{1\tau} \, \lambda^{2\tau} = \big(x^1+\di\, x^2\big)\, V~,
\end{equation}
where $V$ is some constant volume factor (possibly requiring
regularization). This restriction can roughly be interpreted as a
reduction\footnote{Note that more appropriately, one should use the embedding of $\CPP^1$ into $\CK S^3$ which is induced by the preimage of the Hopf fibration $\pi:S^3\rightarrow \CPP^1$. A reduction of the function $f$ on $\CK S^3$ to $\CPP^1$ would then correspond to the pullback  of $f$ along the embedding map. However, the corresponding formulas for $(1,0)$-forms are complicated and resisted our attempts of an analytical treatment.} $\CK S^3\rightarrow S^2\cong \CPP^1$, and the function $f$
correspondingly restricts to the function $z_+:=x^1+\di\, x^2$ which is a building block for global holomorphic sections of the line bundle $\CO_{\CPP^1}(1)\rightarrow \CPP^1$. Powers of $f$ will correspondingly reduce to functions yielding sections of the bundles $\CO_{\CPP^1}(k)$.

We have thus far described the space of holomorphic sections of the
line bundles $\CT\CG_k$ that are transgressions of the prequantum gerbes
$\CG_k$ over $S^3$. From here, one can in principle follow the steps
in the quantization of $S^2$: We define the Hilbert space $\CCH$ to be
the (infinite-dimensional) space of holomorphic sections of $\CT\CG_k$
with some restrictive condition imposed. On this space, we define some
natural inner product, pair ``elementary sections'' with a suitable
set of creation and annihilation operators, and use the outer tensor
product $\CCH\otimes \CCH^*$ as a quantized algebra of functions on
$\CK S^3$. Altogether, we have shown that a quantization of $S^3$
using knot space is fruitful and exhibits some of the desired
features, such as the reduction to the geometric quantization of
$S^2$. The access to a detailed description, however, seems to be
obstructed by technical difficulties and, most importantly, the lack of suitable measures on knot space.

\section*{Acknowledgements}

Parts of this work were carried
out while the authors were visiting the Isaac Newton Institute for Mathematical Sciences in
Cambridge, UK, in
February/March 2012 under the auspices of
the Programme ``Mathematics and Applications of Branes in String and
M-Theory''; we would like to thank David Berman, Neil Lambert and
Sunil Mukhi for the invitation to participate and hospitality during
the programme. This work was supported in part by the Consolidated Grant
ST/J000310/1 from the UK Science and Technology Facilities
Council. The work of CS was supported in part by a Career Acceleration
Fellowship from the UK Engineering and Physical Sciences Research
Council. The work of RJS was supported in part by Grant RPG-404 from the Leverhulme Trust.

\appendices

\subsection{Lie groupoids and Lie algebroids}\label{app:A}

In this appendix we collect a number of definitions and examples
concerning groupoids as they appear for example in
\cite{Hawkins:0612363}. For a book on groupoids see \cite{0521499283},
and for a review of the use of groupoids in quantization see e.g.\ \cite{012185860X}.

\paragraph{Groupoids.} Recall that a group is a small
category\footnote{``Small'' means that both the objects and morphisms in the category form sets.} with one object, in which all morphisms are invertible. The single object corresponds to the unit in the group, while the morphisms correspond to the group elements. Removing the condition that there is a single object, we arrive at the notion of a groupoid: A {\em groupoid} is a small category in which every morphism is an isomorphism.

More explicitly, a groupoid consists of a base set $G_{(0)}=M$ and a set of arrows
$G_{(1)}=G$ between elements of $M$. Each arrow can be projected onto its source and target, yielding two maps $\sfs,\sft:G\rightarrow
M$. We write $G\rightrightarrows M$ and say that $G$ is a groupoid over $M$. Each arrow $g\in G$ has an inverse arrow $g^{-1}\in G$ in the opposite direction. Two arrows with matching head and tail can be concatenated leading to a
partial multiplication on $G\times G$: $(g,h)\mapsto \sfm(g,h)=:g\, h \in
G$. There is an inclusion map $\unit:M\rightarrow G$,
$x\mapsto\unit_x$ of an object $x$ to an arrow loop $x\rightarrow x$. The picture of arrows between elements of the base yields the following consistency relations between the maps $\sfs,\sft,\sfm,\unit$:
\begin{conditions}
 \item[(i)] The concatenation of arrows via the multiplication $\sfm$ is associative when defined;
 \item[(ii)] $\sfs(\sfm(g,h))=\sfs(g)$ and $\sft(\sfm(g,h))=\sft(h)$;
 \item[(iii)] $\sfs(\unit_x)=\sft(\unit_x)=x$;
 \item[(iv)] $g\, \unit_{\sft(g)}=\unit_{\sfs(g)}\, g=g$;
 \item[(v)] The inverse $g^{-1}$ of $g$ is two-sided:
   $\sfs(g^{-1})=\sft(g)$, $\sft(g^{-1})=\sfs(g)$, $g^{-1}\,
   g=\unit_{\sft(g)}$, $g\, g^{-1}=\unit_{\sfs(g)}$;
\end{conditions}
where $g,h\in G$, $x\in M$.

\paragraph{Lie groupoids.} A {\em Lie groupoid} is a groupoid where both sets of objects and
morphisms are smooth manifolds, the source and target maps are
surjective submersions, and all operations are smooth. In this paper, all groupoids are Lie groupoids.

Note that every Lie group $G$ is a Lie groupoid $G\rightrightarrows
pt$ over a one-point manifold $pt$. The source and target maps are
both $\sfs(g)=\sft(g)=pt$ and the unit map is given by the group identity
element $\unit_{pt}=e$.

A non-trivial example of a Lie groupoid important for our discussion is the {\em pair
groupoid} $\Pair(M)=M\times M$ over a smooth manifold $M$. Here an
element $(x,y)$ of $M\times M$ corresponds to an arrow with source $x$
and target $y$: $\sfs(x,y)=x$ and $\sft(x,y)=y$. This interpretation
implies that the inverse is given by $(x,y)^{-1}=(y,x)$, that the multiplication law is $(x,y)\cdot (y,z)=(x,z)$
for $x,y,z\in M$ and that the unit map is $\unit_x=(x,x)$.

\paragraph{Lie algebroids.} A {\em Lie algebroid} is a smooth vector bundle $E\rightarrow M$ over a
manifold $M$ endowed with a Lie bracket on the
$\CC^\infty(M)$-module of smooth
sections $[-,-]_E: \CC^\infty(M,E)\otimes \CC^\infty(M,E)\to \CC^\infty(M,E)$ and a vector bundle map $\rho:E\rightarrow TM$,
$\psi\mapsto \rho_\psi$
called an {\em anchor map} such that:
\begin{conditions}
 \item[(i)] The map $\rho$ is a Lie algebra homomorphism:
\beq
\rho_{[\psi_1,{\psi_2}]_E} =[\rho_{\psi_1},\rho_{\psi_2}]_{TM} \ , \qquad
   {\psi_1},{\psi_2}\in\CC^\infty(M,E) \ .
\eeq
 \item[(ii)] The Leibniz rule is satisfied:
\beq
[{\psi_1},f\, {\psi_2}]_E=f\,[{\psi_1},{\psi_2}]_E+\rho_{\psi_1}(f)\, {\psi_2}~,~~~{\psi_1},{\psi_2}\in\CC^\infty(M,E)~,~~~f\in \CC^\infty(M) \ .
\eeq
\end{conditions}

The tangent bundle $E=TM$ is trivially a Lie algebroid, with the usual commutator
Lie bracket on vector fields over $M$ and the identity anchor
map. Note also that every Lie algebra $\frg$ is a Lie algebroid $\frg\rightarrow
pt$ over a one-point manifold $pt$ with trivial anchor map $\rho =0$.

We are mostly interested in the Lie algebroid arising from the cotangent bundle $T^*M$ of a Poisson
manifold $(M,\pi)$, where the anchor map $\rho:T^*M\rightarrow TM$ is
given by contraction with the Poisson bivector field $\pi$, see \S\ref{sec:PMandLieAlgebroids}.

\paragraph{Differentiation.} Just as a Lie algebra is the linearization of a Lie group at the unit element, a Lie algebroid is obtained by linearizing a Lie groupoid at the units, i.e.\ the objects. The Lie algebroid $A(G)\to M$ of a Lie groupoid $G\rightrightarrows M$ is given by setting:
\begin{conditions}
\item[1.] The vector bundle
\beq
A(G)=\ker \dd\sft\big|_M =\unit^*\ker \dd\sft = \bigsqcup_{x\in M}\,
T_{\unit_x}\sft^{-1}(x) \ \subset \ TG
\eeq
 is the normal
  bundle to the embedding $\unit:M\hookrightarrow G$, with bundle
  projection given by $\sfs$ or $\sft$ (which coincide on $M$).
\item[2.] The anchor map is given by $\rho=\dd\sfs:\ker \dd\sft\to
  TM$, restricted to $A(G)$.
\item[3.] The Lie bracket $[-,-]_{A(G)}$ is inherited from the usual
  commutator on $\CC^\infty(G,TG)$, restricted to left-invariant
  vector fields.
\end{conditions}
This construction defines a functor from the category of Lie groupoids
(and smooth homomorphisms)
to the category of Lie algebroids, which generalizes the classical Lie
functor from Lie groups to Lie algebras: Let $\frg$ be the Lie algebra of a Lie group $G$.
Then $\frg$ is the Lie algebroid of the Lie groupoid $G$.

The tangent bundle $TM$ is the Lie algebroid of the pair groupoid
$\Pair(M)= M\times M$, as $A(M\times M)=\bigsqcup_{x\in M}\, x\times T_xM=TM$.

\paragraph{Integration.} We say that a Lie groupoid $G\rightrightarrows M$ is an
\emph{integration} of a Lie algebroid $E\to M$ if $A(G)\cong E$ as Lie
algebroids. While every Lie algebra can be integrated to a Lie group by the exponential map,
the same is not true in general for Lie groupoids. For a discussion of the corresponding obstructions, see e.g.\
\cite{Crainic:0105033}. However, if a Lie algebroid $E\to M$ does integrate to a Lie
groupoid, then there exists a unique Lie groupoid
$G\rightrightarrows M$ (up to isomorphism) with connected and simply
connected $\sfs$-fibers whose Lie algebroid is
$A(G)=E$. See~\cite[\S2]{Landsman:2000aa} for the construction of an
exponential map $A(G)\to G$ for a general Lie groupoid $G$ whose Lie
algebroid $A(G)$ is endowed with a connection; this generalizes the
exponential map on the tangent bundle $\exp:TM\to M$ of a manifold $M$
with connection.

\paragraph{Symplectic groupoids.} A Lie groupoid $G\rightrightarrows M$ naturally has the structure of a simplicial manifold with 0-nerve
$M$ and 1-nerve $G$. We define the {\em 2-nerve} of a Lie groupoid $G$ as the set of
composable arrows $G_{(2)}\subset G\times G$; whence
$(g_1,g_2)\in G_{(2)}$ if and only if $\sft(g_1)=\sfs(g_2)$. On this set we have the
multiplication map $\sfm:G_{(2)}\rightarrow G$ satisfying
$\sfs(\sfm(g_1,g_2))=\sfs(g_1)$ and $\sft(\sfm(g_1,g_2))=\sft(g_2)$ for $g_1,g_2\in
G$. There are also projections $\pr_1:G_{(2)}\to G$ and $\pr_2:G_{(2)}\to G$
onto the first and second arrow, respectively: $\pr_1(g_1,g_2)=g_1$
and $\pr_2(g_1,g_2)=g_2$. The maps $\sfs$, $\sft$, $\sfm$, $\pr_1$, and
$\pr_2$ are face maps for this simplicial structure. This leads to the definition of the
simplicial coboundary operator
\beq
\dpar^*:=\pr_1^*-\sfm^*+\pr_2^* \, : \,
\Omega^k(G) \ \longrightarrow \ \Omega^k(G_{(2)})~,
\eeq
which commutes with the exterior derivative.

A {\em symplectic groupoid} $\Sigma$ is a Lie groupoid endowed with a
symplectic form $\omega\in \Omega^2(\Sigma)$ which is \emph{multiplicative}, i.e.,\
$\dpar^*\omega:=\pr_1^*\omega-\sfm^*\omega+\pr_2^*\omega=0$. For a detailed discussion of multiplicative two-forms, see e.g.\ \cite{Bursztyn:0303180}. We say that
a symplectic groupoid $\Sigma$ over a Poisson manifold $M$
\emph{integrates} $M$ if $\sft:\Sigma\rightarrow M$ is a Poisson
map\footnote{This means that $(\Sigma,\sft)$ forms a symplectic
  realization of the Poisson manifold $M$.} and the fibers of
$\sfs:\Sigma\rightarrow M$ are connected ($\Sigma$ is
``$\sfs$-connected''); in fact, if $\Sigma\rightrightarrows M$ is a
symplectic groupoid then its base $M$ is always a Poisson manifold. When it exists, an integrating symplectic groupoid for a Poisson
manifold $M$ is unique up to isomorphism.

It is shown in~\cite{Crainic:2002aa} that a Poisson manifold is
integrable in this sense if and only if the Lie algebroid $T^*M$ is
integrable to a Lie groupoid; in this case there is a canonical
isomorphism $A(\Sigma)\cong T^*M$ of Lie algebroids.

Note that on a symplectic groupoid $\Sigma$, there is a unique Poisson
structure on the base such that $\sft$ and $\sfs$ are respectively Poisson and anti-Poisson maps \cite{Weinstein:1991ab}.

\paragraph{Example.}
The cotangent bundle $T^*G$ of a Lie group $G$ can be extended to a symplectic
groupoid over $\frg^*$ where $\frg$ is the Lie algebra of $G$~\cite{Coste:1987aa}, see~\cite{3764350164} for a review. The
embedding $\unit$ is trivially given by $\unit:\frg^*\rightarrow T^*_e
G$. Recall
that the left and right $G$-actions on $G$, $L_h(g):=h\, g$ and
$R_h(g):=g\, h$ with $g,h \in G$, have derivative maps $\dd L_h,\dd
R_h:\frg\rightarrow T_h G$. Given an element $(x,h)\in T^*G$ with $h\in G$ and $x\in T^*_hG\cong \frg^*$, we define the source and target maps
\begin{equation}
 \sfs(x,h):=x \circ \dd R_h \eand \sft(x,h):=x\circ \dd L_h~,
\end{equation}
which live in the dual space of $\frg$. The product
$(x,h):=(x_1,h_1)\cdot(x_2,h_2)$ of two elements of $T^*G$ is defined
if $\sft(x_1,h_1)=\sfs(x_2,h_2)$, i.e.,\ $x_1\circ \dd L_{h_1}=
x_2\circ \dd R_{h_2}$. We then put $(x,h):=x_1\circ \dd
R_{h_2^{-1}}=x_2\circ \dd L_{h_1^{-1}}$, and one has
$\sfs(x,h)=\sfs(x_1,h_1)$ and $\sft(x,h)=\sft(x_2,h_2)$. Moreover, the
target map $\sft$ is a Poisson map with respect to the canonical
symplectic Poisson structure on $T^*G$ and the $+$-KKS Poisson structure on $\frg^*$.

\paragraph{Fibrations of groupoids.} A {\em fibration of Lie groupoids} is a smooth homomorphism
$\sfp:G\rightarrow G'$ between Lie groupoids $G$ and $G'$, such that the base
map $\sfp_{(0)}:G_{(0)} \rightarrow G'_{(0)}$ and the map $F^*:G\rightarrow \sfp_{(0)}^*
G'$ are surjective submersions; here $\sfp_{(0)}^* G'$ denotes the pullback bundle of $G'$ along $\sfp_{(0)}$.

\subsection{$p$-gerbes with connective structures}\label{app:B}

Below we summarize some definitions concerning gerbes with
connective structure. In this paper we use both the concrete description of a
gerbe provided by \v{C}ech cohomology, following the approach of
Hitchin and Chatterjee (see~\cite[\S1.2]{Hitchin:2005uu}
and~\cite{Chatterjee:1998}), and also
Murray's description in terms of bundle
gerbes~\cite{Murray:9407015,Murray:9908135}. See~\cite{Murray:2007ps} for
a comprehensive review with further details and additional references. Smooth functions, line bundles and gerbes are the first three elements in a sequence of $p$-gerbes with $p=-1,0,1$. As it is effortless to generalize most of the notions introduced below to $p$-gerbes, we will do so immediately.

We start by reviewing Cheeger-Simons differential characters and how
they capture a generalized notion of holonomy for $p$-gerbes. We then
come to Deligne cohomology, which gives a cochain model for
Cheeger-Simons differential characters and classifies abelian
$p$-gerbes with connective structure. Finally, we look at the
description for $p=1$ in terms of bundle gerbes and its relation to
differential cohomology, with the $d$-torus serving as an illustrative
example which makes contact with some of our loop space constructions
from~\S\ref{subsec:torus-2pl}.

\paragraph{Cheeger-Simons cohomology.} Let $M$ be a smooth manifold. Let $Z_k(M)$ denote the group of smooth $k$-cycles on $M$, and
$\clidf^k(M)$ the closed $k$-forms on $M$ with integer periods.
Recall~\cite{Alexander:1985aa,0817647309} that the degree $k$
{Cheeger-Simons cohomology} group of differential
characters $\hat{H}^k(M)$ is the infinite-dimensional abelian group
which can described in terms of two exact sequences. Firstly, one has
\beq
0 \ \longrightarrow \ H^{k-1}\big(M\,,\,\sU(1) \big) \ \longrightarrow \ \hat
H^k(M)\ \xrightarrow{ \ \omega \ } \ \clidf^k(M) \ \longrightarrow \ 0 \ ,
\eeq
where the field strength map $\chi\mapsto
\omega_\chi$ defines the
\emph{curvature} of the differential character $\chi$; its kernel is the
group of flat fields on $M$. Secondly, one has
\beq
0 \ \longrightarrow \ \Omega^{k-1}(M)\, \big/\, \clidf^{k-1}(M) \
\longrightarrow \ \hat H^k(M)\ \xrightarrow{ \ c \ } \
H^k(M,\RZ) \ \longrightarrow 0 \ ,
\eeq
where the map $\chi\mapsto c(\chi)$ defines the \emph{characteristic
  class} of the differential character which obeys the compatibility
condition $c(\chi) =[\omega_\chi]$; the kernel
of this map is the torus of topologically trivial fields
whose classes $[\theta]$ have curvature $\dd\theta$.

A differential character $\chi\in \hat H^k(M)$ defines a \emph{holonomy}
\beq
\hol_S(\chi):= \exp\Big(2\pi\, \di\, \oint_S\, \theta_\chi
\Big) \ \in \ \sU(1)
\eeq
for any $k-1$-cycle $S\in
Z_{k-1}(M)$, where the potential $\theta_\chi\in\Omega^{k-1}(S)$ is defined by
$\omega_{\chi|_S}=\dd\theta_\chi$ and we have used $H^k(S,\RZ)=0$. For
flat fields $\omega_\chi=0$, the holonomy defines an element
$\big[\hol(\chi)\big] \in H^{k-1}(M,\sU(1))$.

The Cheeger-Simons cohomology $\hat H^1(M)$ is the space of differentiable
maps $g:M\to\sU(1)$; the characteristic class map is $c(g)=g^*([\dd\phi])\in
H^1(M,\RZ)$ where $[\dd\phi]$ is the fundamental class of $S^1$, the
curvature is the one-form $\omega_g=\dd\log g$, and the holonomy is the
evaluation $\hol_x(g)=g(x)$ of $g$ at $x\in M$.

The differential cohomology $\hat H^2(M)$ is the group of gauge equivalence
classes of line bundles with connection $(E,\nabla)$ on $M$ and gauge
group generated by $\hat H^1(M)$; the
characteristic class map in this case computes the first Chern class
$c_1(L)\in H^2(M,\RZ)$, while the connection $\nabla$ determines a curvature
$F_\nabla$ and a holonomy $\hol_\nabla(\gamma)$ for $\gamma\in Z_1(M)$
which coincide with
the curvature and holonomy of the corresponding differential
character.

The abelian group $\hat H^3(M)$ consists of
differential characters in degree three, which are gauge equivalence
classes of \emph{gerbes $\CG$ over $M$ with connective structure $(A,B)$}
and gauge group generated by the
differential cohomology $\hat H^2(M)$.

Following this pattern, the gauge equivalence classes of {\em
  $p$-gerbes with connective structure} over $M$ are given by the
differential cohomology $\hat H^{p+2}(M)$ with gauge group generated by $\hat H^{p+1}(M)$.

\paragraph{Deligne cohomology.} An explicit cochain model for the Cheeger-Simons groups
is provided by Deligne cohomology. Recall~\cite{0817647309} that the
degree $k$ smooth Deligne cohomology is the $k$-th \v{C}ech
hypercohomology of the truncated sheaf complex $\CD(k)$
\beq
0\ \longrightarrow \ \sU(1)_M \ \xrightarrow{\dd\log} \ \Omega_M^1 \
\xrightarrow{ \ \dd \ } \ \Omega_M^2 \
\xrightarrow{ \ \dd \ } \ \cdots \ \xrightarrow{ \ \dd \ } \
\Omega_M^k \ ,
\eeq
where $\sU(1)_M$ is the sheaf of smooth $\sU(1)$-valued functions on
$M$ and $\Omega_M^k$ is the sheaf of differential $k$-forms on
$M$. Consider thus the double complex with respect to a Stein cover
$U=(U_a)_{a\in I}$ of $M$ given by
\beq
\vcenter{\vbox{\xymatrix{
\vdots & \vdots & & \vdots \\
\CC^2\big(U\,,\,\sU(1)_M\big) \ \ar[u]^\delta \ \ar[r]^{\ \dd\log} & \
\CC^2\big(U\,,\, \Omega_M^1\big) \ \ar[u]^\delta \ar[r]^{ \ \  \ \  \ \dd }  & \
\cdots \ \ar[r]^{\!\!\!\!\!\!\!\!\!\!\!\!\dd } & \
\CC^2\big(U\,,\, \Omega_M^k\big) \ \ar[u]^\delta \\
\CC^1\big(U\,,\,\sU(1)_M\big) \ \ar[u]^\delta \ \ar[r]^{\ \dd\log} & \
\CC^1\big(U\,,\, \Omega_M^1\big) \ \ar[u]^\delta \ar[r]^{  \ \  \ \ \ \dd } & \
\cdots \ \ar[r]^{\!\!\!\!\!\!\!\!\!\!\!\!\dd } & \
\CC^1\big(U\,,\, \Omega_M^k\big) \ \ar[u]^\delta \\
\CC^0\big(U\,,\,\sU(1)_M\big) \ \ar[u]^\delta \ \ar[r]^{\ \dd\log} & \
\CC^0\big(U\,,\, \Omega_M^1\big) \ \ar[u]^\delta \ar[r]^{  \ \  \ \ \ \dd } & \
\cdots \ \ar[r]^{\!\!\!\!\!\!\!\!\!\!\!\!\dd } & \
\CC^0\big(U\,,\, \Omega_M^k\big) \ \ar[u]^\delta
}}}
\eeq
Here $\delta$ is the usual \v{C}ech coboundary operator, and
$\CC^p(U,\sU(1)_M)$ and $\CC^p(U,\Omega^n_M)$ denote the
\v{C}ech $p$-cochains. The degree $n$ Deligne cohomology group $H_{\mathfrak{D}}^n(M,\CD(k))$ is computed from the `diagonal complex'
\begin{equation}
\begin{aligned}
 \CC^0\big(U\,,\,\sU(1)_M\big)~\xrightarrow{~~\mathbf{d}~~}~&~\CC^1\big(U\,,\,\sU(1)_M\big)\oplus\CC^0\big(U\,,\, \Omega_M^1\big)\\&
~\xrightarrow{~~\mathbf{d}~~}~\CC^2\big(U\,,\,\sU(1)_M\big)\oplus\CC^1\big(U\,,\, \Omega_M^1\big)\oplus\CC^0\big(U\,,\, \Omega_M^2\big)~\xrightarrow{~~\mathbf{d}~~}~\cdots~
\end{aligned}
\end{equation}
truncated to sheaves in $\CD(k)$. The differentials $\mathbf{d}$ are defined as
\begin{equation}
 \mathbf{d}\theta:=\left\{
\begin{array}{ll}
\delta \theta+(-1)^p~\frac1{2\pi\,\di}~\dd \log \theta~,&\theta\in \CC^p\big(U\,,\,\sU(1)_M\big)~,~k>0\\
0~,&\theta\in \CC^p\big(U\,,\,\sU(1)_M\big)~,~k=0\\
\delta \theta+(-1)^p~\dd \theta~,&\theta\in \CC^p\big(U\,,\,
\Omega_M^n \big)~,~~n<k~,\\
0~,&\theta\in \CC^p\big(U\,,\,
\Omega_M^n \big)~,~~n\geq k~.
\end{array}\right.
\end{equation}
where we inserted a convenient factor $\frac1{2\pi\,\di}$ into the logarithmic differential. Note that the \v{C}ech coboundary operator acting on \v{C}ech cochains with values in $\sU(1)_M$ has to be regarded as the multiplicative one.

\paragraph{Examples.} Below we denote contractible $n$-fold intersections of open sets of the cover
$U$ by $U_{a_1\dots a_n}:= U_{a_1}\cap\cdots\cap U_{a_n}$.

A degree~$0$ smooth Deligne class $g\in H_{\mathfrak{D}}^0(M,\CD(0))$ is a \v{C}ech cochain $(g_a)\in\CC^0(U,\sU(1)_M)$ satisfying $\mathbf{d} (g_a)=1$ or, equivalently, $\delta (g_a)=g^{-1}_ag_b=1$. Therefore $(g_a)$ defines a smooth map $g:M\to\sU(1)$.

A Deligne class $(g_a,A_{ab})\in H_{\mathfrak{D}}^1(M,\CD(1))$ is a pair
\beq
(g_a,A_{ab}) \ \in \
\CC^1\big(U\,,\,\sU(1)_M\big)\oplus \CC^0\big(U\,,\,
\Omega_M^1\big)
\eeq
satisfying the cocycle conditions $\mathbf{d}(g_a,A_{ab})=(1,0)$, which explicitly reads as
\beq
g_{ab}\, g_{bc}\, g_{ca}&=& 1 \eon U_{abc} \ , \nonumber \\[4pt]
A_a-A_b &=& \frac1{2\pi\,\di}\, \dd \log g_{ab} \eon U_{ab} \ .
\eeq
The equivalence relations, including gauge transformations, are captured by a Deligne one-coboundary $\mathbf{d}(h_a)$ with $(h_a)\in \CC^0(U,\sU(1)_M)$. In formulas, $(g_a,A_{ab})\sim(g_a,A_{ab})+\mathbf{d}(h_a)$ reads as 
\begin{equation}
\begin{aligned}
 g_{ab}&\sim g_{ab}\, (\delta h)_{ab}=g_{ab}\, h^{-1}_a\,
 h_b=h^{-1}_a\, g_{ab}\, h_b~,\\[4pt]
 A_{a}&\sim A_{a}+\mbox{$\frac1{2\pi\,\di}$} \,
\dd\log h_a~.
\end{aligned}
\end{equation}
The $\sU(1)$ \v{C}ech one-cocycle $g_{ab}:U_{ab}\to
\sU(1)$ determines smooth transition functions on overlaps for a hermitian
line bundle $E\to M$. This cocycle represents the first Chern class $c_1(L)=
[g_{ab}]\in H^1(\Sigma,\sU(1))\cong H^2(\Sigma,\RZ)$, where
the canonical isomorphism follows from the exponential sequence
\beq
0 \ \longrightarrow \ \RZ \ \longrightarrow \ \FR \ \xrightarrow{\exp}
\ \sU(1) \ \longrightarrow \ 1 \ ;
\eeq
it is
the obstruction to triviality of the line bundle $E\to M$. The local one-forms
$A_a\in\Omega^1(U_a)$ define a unitary connection $\nabla=\dd+A$ on $E$. When restricted to $U_a$, its curvature $F$ is given by $\dd A_a$.

A Deligne class $(g_{abc},A_{ab},B_a)\in H_{\mathfrak{D}}^2(M,\CD(2))$ is a triple
\beq
(g_{abc},A_{ab},B_a) \ \in \
\CC^2\big(U\,,\, U(1)_M\big)\oplus
\CC^1\big(U\,,\,\Omega_M^1\big)\oplus
\CC^0\big(U\,,\,\Omega^2_M\big)
\eeq
satisfying the cocycle conditions $\mathbf{d}(g_{abc},A_{ab},B_a)=(1,0,0)$. That is,
\beq
g_{abc}\,
g^{-1}_{bcd}\, g_{cda}\,
g^{-1}_{dab} &=& 1 \eon
U_{abcd} \  , \nonumber \\[4pt]
A_{ab}+A_{bc}+ A_{ca} &=& \frac1{2\pi\, \di}\,
\dd \log g_{abc} \eon
U_{abc} \ , \nonumber \\[4pt]
B_a-B_b &=& \dd A_{ab} \eon
U_{ab} \ .
\label{2cocycleconds}\eeq
A Deligne two-coboundary $\mathbf{d}(h_{ab},a_a)$ with $(h_{ab},a_a)\in \CC^1(U,\sU(1)_M) \oplus \CC^0(U,
\Omega_M^1)$ defines a gauge transformation $(g_{abc},A_{ab},B_a)\sim(g_{abc},A_{ab},B_a)+\mathbf{d}(h_{ab},a_a)$, or
\begin{equation}\label{eq:GaugeTrafoGerbe}
 \begin{aligned}
  g_{abc} &\sim g_{abc}\, (\delta h)_{abc}=g_{abc}\,h_{ab}\, h_{bc}\, h_{ca}~,\\[4pt]
  A_{ab} &\sim A_{ab}+\mbox{$\frac1{2\pi\, \di}$}\, \dd\log
h_{ab}+a_a-a_b~,\\[4pt]
  B_{a} &\sim \dd a_b~.
 \end{aligned}
\end{equation}
The $\sU(1)$ \v{C}ech two-cocycle $g_{abc}:U_{abc}\to \sU(1)$ specifies a hermitian
``transition'' line bundle $E_{ab}$ over each overlap
$U_{ab}$, an isomorphism
$E_{ab}\cong E^*_{ba}$, and a trivialization
of the line bundle
$E_{ab}\otimes E_{bc}\otimes E_{ca}$ on
each triple overlap $U_{abc}$; the pair
$\CG=(E_{ab},g_{abc})$ is called a \emph{local gerbe}
on $M$. This cocycle
represents the Dixmier-Douady class $dd(\CG)= [g_{abc}]\in
H^2(M,\sU(1))\cong H^3(M,\RZ)$; it is the obstruction to triviality of the gerbe.
The \v{C}ech one-cochain $A_{ab}$ defines connection one-forms on each line bundle
$E_{ab}\to U_{ab}$ such that the section
$g_{abc}$ is covariantly constant with respect to the
induced connection on $E_{ab}\otimes E_{bc}\otimes
E_{ca}$. The collection of two-forms $B_a\in\Omega^2(U_a)$ are called the {\em curving} of the \v{C}ech one-cochain $A_{ab}$. Together, the pair $(A_{ab},B_a)$ defines a \emph{connective structure} on the gerbe $\CG=(E_{ab},g_{abc})$. When restricted to $U_a$, its curvature is given by $H=\dd B_a$.
The \emph{gauge group} of the gerbe is generated by line bundles
$\eta \to M$ with connection $\nabla=\dd+a$ and curvature
$f_\nabla=\dd a$ with $\omega=-\frac1{2\pi\,
  \di}\,f_\nabla\in \clidf^2(M)$ through the gauge transformations
\beq
E \ \longmapsto \ E_{ab}\otimes \eta \big|_{U_{ab}} \ ,
  \qquad A_{ab} \ \longmapsto \ A_{ab}+
  a\big|_{U_{ab}} \eand B_a \
  \longmapsto \ B_a+f_\nabla \ .
\eeq

In general, a Deligne $p+1$-cocycle defines the connective structure
of a $p$-gerbe. The details of the construction are evident from the above examples.

\paragraph{Holonomy and curvature.} The construction of holonomy and curvature of a Deligne class defines
an isomorphism $H_{\mathfrak{D}}^{p+1}(M)\to \hat H^{p+2}(M)$.

Given a degree~$1$ smooth Deligne class $[(g_{ab},A_a)]$
represented by a hermitian line bundle $E\to M$ with unitary
connection $\nabla=\dd+A$, the
curvature is the globally defined two-form given by $F_\nabla=\dd
A_a$ on $U_a$ with $\omega=-\frac1{2\pi\,
  \di}\,F_\nabla\in \clidf^2(M)$.
By Stokes' theorem, the holonomy of $\nabla$ around any one-cycle
$\gamma\subset M$ is then obtained from the product formula~\cite{Gawedzki:1987ak,Kiyonori:2001aa}
\beq
\hol_\gamma(A) =\prod_{a\in I} \, \exp\Big(2\pi\, \di\,
\int_{\gamma_a}\, A_a\Big) \ \prod_{a,b\in I}\,
g_{ab}(\gamma_{ab}) \ ,
\eeq
where $\gamma_a\subset U_a$ is a path in a subdivision of
the loop $\gamma$ into segments and $\gamma_{ab}=\gamma_a\cap
\gamma_b$ is a point in $U_{ab}$. This
definition agrees with the general definition of holonomy in terms of
differential characters.

Given a degree~$2$ smooth Deligne class
$[(g_{abc},A_{ab},B_a)]$ represented by a
gerbe $\CG$ over $M$ with connective structure $(A,B)$, the curvature
of the corresponding differential character is the globally defined
closed three-form $H=H_{(A,B)}$ given by $H=\dd B_a$ on
$U_a$ with $\varpi=-\frac1{2\pi^2\,\di}\, H\in \clidf^3(M)$,
while its characteristic class is the
Dixmier-Douady class $dd(\CG) \in H^3(M,\RZ)$ of the gerbe $\CG$.
Its holonomy around a two-cycle $S\subset M$ is obtained
by choosing a triangulation $\{S_a\}_{a\in I}$ of $S$ subordinate to the open cover
$S\cap U$. Keeping careful track of orientations, by repeated
application of Stokes' theorem one arrives at the product
formula~\cite{Gawedzki:1987ak,0817647309,Kapustin:1999di,Kiyonori:2001aa}
\beq
\hol_S(B)=\prod_{a\in I}\, \exp\Big(2\pi\, \di\,\int_{S_a}\,
B_a\Big) \ \prod_{a,b\in I}\,
\exp\Big(2\pi\,\di\,\int_{S_{ab}}\, A_{ab}\Big) \
\prod_{a,b,c\in I}\,
g_{abc}(S_{abc})~,
\eeq
where $S_{ab}$ is the common boundary edge of the surfaces $S_a$ and
$S_b$, and $S_{abc}=S_{ab}\cap
S_{bc}\cap S_{ca}$ are vertices of the
triangulation of $S$. The coincidence between this expression and the
general formula in terms of differential characters is shown explicitly in~\cite{Carey:2002xp}.

\paragraph{Bundle gerbes.} A \emph{bundle gerbe} on $M$ \cite{Murray:9407015} is a pair $\CG= (E,X)$, where $\phi:X\to M$ is
a surjective submersion and $E$ is a hermitian line bundle\footnote{In the case of a bundle $p$-gerbe, $E$ would be a bundle $p-1$-gerbe. For simplicity, we restrict ourselves here to bundle (1-)gerbes.} over the
fiber product $X^{[2]}$ equipped with an associative fiber multiplication
\beq
E_{(x_1,x_2)}\otimes E_{(x_2,x_3)} \ \longrightarrow \
E_{(x_1,x_3)}
\eeq
for all $(x_1,x_2), (x_2,x_3)\in X^{[2]}$. Recall that the fiber product $X^{[2]}:=
X\times_M X=\{(x_1,x_2)\in X\times X\ |\ \phi(x_1)=\phi(x_2)\}$ is a
pair groupoid $X^{[2]}\rightrightarrows X$ with base $X$, and source and target maps
$\pr_1:(x_1,x_2)\mapsto x_1$ and $\pr_2:(x_1,x_2)\mapsto x_2$,
respectively. Note that the map $\phi$ admits local sections, and hence
defines a ``quasi-cover'' of $M$. For each $k$ there is a linear map
\beq
\delta:=\pr_1^*-\pr_2^*\, :\, \Omega^k(X) \ \longrightarrow \
\Omega^k\big(X^{[2]}\big)
\eeq
which commutes with the exterior derivative. The kernel of $\delta$ is the image of the injection
$\phi^*:\Omega^k(M)\hookrightarrow \Omega^k(X)$, while its image is the
subspace of forms in $\Omega^k(X^{[2]})$ which commute with the
multiplication in the groupoid $X^{[2]}\rightrightarrows X$.
Due to the bundle gerbe product,
the bundle $E$ yields a $\sU(1)$-groupoid extension of $X^{[2]}$ and
there are isomorphisms $E_{(x,x)}\cong \FC$ and $E_{(x_1,x_2)}\cong
E^*_{(x_2,x_1)}$.

A \emph{connective structure} $(\nabla,B)$ on a bundle gerbe
\beq
\xymatrix@C=10mm{
E \ar[d] & \\
X^{[2]} \ \ar@< 2pt>[r]^{\pr_1} \ar@< -2pt>[r]_{\pr_2} & \ X \ar[d]^\phi \\
 & M
}
\eeq
is given by a unitary connection $\nabla=\dd +A$ on
$E$ which commutes with the multiplication on $E$. Then there
exists a (not unique) curving $B\in\Omega^2(X)$ such that
\beq
F_\nabla= \delta(B) =\pr_1^*(B)-\pr_2^*(B) \ \in \
\Omega^2\big(X^{[2]} \big) \ .
\label{FnablaX2}\eeq
Since $\delta(\dd B)=\dd\, \delta(B)=\dd F_\nabla=0$, it satisfies $\dd B=\phi^*(H)$ for
a unique three-form $H= H_{(\nabla,B)}$ with
$\varpi=-\frac1{2\pi^2\,\di}\, H\in\Omega^3_{{\rm cl},\RZ}(M)$. The curvature $\varpi$ is
again the de~Rham representative of the Dixmier-Douady class
$dd(\CG)\in H^3(M,\RZ)$, which is
the obstruction to triviality of the
bundle gerbe.

For the Deligne cohomology of bundle gerbes with connective structure, choose a Stein cover
$U=(U_a)_{a\in I}$ of $M$ with local sections
$\psi_a:U_a\to X$ of $\phi:X\to M$, and let
$X_U:=\bigsqcup_a\, U_a$ be the nerve of $U$ with the obvious surjective
submersion $X_U\to M$. Then
$X_U^{[2]}=\bigsqcup_{a,b}\, U_{ab}$, and the sections $\psi_a$
define a fiber preserving map $\psi:X_U\to X$. We use this map to
pullback the bundle gerbe $(E,X)$ to a stably
isomorphic\footnote{The notion of stable isomorphism of bundle gerbes is explained below.} bundle gerbe
$(\psi^*(E), X_U)$, which is just a collection of line bundles
$\psi^*(E)_{ab}\to U_{ab}$ satisfying the cocycle conditions
for a
local gerbe. We can trivialize the line bundles by unit norm
sections $\sigma_{ab}:U_{ab}\to \psi^*(E)_{ab}$ over each overlap. Then we may multiply $\sigma_{ab}$ and
$\sigma_{bc}$ together using the bundle gerbe product; on
$U_{abc}$ we have $\sigma_{ab}\sigma_{bc}= g_{abc}\, \sigma_{ac}$,
and associativity implies that
$g_{abc}:U_{abc}\to \sU(1)$ is a \v{C}ech two-cocycle. We also
define $A_{ab}\in\Omega^1(U_{ab})$ by
$A_{ab}=(\psi_a,\psi_b)^*(A)$, so that
$\nabla\sigma_{ab}=A_{ab}\otimes \sigma_{ab}$,
and $B_a\in\Omega^2(U_a)$ by
$B_a=\psi_a^*(B)$. From (\ref{FnablaX2}) the cocycle
conditions (\ref{2cocycleconds}) then easily follow.

\paragraph{Example.} Let $M=\FT^d=V/\Lambda$ be a $d$-dimensional torus, where $V=\FR^d$ is a
real vector space of dimension $d$ and $\Lambda=\RZ^d$ a lattice in
$V$ of maximal rank. Let $X=V$, and let $\phi:V\to\FT^d$ be the
universal cover of $\FT^d$; it has disconnected fibers whose connected components
are labeled by ``winding numbers'' $w=(w^1,\dots,w^d)\in\Lambda$. We
write $v=(v^1,\dots,v^d)$ for a vector in $V$. Then $(u,v)\in V^{[2]}$
if and only if $w:= u-v\in\Lambda$, so there is a disconnected union
\beq
V^{[2]}=\bigsqcup_{w\in\Lambda}\, V_w
\eeq
with connected components $V_w\cong V$ for all $w\in\Lambda$. Define a
line bundle $E\to V^{[2]}$ by $E:=\bigsqcup_{w\in\Lambda}\, E_w$, with
$E_w\cong V\times\FC$ the trivial line bundle over each connected
component endowed with trivial fiberwise product induced by the
multiplication in $\FC$. Then
$\CG(\FT^d)=(E,V)$ is a bundle gerbe on $\FT^d$. If $\dd x^i$, $i=1,\dots,d$, denotes the basis for $H^1(\FT^d,\RZ)=\Lambda^*$ dual to
a canonical basis of one-cycles $\gamma_i$ for
$H_1(\FT^d,\RZ)=\Lambda$, i.e., $\oint_{\gamma_i}\, \dd x^j=
\delta_i{}^j$, then the pullback by either of the projections
$\pr_1,\pr_2: V^{[2]}\to V$ of the basis forms $\dd v^i:=\phi^*(\dd
x^i)$ will also be denoted $\dd v^i$ on each connected component $V_w$
(note that $\dd u^i=\dd v^i$ for $(u,v)\in
V^{[2]}$). For each winding sector $w\in\Lambda$, endow the line bundle
$E_w\to V$ with connection $\nabla_w=\dd+A_w$ given by
\beq
A_w=\tfrac{1}{3!}\, \varpi_{ijk}\, w^i\, v^j\, \dd v^k
\eeq
where $\varpi_{ijk}$ is a totally anti-symmetric constant three-tensor.
Then the collection of one-forms $A=(A_w)_{w\in\Lambda}$ are part of a connective structure for the bundle gerbe $\CG(\FT^d)$. One has $F_{\nabla_w}=\dd A_w=\delta_w(B)$, where $\delta_w$ is the
restriction of the image of the map $\delta$ to $\Omega^2(V)\to
\Omega^2(V_w)$, and the curving $B\in\Omega^2(V)$ is given by
\beq
B=\tfrac{1}{3!}\, \varpi_{ijk}\, v^i\, \dd v^j\wedge \dd v^k \ .
\eeq
Since $\dd B=\tfrac{1}{3!}\, \varpi_{ijk}\, \dd v^i \wedge\dd v^j\wedge
\dd v^k$, the curvature $H\in\Omega^3(\FT^d)$ reads as
\beq
H=\tfrac{1}{3!}\, \varpi_{ijk}\, \dd x^i \wedge \dd x^j \wedge \dd x^j \ .
\label{T3Hflux}\eeq
When $d=3$ and $\varpi_{ijk}=\varepsilon_{ijk}$, then
$H\in \clidf^3(\FT^3)$ and the Dixmier-Douady class of the bundle gerbe
$\CG(\FT^3)$ is the natural
generator of $H^3(\FT^3,\RZ)=\RZ$.

\paragraph{Stable isomorphism.} For any line bundle $\mbf T\to X$ we define a bundle gerbe
\beq
\xymatrix@C=10mm{
\delta(\mbf T) \ar[d] & {\mbf T} \ar[d] \\
X^{[2]} \ \ar@< 2pt>[r]^{\pr_1} \ar@< -2pt>[r]_{\pr_2} & \ X \ar[d]^\phi \\
 & M
}
\eeq
where $\delta(\mbf T)=\pr_1^*(\mbf T)\otimes \pr_2^*(\mbf T^*)$, i.e.,
$\delta(\mbf T)_{(x_1,x_2)}= \mbf T_{x_2}\otimes \mbf T^*_{x_1}$. A
bundle gerbe $\CG$ over $M$ which is isomorphic to a bundle gerbe of the form
$(\delta(\mbfT),X)$ has vanishing Dixmier-Douady class $dd(\CG)=0$ and is said to be
\emph{trivial}. The gerbe $(\delta(\mbfT),X)$ has a natural connective structure
provided by choosing a connection $\nablatr$ on $\mbfT\to X$, and
taking the induced connection
$\nabla:=\delta(\nablatr)=\pr_1^*(\nablatr)-\pr_2^*(\nablatr)$ on
$\delta(\mbfT)\to X^{[2]}$ with curving $B=F_\nablatr\in\Omega^2(X)$; then $\dd
B=0$ so $(\delta(\mbfT),X)$ is a flat gerbe, i.e., $H=0$. More generally, a
connective structure on a trivial gerbe is specified by a two-form
$b\in\Omega^2(M)$ such that
\beq
B=F_\nablatr + \pi^*(b) \ .
\eeq
Its curvature is
\beq
H=\dd b \ ,
\eeq
and the Deligne two-cocycle $(g_{abc},A_{ab}, B_a)$ of
this connective structure is a gauge transform of the trivial two-cocycle
$(1,0,b)$ by a Deligne two-coboundary $\mathbf{d}(h_{ab},a_a)$, cf.\ \eqref{eq:GaugeTrafoGerbe},
\beq
g_{abc}= h_{ab}\, h_{bc}\,
h_{ca} \ , \qquad
A_{ab}=a_a-a_b+\frac1{2\pi\, \di}\, \dd
\log h_{ab} \eand B_a=b+\dd a_a~,
\eeq
where $(h,a)$ is the Deligne one-cocycle of the line
bundle $(\mbfT,\nablatr)$. The gauge equivalence classes of connective structures on
the trivial gerbe thus form the group $\Omega^2(M)/\clidf^2(M)$ of
topologically trivial $B$-fields on $M$.

A \emph{stable isomorphism} between bundle gerbes $\CG=(E,X)$
and $\CG'=(E',X'\, )$ with connective structure is a trivialization of the product
$\CG^*\otimes \CG' := (E^*\otimes E',X\times_M X'\,)$ as a gerbe with
connective structure; their connections coincide $\nabla'=\nabla$, while their
$B$-fields and $H$-fluxes are related by $B'=B+F_\nablatr+ \phi^*(b)$ and
$H'=H+\dd b$ for some two-form $b\in\Omega^2(M)$. The set of stable isomorphism classes of bundle gerbes
with connective structure is the Cheeger-Simons differential cohomology group~$\hat
H^3(M)$.

\subsection{Courant algebroids}\label{app:C}

In this appendix we collect relevant definitions and results
concerning Courant algebroids, which are the appropriate
generalizations of Lie algebroids arising in higher quantization. A more detailed overview of this material can
be found e.g.\ in~\cite{Roytenberg:1999aa}.

\paragraph{Courant algebroids.} Courant algebroids are symplectic Lie 2-algebroids \cite{Roytenberg:0712.3461}. Explicitly, consider a vector bundle $E\rightarrow M$ over a smooth manifold $M$
equipped with a non-degenerate symmetric bilinear form $\langle-,-\rangle$ and a skew-symmetric bracket
\begin{equation}
[-,-]_E \,:\, \CC^\infty(M,E)\otimes \CC^\infty(M,E) \ \longrightarrow
\ \CC^\infty(M,E)~,
\end{equation}
together with an anchor map $\rho:E\rightarrow TM$. We define further the {\em Jacobiator} $J:\CC^\infty(M,E)\otimes\CC^\infty(M,E)\otimes\CC^\infty(M,E)\rightarrow\CC^\infty(M,E)$ by
\begin{equation}
 J(\psi_1,\psi_2,\psi_3):=\big[[\psi_1,\psi_2]_E\,,\,\psi_3\big]_E+\big[[\psi_2,\psi_3]_E \,,\,\psi_1\big]_E+\big[[\psi_3,\psi_1]_E \,,\,\psi_2\big]_E~,
\end{equation}
a ternary map $T:\CC^\infty(M,E)\otimes\CC^\infty(M,E)\otimes\CC^\infty(M,E)\rightarrow \CC^\infty(M)$ by
\begin{equation}
 T(\psi_1,\psi_2,\psi_3)=\big\langle[\psi_1,\psi_2]_E\,,\,\psi_3\big\rangle+\big\langle[\psi_2,\psi_3]_E\,,\,\psi_1\big\rangle+\big\langle[\psi_3,\psi_1]_E\,,\, \psi_2\big\rangle~,
\end{equation}
and the pullback $\CD:\CC^\infty(M)\to \CC^\infty(M,E)$ of the exterior derivative $\dd$ via the
adjoint map $\rho^*$ by
\begin{equation}
 \langle \CD f,\psi \rangle=\rho_\psi(f) \ ,
\end{equation}
where $f\in \CC^\infty(M)$ and $\psi,\psi_i\in\CC^\infty(M,E)$.

Such a vector bundle is called a {\em Courant algebroid}~\cite{Liu:1997aa} if the following conditions are satisfied:
\begin{conditions}
 \item[(i)] The Jacobi identity holds up to an exact expression: \ $J(\psi_1,\psi_2,\psi_3)=\CD T(\psi_1,\psi_2,\psi_3)$;
 \item[(ii)] The anchor map $\rho$ is compatible with the bracket: \
   $\rho_{[\psi_1,\psi_2]_E} =[\rho_{\psi_1} ,\rho_{\psi_2} ]_{TM} $;
 \item[(iii)] There is a Leibniz rule: \ $[\psi_1,f\, \psi_2]_E=f\,
   [\psi_1,\psi_2]_E+\rho_{\psi_1}( f)\, \psi_2-\tfrac{1}{2}\, \langle
   \psi_1,\psi_2\rangle \, \CD f$;
 \item[(iv)] $\langle \CD f,\CD g\rangle=0$;
 \item[(v)] $\rho_\psi\big(\langle \psi_1,\psi_2\rangle \big) =\big\langle[\psi
   ,\psi_1]_E+\tfrac{1}{2}\, \CD\langle \psi
   ,\psi_1\rangle\,,\,\psi_2 \big\rangle+ \big\langle \psi_1 \,,\,[\psi
   ,\psi_2]_E+\tfrac{1}{2}\, \CD\langle \psi ,\psi_2\rangle \big\rangle$;
\end{conditions}
where $\psi,\psi_i \in\CC^\infty(M,E)$ and
$f,g\in\CC^\infty(M)$.

An involutive Lagrangian sub-bundle $D\subseteq E$, i.e., $D\cong
D^\perp$ with respect to the metric $\langle-,-\rangle$, is called a
\emph{Dirac structure} along $M$. The integrability
condition for a Dirac structure is the requirement that its space of
sections $\CC^\infty(M,D)$ is closed under the bracket $[-,-]_E$ of the Courant
algebroid. Dirac structures provide a generalized and unified
framework in which to treat symplectic and Poisson structures on manifolds.

\paragraph{Exact Courant algebroids.} The Courant algebroid we are
exclusively interested in is given by the {\em standard Courant algebroid}
structure on the Pontryagin bundle $C=TM\oplus T^*M$.
On sections of $C$, we define the {\em Dorfman bracket}
\begin{equation}
 (X_1,\alpha_1)\circ(X_2,\alpha_2):=[X_1,X_2]_{TM}
 +\CL_{X_1}\alpha_2-\iota_{X_2}\, \dd\alpha_1~.
\end{equation}
This bracket is convenient, as it satisfies a Leibniz rule:
$\psi_1\circ(\psi_2\circ \psi_3)=(\psi_1 \circ \psi_2)\circ \psi_3
+\psi_2\circ(\psi_1 \circ \psi_3)$ for $\psi_i \in
\CC^\infty(M,C)$. We are, however, more interested in the {\em
  Courant(-Dorfman) bracket} which is the skew-symmetrization of the
Dorfman bracket given by
\begin{equation}
 \big[(X_1, \alpha_1)\,,\,(X_2,\alpha_2)\big ]:=\big([X_1,X_2]_{TM} \,,\,
 \CL_{X_1}\alpha_2-\CL_{X_2}\alpha_1-\tfrac{1}{2}\, \dd
 (\iota_{X_1}\alpha_2-\iota_{X_2}\alpha_1) \big)~.
\end{equation}
The Courant algebroid structure on $C$ is given by the Courant bracket, the metric induced by the natural pairing between $TM$ and $T^*M$,
\begin{equation}
 \big\langle (X_1,\alpha_1)\,,\, (X_2,\alpha_2)\big\rangle=\iota_{X_1}\alpha_2+\iota_{X_2}\alpha_1~,
\end{equation}
and the anchor map is the trivial projection $\rho:C\rightarrow
TM$; the map $\CD:\CC^\infty(M)\to \CC^\infty(M,C)$ is given by $\CD
f=\frac12\, \dd f$. This is an {\em exact Courant algebroid}, i.e., it fits into the short exact sequence
\begin{equation}
 0\ \longrightarrow \ T^*M \ \stackrel{\rho^*}{\longrightarrow} \ C\
 \stackrel{\rho}{\longrightarrow} \ TM\ \longrightarrow \ 0~.
\end{equation}
The bundle metric on $C$ is of split signature, and the cotangent
bundle $T^*M$ is a Lagrangian sub-bundle defining a Dirac structure
along $M$.

Conversely, any closed three-form $\varpi$ on $M$ yields an exact Courant
algebroid with a Lagrangian splitting: We endow $C=TM\oplus T^*M$
with the structure of a Courant algebroid with the $\varpi$-twisted
Courant bracket given by~\cite{Severa:2001qm}
\beq
 \big[(X_1, \alpha_1)\,,\,(X_2,\alpha_2)\big ]_\varpi &:=& \big([X_1,X_2]_{TM}
 \,,\,
 \CL_{X_1}\alpha_2-\CL_{X_2}\alpha_1 \\ && \qquad \qquad
 \qquad -\, \tfrac{1}{2}\, \dd
 (\iota_{X_1}\alpha_2-\iota_{X_2}\alpha_1)+\iota_{X_1}\, \iota_{X_2} \varpi
 \big) \ , \nonumber
\eeq
and the remaining structure maps given as above. Every exact Courant algebroid on $M$ is
isomorphic to one of this form; given a Lagrangian splitting
$\lambda:TM\to C$, the closed three-form
$\varpi\in Z^3(M)$ is obtained as
\beq
\varpi(X_1,X_2,X_3)=\big\langle [\lambda(X_1),\lambda(X_2)]_C \,,\,\lambda(X_3)\big\rangle \ .
\eeq
Given two closed three-forms $\varpi$ and
$\varpi'$, the associated exact Courant algebroids are isomorphic if and
only if $[\varpi]=[\varpi'\,]$ in $H^3(M,\FR)$. Thus exact Courant algebroids over a
manifold $M$ are classified by the degree three cohomology
$H^3(M,\FR)$. In particular, given a bundle gerbe with connective
structure encoded in a Deligne two-cocycle $[(g_{abc},A_{ab}, B_a)]$
with respect to a Stein cover $U=(U_a\to M)$ of $M$,
we put the standard Courant algebroid $C_a:= TU_a\oplus T^*U_a$ on
each patch $U_a$ and glue them together using $\dd A_{ab}$; this gives
an exact Courant algebroid $C\to M$ with a splitting $\lambda:TM\to C$
given by the curving $(B_a)$. Under this correspondence, a Dirac
structure $D\subset C$ along $M$ corresponds to a canonical flat
$D$-connection on the associated gerbe.

For details on how to integrate exact Courant algebroids, see
\cite{LiBland:2011aa,Sheng:1103.5920} and references therein. Polarization in this
context should be described as a lift of a Dirac structure $D$ of the
standard Courant algebroid $C=TM\oplus T^*M$ to an integrating Courant
groupoid $TG\oplus T^*G$ over $TM\oplus A^*(G)$, where
$G\rightrightarrows M$ is a Lie groupoid. Note that the vector
bundle $D\to M$ is naturally a Lie algebroid with bracket given by the
Courant bracket and anchor map the restriction to $D$ of the canonical
projection $E\to TM$; the general problem of integration of Dirac
structures was solved in~\cite{Bursztyn:0303180}.

\paragraph{Lie 2-algebras.} Given an exact Courant algebroid $C$
associated to a 2-plectic manifold, we can define an associated Lie
2-algebra\footnote{Recall that a (semistrict) Lie 2-algebra is a 2-term $L_\infty$-algebra.} $L_\infty(C)$ as
follows~\cite{Roytenberg:0203110,Roytenberg:0712.3461}: The underlying
graded vector space is
\begin{equation}
 L_\infty(C)=L_{-1}\oplus L_0\ewith L_0=\CC^\infty(M,C) \eand
 L_{-1}=\CC^\infty(M) ~,
\end{equation}
endowed with maps
\begin{equation}
\begin{aligned}
 \mu_1(f+\psi)&:=\CD f~, \\[4pt]
 \mu_2(f_1+\psi_1,f_2+\psi_2)&:=[\psi_1,\psi_2]_E+\tfrac{1}{2}\,
 \big(\langle \psi_1,\CD f_2\rangle-\langle \CD
 f_1,\psi_2\rangle\big)~, \\[4pt]
 \mu_3(f_1+\psi_1,f_2+\psi_2,f_3+\psi_3)&:=-T(\psi_1,\psi_2,\psi_3)~,
\end{aligned}
\end{equation}
where $f,f_i\in \CC^\infty(M)$ and $\psi, \psi_i\in \CC^\infty(M,C)$.

\paragraph{Example.} Let $M=\FT^d$ be the $d$-dimensional torus with
the constant
$H$-flux (\ref{T3Hflux}). In the
local coordinates $(x^i)$ for $\FT^d$ used in \eqref{T3Hflux}, a natural frame for
$TM\oplus T^*M$ is given by
\beq
X_i=\frac\partial{\partial x^i} \eand P^i=\dd
x^i \ .
\eeq
Writing $X_i$ for $(X_i,0)$ and $P^i$ for
$(0,P^i)$ for simplicity, the metric is given by
\beq
\langle X_i,P^j\rangle=\delta_i{}^j \ .
\eeq
The corresponding twisted Courant-Dorfman algebra is isomorphic to the
$d$-dimensional Heisenberg algebra, realized as the 2-step
nilpotent Lie algebra of rank~$d$ with the non-trivial brackets
\beq
[X_i,X_j]_H=\varpi_{ijk}\, P^k \ .
\eeq
Together with the remaining non-trivial structure map
\beq
T(X_i,X_j,X_k)= 3\,\varpi_{ijk} \ ,
\eeq
this yields the Lie 2-algebra describing the
nonassociative deformation of the closed string phase space by a constant
background $H$-flux in the $R$-space duality frame~\cite{Lust:2010iy,Condeescu:2012sp,Mylonas:2012pg}.

\paragraph{Generalized geometry.}

In \emph{generalized geometry}, the standard Courant algebroid
$C= TM\oplus T^*M$ is
called the \emph{generalized tangent bundle} and it is a replacement for
the ordinary tangent bundle. The Courant-Dorfman bracket is similarly
thought of as a replacement for the ordinary commutator of vector
fields. Just as the group of diffeomorphisms ${\sf Diff}(M)$ acts as
bundle maps of $TM$ preserving the Lie bracket $[-,-]_{TM}$, the group
${\sf Diff}(M)\ltimes \Omega_{\rm cl}^2(M)$ acts as bundle morphisms of $C$
preserving the Courant algebroid structure.

More generally, for any Courant algebroid $E\to M$ we define a
\emph{generalized complex structure} to be a vector bundle map $J:E\to
E$ with $J^2=-\unit_E$ which preserves the fiberwise metric and
satisfies an integrability condition. The map $J$ can be equivalently
encoded by a complex Dirac structure $D\subset E\otimes\FC$ which is transverse
to its complex conjugate $\overline{D}$: The bundle $D$ corresponds to
the $+\di$-eigenbundle of the complexification of $J$ and
$\overline{D}$ to the $-\di$-eigenbundle. Then $E\otimes\FC$ is a
complex Courant algebroid with a splitting
$E\otimes\FC=D\oplus\overline{D}$ into Dirac sub-bundles. This induces
flat $\overline{D}$-connections on the corresponding gerbes which may
be thought of as generalized holomorphic structures in
prequantization, see~\cite{Gualtieri:1007.3485}.

\bibliographystyle{latexeu}

\end{document}